\documentclass[aps,amsmath,amssymb,prd,showpacs,floatfix,preprint,superscriptaddress,nofootinbib,12pt]{article}
\pdfoutput=1
\usepackage{jheppub}
\usepackage{float}
\usepackage{verbatim}
\usepackage[section]{placeins}
\usepackage{bm}
\usepackage{mathtools}
\usepackage{tcolorbox}
\usepackage{cancel}
\usepackage[utf8]{inputenc}
\usepackage{slashed,verbatim}
\pdfoutput=1
\usepackage{comment}
\usepackage{epigraph,lipsum}
\usepackage{fancyhdr}
\usepackage{epigraph}
\usepackage{cancel}
\usepackage{graphicx}
\makeatletter
\def\@fpheader{\relax}
\makeatother

\usepackage{lipsum}

\usepackage{wasysym}
\usepackage{amssymb}

\newcommand\blfootnote[1]{%
  \begingroup
  \renewcommand\thefootnote{}\footnote{#1}%
  \addtocounter{footnote}{-1}%
  \endgroup
}

\usepackage{bbding}
\usepackage{pifont}
\usepackage{gensymb}

\usepackage[framemethod=tikz]{mdframed}

\usepackage{amsmath,epsfig}
\usepackage{amssymb,amsfonts}
\usepackage{latexsym}
\usepackage{graphicx}

\usepackage{subeqnarray}
\usepackage{xcolor}

\usepackage{graphicx}
\usepackage{ulem}

\def\be{\begin{equation}}
\def\ee{\end{equation}}
\def\bea{\begin{eqnarray}}
\def\eea{\end{eqnarray}}

\newcommand\fverb{\setbox\pippobox=\hbox\bgroup\verb}
\newcommand\fverbdo{\egroup\medskip\noindent%
                        \fbox{\unhbox\pippobox}\ }
\newcommand\fverbit{\egroup\item[\fbox{\unhbox\pippobox}]}

\newcommand{\bear}{\begin{eqnarray}}

\newcommand{\eear}{\end{eqnarray}}

\newcommand{\bsea}{\begin{subeqnarray}}
\newcommand{\esea}{\end{subeqnarray}}
\newbox\pippobox

\def\6{\partial}

\newcommand{\comments}[1]{}

\input Starburst.fd
\usepackage{txfonts}

\newcommand{\heart}{\ensuremath\varheartsuit}

\allowdisplaybreaks[3]

\preprint{IFT-UAM/CSIC-20-50}

\begin{document}

\title{%
  \Huge  Magnetophonons   \& type-B Goldstones\\ from Hydrodynamics to Holography}
\author[\Large \heart]{Matteo Baggioli}
\affiliation[\heart]{Instituto de Fisica Teorica UAM/CSIC,
c/ Nicolas Cabrera 13-15, Cantoblanco, 28049 Madrid, Spain}
\author[\Diamond]{, Sebastian Grieninger}

\affiliation[\Diamond]{Theoretisch-Physikalisches Institut, Friedrich-Schiller-Universit\"at Jena,
Max-Wien-Platz 1, D-07743 Jena, Germany.}
%,\ddag\affiliation[\ddag]{Institut für Theoretische Physik, Technische Universität Wien, Wiedner Hauptstr. 8-10, A-1040 Vienna, Austria}

\author[\bigvee, \clubsuit]{, Li Li}

\affiliation[\bigvee]{CAS Key Laboratory of Theoretical Physics, Institute of Theoretical Physics,
Chinese Academy of Sciences, Beijing 100190, China}

\affiliation[\clubsuit]{School of Physical Sciences, University of Chinese Academy of Sciences,
No.19A Yuquan Road, Beijing 100049, China}

\emailAdd{matteo.baggioli@uam.es}
\emailAdd{sebastian.grieninger@gmail.com}
\emailAdd{liliphy@itp.ac.cn}

\blfootnote{{\large \heart}\,\,\,\url{https://members.ift.uam-csic.es/matteo.baggioli}}

\abstract{We perform a detailed analysis of a large class of effective holographic models with broken translations at finite charge density and magnetic field. We exhaustively discuss the dispersion relations of the hydrodynamic modes at zero magnetic field and successfully match them to the predictions from charged hydrodynamics. At finite magnetic field, we identify the presence of an expected type-B Goldstone boson $\mathrm{Re}[\omega]\sim k^2$, known as magnetophonon and its gapped partner -- the magnetoplasmon. We discuss their properties in relation to the effective field theory and hydrodynamics expectations. Finally, we compute the optical conductivities and the quasinormal modes at finite magnetic field. We observe that the pinning frequency of the magneto-resonance peak increases with the magnetic field, in agreement with experimental data on certain 2D materials, revealing the quantum nature of the holographic pinning mechanism.}

\maketitle\thispagestyle{empty}
\newpage
 \setcounter{page}{1}
\section{Introduction}
\begin{figure}
    \centering
    \includegraphics[height=5cm]{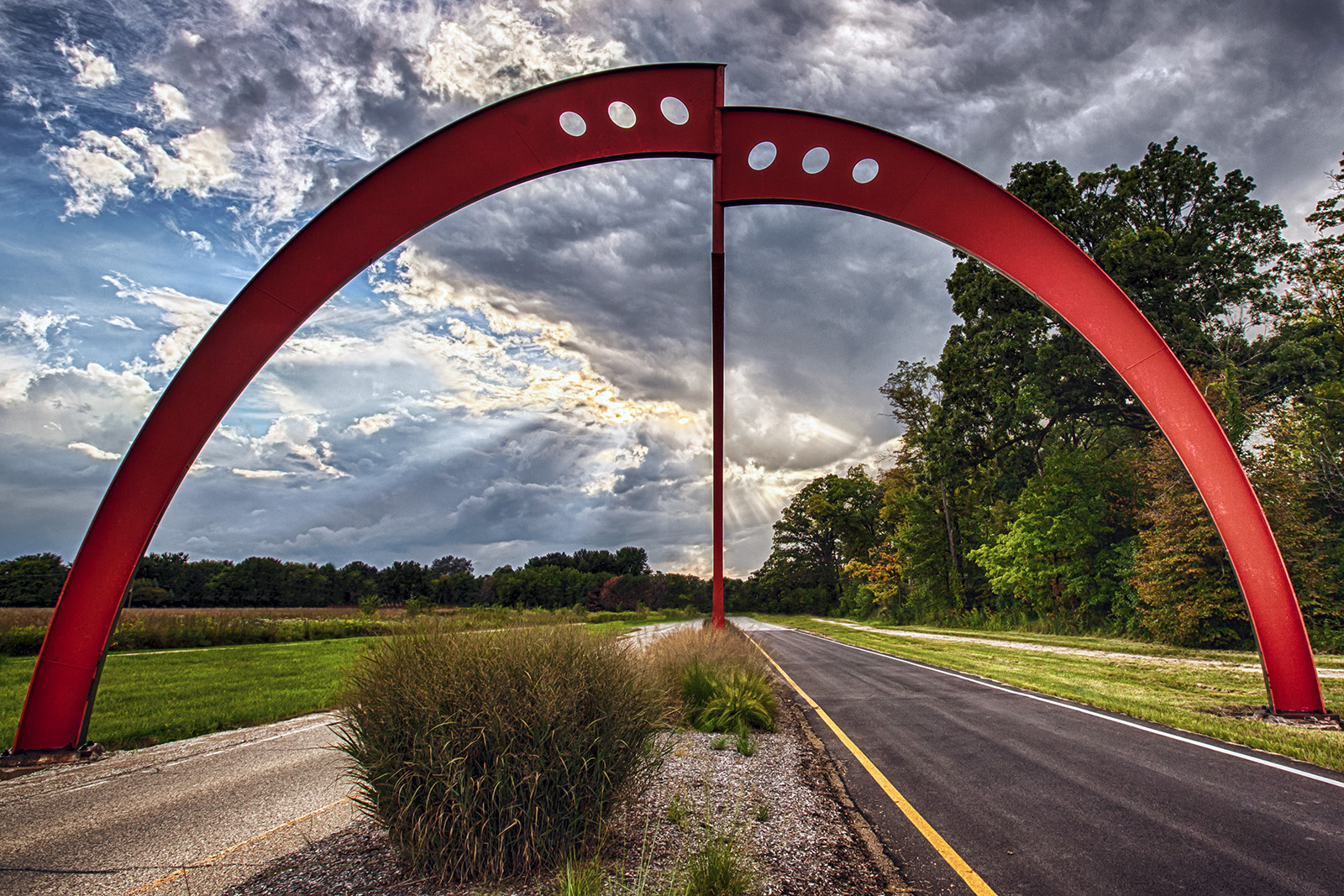}
    \quad
    \includegraphics[height=5cm]{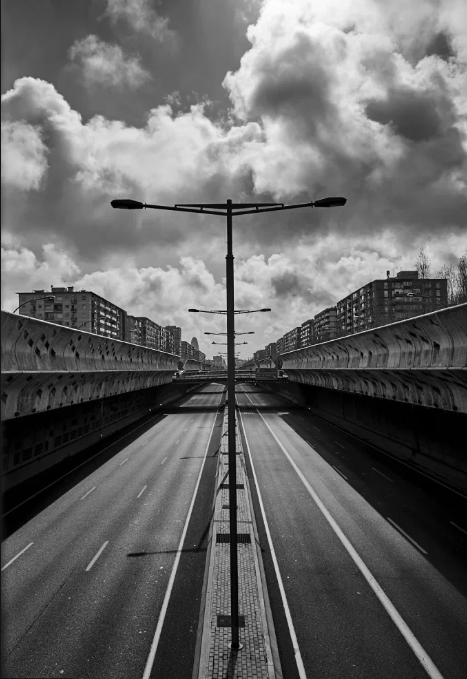}
    \caption{\textbf{Left: }Soft explicit breaking of parity symmetry. The arcs are slightly asymmetric -- soft explicit breaking. \textbf{Right: }Spontaneous breaking of parity symmetry. A day with clouds breaks the parity symmetry of the landscape.}
    \label{fig0}
\end{figure}

\epigraph{Since the beginning of physics, symmetry considerations have provided us with an extremely powerful and useful tool in our effort to understand nature. Gradually they have become the backbone of our theoretical formulation of physical laws.}{\textit{Tsung-Dao Lee}}\noindent
This quote by Tsung-Dao Lee nicely summarizes the fundamental role of symmetries in nature and more importantly in its theoretical description pursued by Physics. Surprisingly enough, the power of symmetries is also extended to situations where symmetries appear to be spontaneously broken~\cite{Beekman:2019pmi} -- non-linearly realized, and %situations where they are 
softly broken~\cite{Burgess:1998ku} (see figure~\ref{fig0}). In these cases, the imprints left by the (broken) symmetries are encoded in the appearance of new, dynamical low energy excitations, known as Goldstone and pseudo-Goldstone modes, respectively. The original argument behind the Goldstone theorem goes back to 1961 and reads: ``\textit{if there is continuous symmetry transformation under which the Lagrangian is invariant, then either the vacuum state is also invariant under the transformation, or there must exist spinless particles of zero mass}''~\cite{PhysRev.127.965}. This idea has been verified and exploited in all possible branches of physics starting from the well-known BCS theory for superconductivity~\cite{PhysRev.117.648} and ending with the physics of the God particle -- the Higgs Boson~\cite{Cho141,Higgs:1964ia,PhysRevLett.13.321}.\\
Importantly, the theorem itself only ensures the existence of a gapless mode:
\begin{equation}
   \lim_{k \,\rightarrow \,0}\, \omega(k)\,\rightarrow\,0\,,
\end{equation}
with $\omega$ and $k$ the frequency and wave-number, respectively. Only the addition of the following assumptions:
\begin{enumerate}
    \item Poincar\'e Invariance,
    \item Internal continuous global symmetries,
    \item Non dissipative systems,
\end{enumerate}
implies a stronger version, which can be formulated as: ``\textit{the breaking of an internal continuous global symmetry guarantees the existence of gapless modes; the number of those modes coincides exactly with the number of broken symmetry generators and their dispersion relation is linear $\omega(k)\sim k$}''.
Within this stronger scenario, the number of Goldstone modes is given by the dimension of the coset space $G/H$, where $G$ is the broken group and $H$ the preserved one:
\begin{equation}
    n_{GB}\,=\,\mathrm{dim}\, G/H\,=\,\mathrm{dim}\,G\,-\,\mathrm{dim}\,H\,. \label{pp}
\end{equation}
This number is nothing else than the counting of the ``flat directions'' of fluctuations of the order parameters.\\

\begin{figure}
    \centering
    \includegraphics[width=0.9\linewidth]{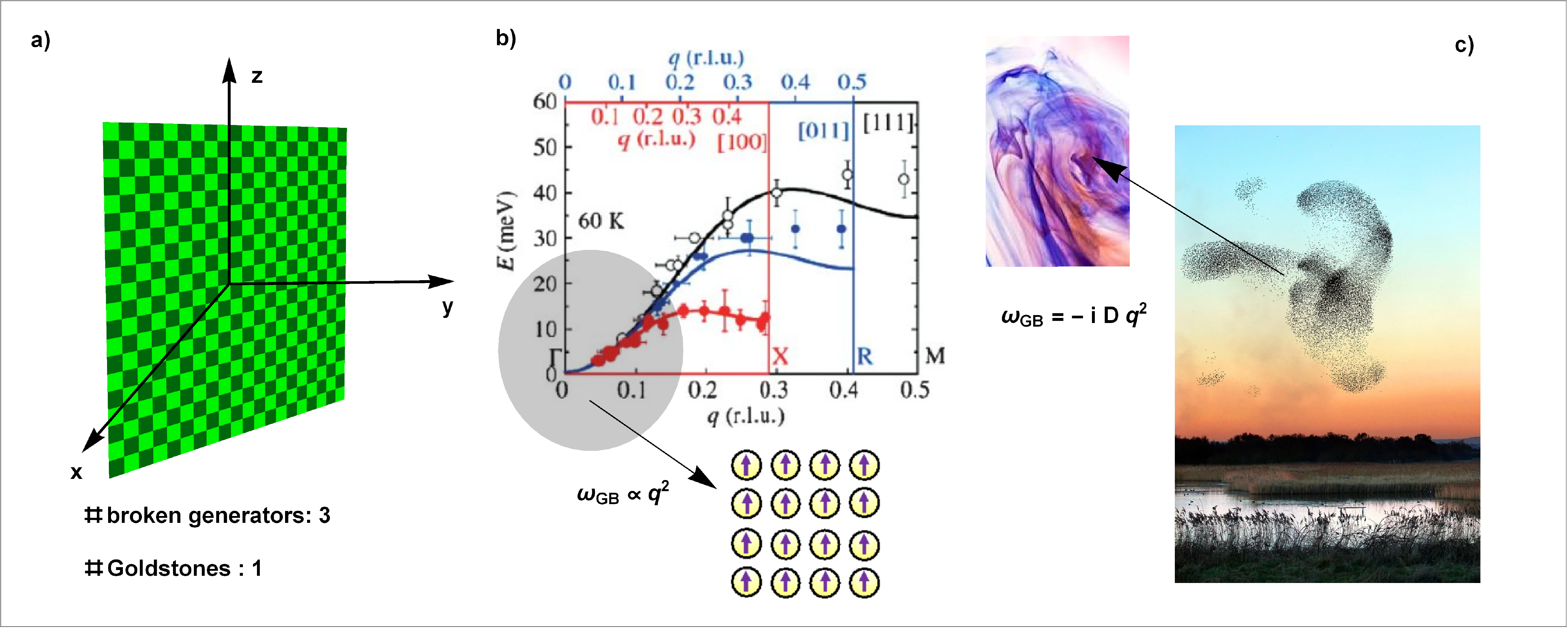}
    \caption{Three different cases of ``anomalous Goldstones''. \textbf{a)} A plane in a fixed position in $3$-dimensional space. There are three broken generators but only one Goldstone mode. \textbf{b)} The case of a ferromagnet. There are two broken generators and only one Goldstone which is quadratic. The data shown are taken from \cite{doi:10.1143/JPSJ.75.111002}. \textbf{c)} A flock of birds. The Goldstone boson has a diffusive dispersion relation \cite{TONER2005170}.}
    \label{fig1}
\end{figure}

Clearly, there is a plethora of physical systems which do not satisfy the requirements above. This leads to very interesting phenomena, which can be summarized as:
\begin{itemize}
    \item The number of Goldstone modes appearing is less than that of broken generators.
    \item The dispersion relation of the Goldstone modes is not linear.
    \item The Goldstone modes are not propagating but rather diffusive.
\end{itemize}
Let us briefly give some explicit examples for each of these situations (see also figure~\ref{fig1}). One setting in which the number of Goldstone modes are less than the broken generators is that of spacetime symmetries \cite{Low:2001bw}, where this anomaly is technically due to what is known as the \textit{Inverse Higgs Constraint}~\cite{Nicolis:2013sga,Endlich:2013vfa}. In simple words, this phenomenon arises since the would be independent Goldstone modes may actually be written as derivatives of other Goldstones and we may integrate them out from the effective low energy description. From a more general point of view, a reduced number of Goldstones, compared to the number of broken generators, is due to the fact that the broken generators $\mathcal{Q}_\alpha,\mathcal{Q}_\beta$ commute with the Hamiltonian but not with one another:
\begin{equation}
    \left[H,\mathcal{Q}_\alpha\right]\,=\,0\,,\quad \left[H,\mathcal{Q}_\beta\right]\,=\,0\,,\quad \text{but}\quad  \left[\mathcal{Q}_\alpha,\mathcal{Q}_\beta\right]\,\neq\,0\,;
\end{equation}
thus the symmetries cannot be thought of as independent.
The emblematic case is the simultaneous breaking of spacetime rotations and translations. The broken generators obey the Poincar\'e algebra:
\begin{equation}
    \left[J_m\,,P_n\right]\,=\,i\,\epsilon_{mnk}\,P_k\,,
\end{equation}
which forces them to not commute. This effect has important consequences, and it is exactly the reason why we do not observe any Goldstone mode for rotations in a crystal\,\footnote{There is an interesting analogous story with the Goldstones for boosts. In the same way, they are ``almost never'' observed. Apparently, if they were, they would be quite different from standard ones (i.e. a continuum)~\cite{Alberte:2020eil}.}. Other simple examples are: (I) a plane in a fixed position
in $3$-dimensional space, $3$ broken generators but only one Goldstone (panel \textbf{a)} of figure~\ref{fig1}); (II) the breaking of conformal invariance in $4$-dimensional spacetime, $5$ broken generators but only one Goldstone -- the dilaton.\\

The second situation (which is strongly connected to the first) refers to physical systems in which the Goldstone bosons have a dispersion relation of the type:
\begin{equation}
    \omega(k)\,\sim\,k^{n}\,\,,\quad \text{with}\quad n\,\neq\,1\,,
\end{equation}
where, in other terms, the Goldstone modes are not linear.
These kind of Goldstone bosons are typical of non-relativistic systems and they have been recently rigorously classified and labelled as type-II or type-B Goldstone modes ~\cite{Watanabe:2011ec,PhysRevLett.108.251602,PhysRevLett.110.091601,Lenz:2020bxk,Pannullo:2019prx}. The fundamental point in this discussion is that the effective low energy description can be written as~\cite{Watanabe:2014fva}:
\begin{equation}
    \mathcal{L}\,=\,\frac{1}{2}\,\rho_{ab}\,\partial_t \pi^a \pi^b\,+\,\frac{1}{2}\,\bar{g}_{ab}\,\partial_t \pi^a \partial_t\pi^b\,-\,\frac{1}{2}\,g_{ab}\,\nabla \pi^a\,\cdot\nabla\pi^b\,+\,\dots\,, \label{et}
\end{equation}
where $\pi^I$ are the Goldstone fluctuations, and $\bar{g}_{ab}$ and $g_{ab}$ are symmetric with respect to $a$ and $b$. The $\rho_{ab}$ is an anti-symmetric matrix referred to as the Watanabe-Brauner matrix and the corresponding first term in \eqref{et} cannot appear in Lorentz invariant systems \cite{Watanabe:2014fva}. More fundamentally, such matrix is given by the commutator of the broken generators:
\begin{equation}
    \rho_{ab}\,=\,-\,i\,\left[\mathcal{Q}_a,\mathcal{Q}_b\right]\,.
\end{equation}
The rank of such matrix determines the number of the different types of Goldstone bosons:
\begin{align}
    & \omega(k)\,=\,k^{2\,n\,-\,1}\,, \quad \text{TYPE A}\,,\\
    & \omega(k)\,=\,k^{2\,n}\,, \quad\quad \text{TYPE B}\,,
\end{align}
where $n$ is an integer\,\footnote{Beforehand, the two type of excitations were labelled with ``type I'' and ``type II''.}.
Within this scenario, the naive counting in eq.\eqref{pp} is violated. More precisely
\begin{equation}
    \mathfrak{n}\,=\,n_{GB}\,-\,\frac{1}{2}\,\mathrm{rank}\,\rho\,,\quad n_B\,=\,\frac{1}{2}\,\mathrm{rank}\,\rho\,,\quad n_A\,=\,n_{GB}\,-\,\mathrm{rank}\,\rho \label{ww}\,.
\end{equation}
For an introductory review about this generalized counting criterion see~\cite{Watanabe:2019xul} .
The most famous example in this category is that of the ferromagnet, in contrast to the antiferromagnet. Both systems break SO$(3)\rightarrow $SO$(2)$ but in the first case we have only one Goldstone mode -- the magnon -- which is quadratic, while in the second case two standard linear Goldstones. The difference is due to the fact that a ferromagnet has a ground state with all the spins aligned and a finite spin density, and it can be explained exactly with the formalism illustrated above (namely the rank of the Watanabe-Brauner matrix being $\neq 0$).

The last case, that of diffusive Goldstones, is more recent. It has been noticed, and then formalized using out-of-equilibrium effective field theory (EFT) methods~\cite{Minami:2018oxl,Hidaka:2019irz,Landry:2019iel}, that in dissipative systems (e.g. open systems) diffusive Goldstone modes can appear:
\begin{equation}
    \omega\,=\,-\,i\,D\,k^2\,.
\end{equation}
These modes can be explained using the same effective field theory formalism as in eq.\eqref{et}, and they are apparently observed in several physical systems~\cite{PhysRevLett.75.4326}\footnote{Recently, diffusive goldstone bosons have been also observed in holographic models \cite{Donos:2019txg,Amoretti:2018tzw,Ammon:2020xyv}. The nature of such modes is still unclear~\cite{Baggioli:2020nay}.}. More exotic subdiffusive modes  (e.g. $\omega\,=\,-\,i\,D\,k^4\,$) can arise within the hydrodynamic theory of fractons \cite{Gromov:2020yoc}.

As hydrodynamics and effective field theory, holography, in particular its bottom-up version, is founded on symmetry principles and the slogan that ``\textit{the global symmetries of the dual field theory corresponds to gauge symmetries / isometries in the bulk gravitational description}'' ~\cite{Baggioli:2019rrs}. Simultaneously, the spontaneous and pseudo-spontaneous symmetry breaking has been discussed and studied in several holographic models, starting from the famous holographic superconductor setup~\cite{Hartnoll:2008vx} to the more recent holographic systems with broken translations~\cite{Alberte:2017oqx}. The topic of anomalous Goldstone bosons has also been investigated in the context of type B modes in~\cite{Amado:2013xya} and in that of diffusive Goldstones in~\cite{Donos:2019txg}.

In this work, we aim at studying a different situation using holographic techniques. In particular, we consider the dynamics of magnetophonon resonances -- Goldstone bosons which appear in systems with spontaneously broken translations at finite magnetic field ~\cite{PhysRevB.18.6245,PhysRevB.46.3920}. The interest is twofold; first, these modes are interesting \textit{per se} because they are another example of type B Goldstone modes with dispersion relation:
\begin{equation}
    \omega\,\sim\,k^2\,.
\end{equation}
In particular, by switching on a magnetic field $B$ we can observe the hybridization between the two linearly propagating Goldstone bosons -- the longitudinal and transverse phonons:
\begin{equation}
    \omega_{\parallel,\perp}\,=\,v_{\parallel,\perp}\,k\,,
\end{equation}
to a single quadratic mode -- the magnetophonon.

On the other side, from a condensed matter perspective, the physics of magnetophonon resonances is particularly appealing in the presence of small explicit breaking of translations. In such case, at zero magnetic field, the would be phonons acquire a pinning frequency $\omega_0$, which follows the Gell-Mann-Oakes-Renner (GMOR) relation~\cite{PhysRev.175.2195}. This pinning frequency manifests itself in a mid-IR peak in the longitudinal conductivity $\mathrm{Re}[\sigma_{xx}]$ (the typical case is that of pinned charge density waves, see~\cite{RevModPhys.60.1129}). At finite (and large) magnetic field, the pinned magnetophonon peak can survive at very low frequencies even in presence of a strong pinning mechanism~\cite{Delacretaz:2019wzh}. Using classical hydrodynamic arguments ~\cite{PhysRevB.18.6245}, the position of the peak in large magnetic fields gets shifted to
\begin{equation}
    \omega_{pk}\,\sim\,\frac{\omega_0^2}{\omega_c}\,,
\end{equation}
where $\omega_c \sim B$ is the cyclotron frequency. Hence, its position decreases linearly with $1/B$, as observed experimentally in certain (but importantly not all) compounds~\cite{Delacretaz:2019wzh}.

Unfortunately, there is no consensus on the scaling of $\omega_{pk}$ as a function of $B$, and several scalings are indeed seen in experiments~\cite{Chen2005QuantumSO}. Moreover, it is not  a universal fact that the magnetophonon frequency $\omega_{pk}$ always decreases with the magnetic field $B$. As we will see, holography is indeed one of the cases where this does not happen. More interestingly and broadly, the dynamics of the magnetophonon peak as a function of the magnetic field can reveal the fundamental nature of the ``disorder'' responsible for its pinning~\cite{Chen2005QuantumSO,PhysRevLett.89.176802,PhysRevLett.93.206805,Chen_2006,2007IJMPB..21.1379C,PhysRevB.89.075310,PhysRevB.92.035121}, and it thus encodes a very valuable insight into the system at hand. In the following section~\ref{magn}, we will provide more details about the physics of magnetophonon resonances.

In summary, in this work, we utilize the recently discussed homogeneous holographic models with broken translations~\cite{Baggioli:2014roa,Alberte:2015isw} to study the dynamics of magnetophonon resonances. First, we study the appearance of this mode and its type B nature, and we compare the holographic results with the hydrodynamic predictions. Secondly, we introduce a small source of explicit breaking. We study in detail the longitudinal and transverse conductivities, and the magnetophonon peak as a function of the magnetic field, charge density and translational breaking strength.
As we will describe in detail, our analysis and, in particular, the dependence of the magnetophonon peak frequency as a function of the external magnetic field $B$ could shed light on the nature of the ``disorder'' introduced by the homogeneous holographic models such as the well-known ``linear-axions model''~\cite{Andrade:2013gsa}. Despite a lot of work on this model and generalizations~\cite{Baggioli:2014roa,Alberte:2015isw,Alberte:2016xja,Baggioli:2016oqk,Amoretti:2016cad,Baggioli:2016rdj,Baggioli:2016pia,Baggioli:2017ojd,Cremonini:2017qwq,Alberte:2017cch,Blauvelt:2017koq,Cremonini:2018kla,Baggioli:2018vfc,Baggioli:2018nnp,Baggioli:2019abx,Esposito:2017qpj,Andrade:2013gsa,Amoretti:2019kuf,Amoretti:2017frz,Amoretti:2018tzw,Amoretti:2019cef,Donos:2018kkm,Donos:2019tmo,Donos:2020viz,Gouteraux:2014hca}, the physical nature of the dual field theories is still not well understood\,\footnote{See, for example, the controversy about the hydrodynamic description of the dual viscoelastic field theory in~\cite{Ammon:2019apj,Ammon:2020xyv} and the discussion in~\cite{Baggioli:2020nay}.}.\\[0.2cm]

\noindent \textbf{Structure of the paper} \\[0.2cm]
The paper is organized as follows: in section \ref{magn} we review the fundamental features of magnetophonon resonances from an EFT and condensed matter point of view; in section \ref{sec:model} we briefly present the holographic model which we consider throughout this work; in section \ref{sec:hydro} we discuss the hydrodynamic description of our holographic model in presence of the spontaneous breaking of translations and finite charge density and zero magnetic field; in section \ref{sec:typeB} we analyze the dispersion relation of the magnetophonons in the absence of pinning, and in particular we focus on their type-B nature; in section \ref{sec:peak} we study the electric conductivities and the dynamics of the pinned magnetophonons in the presence of a small source of explicit breaking of translations; finally, in section \ref{conc} we conclude and discuss the importance of our results and the comparison with experimental data. In addition, in appendix \ref{app0} we review the hydrodynamic framework of~\cite{PhysRevB.18.6245} and in appendix \ref{appHydro} that of~\cite{Armas:2020bmo,Armas:2019sbe}, and in appendices \ref{app1} and \ref{app2} we provide further technical details on the computations.  

\section{A brief history of magnetophonons}
\label{magn}
In this section, we review the fundamental aspects of magnetophonon resonances. For an excellent and more detailed discussion see~\cite{Chen2005QuantumSO}.

\begin{figure}
    \centering
    \includegraphics[width=0.55\linewidth]{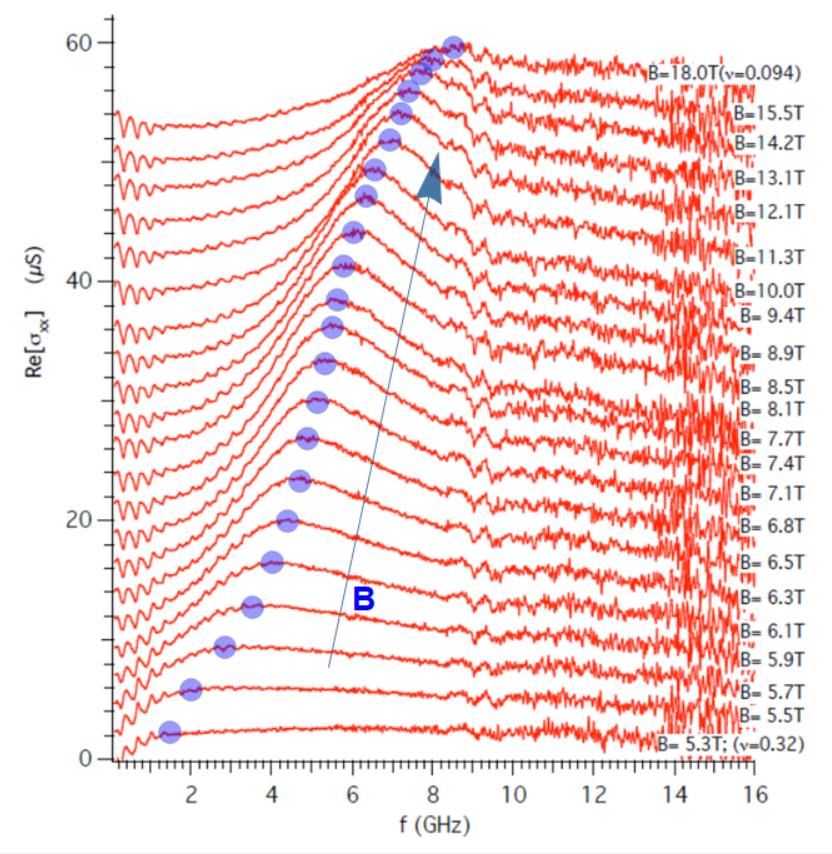}
    \caption{Experimental data for a 15nm wide AlGaAs/GaAs/AlGaAs
quantum well (QW) taken from~\cite{Chen2005QuantumSO}. The behavior as a function of the magnetic field is highlighted. Measured data, and in particular the scaling $\omega_{pk}(B)$ indicate that the disorder
that dominates the pinning in this material is most likely some dilute disorder. We will come back to this point in the conclusions, section~ \ref{conc}. The data and the figure are taken and adapted with permission from \cite{Chen2005QuantumSO}.}
    \label{fig2}
\end{figure}

The history of magnetophonons started in the late 70s when the authors of~\cite{1975JETPL..22...11L,PhysRevB.19.5211} suggested that the presence of a strong magnetic field $B$ in two-dimensional structures would facilitate the formation of Wigner Crystals which are ordered electronic structures appearing at low temperature and low density due to the strong Coulomb interactions. Most of the interest has now shifted to the study of the so called ``pinning mode” resonance, which has been experimentally observed in a plethora of pinned solid phases of two dimensional electron system (2DES)~\cite{Chen2005QuantumSO} (see figure~\ref{fig2} for one concrete example). For a summary of all the experimental observations see references in~\cite{Delacretaz:2019wzh}.

The first point that we should clarify is why the magnetophonon resonances have a quadratic dispersion relation. The reason is simple. In absence of any magnetic field, $B=0$, the momenta (intended as operators), obey the standard Poincar\'e algebra:
\begin{equation}
    \left[P_i\,,\,P_j\right]\,=\,0\,,
\end{equation}
and they commute. This is the reason why longitudinal and transverse phonons decouple. On the contrary, in the case of a finite magnetic field, the algebra is modified and it becomes:
\begin{equation}
     \left[P_i\,,\,P_j\right]\,=\,-\,i\,\epsilon_{ij}\,B\,\mathcal{Q}\,,
\end{equation}
where $\mathcal{Q}$ is the electric charge operator. At the level of the effective action for the Goldstone fluctuations $\pi^I$, the presence of a finite magnetic field allows the appearance of a new term:
\begin{equation}
    \mathcal{L}\,=\,\epsilon^{ij}\,\pi_i\,\partial_t\pi_j\,+\,\dots\,,
\end{equation}
exactly like the one in the formalism of~\cite{Watanabe:2014fva}.
In two spatial dimensions, this implies that the corresponding Watanabe-Brauner matrix $\rho_{ij}$ is now non-trivial, and in particular has $\mathrm{rank}(\rho)=2$. Following the counting rules explained in the previous section, we immediately obtain that:
\begin{equation}
    n_A\,=\,0\,,\quad n_B\,=\,1\,,\quad \longleftarrow \quad \text{magnetophonon}\,.
\end{equation}

Importantly, as we will show explicitly in our holographic theory, the appearance of a type-B goldstone mode is always accompanied by the presence of the so-called ``gapped partner''. More precisely, at finite magnetic field, the two linear propagating sound modes -- Goldstones of translations -- combine into a type-B mode, the \textit{magnetophonon}, and a gapped mode sometimes referred to as the \textit{magnetoplasmon}. Interestingly, under some non-degeneracy assumptions, the number of type-B phonons and the number of gapped partners (sometimes called ``almost-Goldstone bosons'') sum up to the number of broken generators \cite{Kapustin:2012cr}. In our case, the broken generators are the two momenta, $P_x,P_y$. At zero magnetic field, we have two type-A linear Goldstone bosons. At finite magnetic field, we have one type-B magnetophonon and one gapped partner. Either ways, $1+1=2$.
Out of curiosity, a very similar situation arises in the context of vortex-lattices in superfluid, where the quadratic mode is known as the \textit{Tkachenko mode} and the gapped partner as the \textit{Kohn mode} (see figure 2 in~\cite{Moroz:2018noc}).

At this point, we can go a step further and try to understand the dispersion relation of the magnetophonon resonance using hydrodynamic methods~\cite{PhysRevB.18.6245,Delacretaz:2019wzh}. We review the basics of the hydrodynamic description in appendix~\ref{app0}. Here we limit ourselves to present only the main results necessary for our discussion.
At finite magnetic field, the transverse and longitudinal phonons couple together, in a way that the resulting frequencies become
~\cite{PhysRevB.46.3920}:
\begin{equation}
    \omega_{\pm}^2\,=\,\frac{1}{2}\,\left(\omega_c^2\,+\,\omega_\parallel^2\,+\,\omega_\perp^2\right)\,\pm\,\frac{1}{2}\sqrt{\left(\omega_c^2\,+\,\omega_\parallel^2\,+\,\omega_\perp^2\right)^2\,-\,4\,\omega_\perp^2\,\omega_\parallel^2}\,,
\end{equation}
where $\omega_c$ is the cyclotron frequency and $\omega_{\perp,\parallel}$ the frequencies of the linear decoupled phonons.
At zero momentum, $k=0$, $\omega_{\perp,\parallel}=0$ and we are left with two modes:
\begin{equation}
    \omega_{-}\,=\,0\,,\quad \quad  \omega_+\,=\,\omega_c\,.
\end{equation}
The root with the negativ sign is the massless type-B magnetophonon, while the root with the plus sign is the gapped partner -- the magnetoplasmon. At small momentum, $k/T \ll 1$, we get:
\begin{equation}
    \mathrm{Re}\,[\omega_+]\,=\,\omega_c\,+\,\frac{(v_\parallel^2+v_\perp^2)}{2\,\omega_c}\,k^2\,+\,\dots\,,\quad \quad \mathrm{Re}\,[\omega_-]\,=\,\frac{v_\perp\,v_\parallel}{\omega_c}\,k^2\,+\,\dots \label{vv2}\,,
\end{equation}
and we observe the quadratic behavior which we have mentioned.

In the presence of small explicit breaking (e.g. impurities), the dispersion relation of the magneton-phonon gets modified into:
\begin{equation}
  \mathrm{Re}\, [\omega_-(k)]\,=\,\frac{\sqrt{\left(\omega^2_0\,+\,\omega_\perp^2(k)\right)\left(\omega_0^2\,+\,\omega^2_\parallel(k)\right)}}{\omega_c}\,,
\end{equation}
with $\omega_0$ the pinning frequency. This mode acquires a finite gap $\omega_-(k=0)\equiv \omega_{pk}$. We will come back to these dispersion relations in much more detail in section \ref{sec:typeB}.

A last fundamental point, from a more phenomenological perspective, is related to the dependence of the magnetophonon peak $\omega_{pk}$ with respect to the external magnetic field $B$. This observable has a privileged role since it is the easiest to measure accurately and since it can give important information on ``the type'' of disorder in the material. Using hydrodynamics~\cite{PhysRevB.18.6245,PhysRevB.46.3920}, we can infer that the position of the resonance peak is given by:
\begin{equation}
    \omega_{pk}\,=\,\frac{\omega_0^2}{\omega_c}\,.
\end{equation}
This means that a classical treatment of the pinning mechanism~ \cite{PhysRevB.18.6245,PhysRevB.46.3920} would lead to:
\begin{equation}
    \omega_{pk}\,\sim\,\frac{1}{B}\,.
\end{equation}

Unfortunately or interestingly, recent experimental results~\cite{PhysRevLett.79.1353} are in disagreement with this prediction. They instead observed a peak increasing with the magnetic field (see one example in figure~\ref{fig2}). In order to understand these experimental results, in which the peak increases with the magnetic field (and it decreases with the density) one needs to go beyond the ``classical treatment''. To this end, some of the most famous models are those of~\cite{PhysRevB.59.2120,PhysRevB.62.7553,PhysRevB.65.035312}. Very interestingly, the scaling of $\omega_{pk}$ with the magnetic field $B$ depends crucially on the nature of the disorder which produces the pinning, and it can be used as an efficient tool to disentangle various types of disorder. In more detail, the dependence is very sensitive to whether the system is in a classical or quantum regime. This can be quantified using the concepts of magnetic length $l_b=\sqrt{\hbar/e B}$ and disorder correlation length $\xi$. In terms of these quantities:
\begin{align}
    &l_b\,\gg\,\xi\quad \quad \text{quantum regime}\,,\\
    &l_b\,\ll\,\xi\quad \quad \text{classical regime}\,.
\end{align}
Note that the result $\omega_{pk}\sim 1/B$ holds only in the classical regime. In the opposite regime, the results can be very different and the peak can increase with the magnetic field. As an example, in the quantum regime, a model for dilute disorder would give $\omega_{pk}\,\sim\,B$~\cite{PhysRevB.59.2120}, and if specific corrections are taken into account (mainly via numerical simulations) the theory predicts a sublinear increase of the peak frequency:
\begin{equation}
    \omega_{pk}\,\sim\,B^\gamma\,, \quad \quad \text{with}\quad \quad 0<\gamma<1\,.
\end{equation}
For more options and theoretical models see~\cite{Chen2005QuantumSO,Kim_2012,Goerbig_2007,PhysRevB.88.165407,PhysRevLett.110.227402,Ploch_2007,HAMAGUCHI199085,Greenaway_2019,Kumaravadivel_2019} and references therein.

In the following, we will consider a specific holographic model which contains magnetophonon resonances. We will study their features in the case of spontaneously broken translations, and once the mode becomes pinned.

\section{The holographic model}\label{sec:model}
We consider the large class of holographic models introduced in~\cite{Baggioli:2014roa,Alberte:2015isw} and defined by the following four-dimensional bulk action:
\begin{equation}\label{action}
S\,=\, M_P^2\int d^4x \sqrt{-g}
\left[\frac{R}2+\frac{3}{\ell^2}- \, V(X)\,-\,\frac{1}{4}\,F^2\right]\, ,
\end{equation}
where $X \equiv \frac12 \, g^{\mu\nu} \,\partial_\mu \phi^I \partial_\nu \phi^I$ and $F^2\equiv F_{\mu\nu}F^{\mu\nu}$ with the field strength being $F=dA$. The AdS radius $\ell$ and the Planck mass $M_P$ will set to be unity. Given the large amount of works discussing this model, we will be brief (see~\cite{Baggioli:2016rdj,Baggioli:2019rrs} for more details and \cite{Ammon:2015wua} for an introduction to holography).

In order to work in a 2+1 dimensional field theory at finite temperature, we consider a black bran in an asymptotically AdS$_4$ bulk geometry. The metric reads in infalling Eddington–Finkelstein coordinates
\begin{equation}
\label{backg}
ds^2=\frac{1}{u^2} \left[-f(u)\,dt^2-2\,dt\,du + dx^2+dy^2\right]\, ,
\end{equation}
with $u\in [0,u_h]$ the radial holographic direction ranging from the boundary $u=0$ to the horizon, $f(u_h)=0$.
The bulk profile for the scalars is
\begin{equation}
    \phi^I\,=\,\kappa\,x^I\,,\label{eq:bulkscalar}
\end{equation}
which is trivially a solution of the system because of the global shift symmetry $\phi^I\rightarrow \phi^I+b^I$ of the action \eqref{action}.
We introduce both a finite charge density $\rho$ and an external magnetic field $B$ via the gauge field $A_\mu$ in radial gauge $A_u=0$:
\begin{equation}
    A_t\,=\,\mu\,-\,\rho\,u\,,\quad A_x\,=\,\,-\,\frac{B}{2}\,y\,,\quad A_y\,=\,\,\frac{B}{2}\,x\,.
\end{equation}
We furthermore require the temporal component of the gauge field to vanish at the horizon, which implies in the case $u_h=1$ that $\rho=\mu$.
The emblackening factor takes the simple form:
\begin{equation}\label{backf}
f(u)= u^3 \int_u^{u_h} dv\;\left[ \frac{3}{v^4} -\frac{V(\kappa^2\,v^2)}{v^4}\,-\,\frac{(\rho^2+B^2)}{2} \right] \, .
\end{equation}
The corresponding temperature of the dual theory reads:
\begin{equation}
T=-\frac{f'(u_h)}{4\pi}=\frac{6 -  2  V\left(\kappa^2\,u_h^2 \right)\,-\,(\rho^2+B^2)\,u_h^4}{8 \pi\, u_h}\, ,\label{eq:temperature}
\end{equation}
while the entropy density is simply $s=2\pi/u_h^2$ (from now on, $M_p=1$).
In section \ref{sec:hydro} and \ref{sec:typeB}, we consider potentials of the form 
\begin{equation}
    V(X)=m^2\,X^3\,,\label{eq:potspont}
\end{equation}
corresponding to spontaneously broken translations in the dual field theory~\cite{Alberte:2017oqx}.
In section \ref{sec:peak}, we will consider the polynomial potential:
\begin{equation}
    V(X)\,=\,\underbrace{\alpha\,X}_{explicit}\,+\,\underbrace{\beta\,X^3}_{spontaneous}\,.
\end{equation}
The first part of such potential corresponds to an explicit breaking of translational invariance~\cite{Andrade:2013gsa,Baggioli:2018vfc}, while the second part implements its spontaneous breaking~\cite{Alberte:2017oqx,Andrade:2019zey,Ammon:2019apj,Baggioli:2019elg,Baggioli:2019mck,Baggioli:2018bfa}. The combination of the two terms allows to study the pseudo-spontaneous regime, where the breaking is mostly spontaneous~\cite{Alberte:2017cch,Ammon:2019wci,Baggioli:2019abx}. More rigorously, we will always work in the limit:
\begin{equation}
    \text{pseudo-spontaneous regime:}\quad \alpha\ll 1\,,\quad \beta \gg \alpha\,. \label{ps}
\end{equation}
For completeness, let us write down the temperature of the field theory considered under this choice of potential:
\begin{equation}
    T\,=\,\frac{3}{4\,\pi}\,-\,\frac{(\rho^2\,+\,B^2)\,u_h^3}{8\,\pi}\,-\,\frac{\alpha\,\kappa^2\,u_h^2\,+\,\beta\,\kappa^6\,u_h^6}{4\,\pi\,u_h}\,.
\end{equation}
Before proceeding, let us summarize our dimensionless parameters:
\begin{equation}\label{param}
    \left\{\frac{\rho}{T^2}\,,\frac{\kappa}{T}\,,\frac{B}{T^2},\,\alpha\,,\beta\right\}\,.
\end{equation}
Notice that the results in the explicit regime, where $\beta=0$, or eventually $\beta \ll \alpha$, can be found in~\cite{Kim:2015wba}, and they represent a very good check for our numerics.

\section{A hydrodynamic warm-up}\label{sec:hydro}
%%%
Before proceeding to the main points of our paper, we consider a simpler problem which to-date remains still unsolved. More precisely, we investigate the hydrodynamic description of our homogeneous holographic model with spontaneously broken translations at finite charge density. At zero charge density (or alternatively zero chemical potential $\mu=0$), the problem was recently solved in ~\cite{Ammon:2020xyv} after realizing explicitly in~\cite{Ammon:2019apj} that the hydrodynamic description of~\cite{PhysRevB.22.2514,chaikin2000principles,PhysRevA.6.2401,Delacretaz:2017zxd} was lacking fundamental terms to match the holographic results\,\footnote{In order to avoid any misunderstanding, let us be precise and explain this in detail. The hydrodynamic framework of~\cite{PhysRevB.22.2514,chaikin2000principles,PhysRevA.6.2401,Delacretaz:2017zxd} considered systems which are thermodynamically favored, i.e. the crystal pressure $\mathcal{P}=0$. This is the main reason of the disagreement found in~\cite{Ammon:2019apj}. However, even in unstrained configurations with $\mathcal{P}=0$ the temperature derivative of $\mathcal{P}(T)$, neglected in ~\cite{Delacretaz:2017zxd}, is fundamental to achieve the correct hydrodynamic description~\cite{Ammon:2020xyv}. 
}. In this section, we plan to check explicitly if the improved hydrodynamic description of~\cite{Armas:2020bmo} matches the holographic results from our model in the presence of finite charge density. Skipping the technical details which can be found in the original paper~\cite{Armas:2020bmo} and are summarized in our appendix \ref{appHydro}, we immediately jump to the hydrodynamic modes which are expected in the system. To avoid any confusion, in the rest of the paper, we will indicate with the suffix $\perp$ the quantities (e.g. hydrodynamic modes) which belong to the transverse sector (transverse with respect to the choice of the wave-vector $k$) and with $\parallel$ those in the longitudinal spectrum.

Under these assumptions, our system will exhibit the following hydrodynamic modes:
\begin{align}
    & \text{transverse sector:}\quad \omega_\perp\,=\,v_\perp\,k\,-\,\frac{i}{2}\,\Gamma_\perp\,k^2\,,\label{tmode}\\
    &\text{longitudinal sector:}\quad\omega_\parallel\,=\,v_\parallel\,k\,-\,\frac{i}{2}\,\Gamma_\parallel\,k^2\,,\quad \omega^\parallel_{1,2}\,=\,-\,i\,D^\parallel_{1,2}\,k^2\,.\label{lmode}
\end{align}
In the transverse sector we have a single propagating shear sound mode with speed $v_\perp$ together with its attenuation constant $\Gamma_\perp$. The longitudinal sector is more complicated. We have a longitudinal propagating sound mode with speed $v_\parallel$ and attenuation constant $\Gamma_\parallel$ and two additional diffusive modes $\omega^\parallel_{1,2}$ with diffusion constants $D^\parallel_{1,2}$. These last two modes are a combination of the crystal diffusion $D_\phi$ discussed in~\cite{Ammon:2020xyv,Baggioli:2020nay} and the standard charge diffusion $D_q$ (see, for example,~\cite{Ge:2008ak}).

According to the results of~\cite{Armas:2020bmo}, for conformal field theories, the various transport coefficients may be written as:
\begin{align}
    & v_\perp^2\,=\,\frac{G}{\chi_{\pi\pi}}\,,\quad \Gamma_\perp\,=\,\frac{\eta}{\chi_{\pi\pi}}\,+\,\frac{G\,\Pi_f^2}{\sigma\,\chi_{\pi\pi}^2}\,,\\
    & v_\parallel^2\,=\,\frac{1}{2}\,+\,v_\perp^2\,,\quad \Gamma_\parallel\,=\,\frac{\eta}{\chi_{\pi\pi}}\,+\,\frac{\Pi_f^2\,G^2}{\sigma\,\chi_{\pi\pi}^3\,v_L^2}\,,
\end{align}
while the diffusion constant $D^\parallel_{1,2}$ can be found as solutions of the quadratic equation:
\begin{align}
    &\left(D\,-\,\frac{\Pi_f^2}{\sigma}\,\frac{G\,+\,\mathfrak{B}\,-\,\mathcal{P}}{2 \,\chi_{\pi\pi}\,v_\parallel^2\,\left(\Pi_f\,+\,\Pi_l\right)}\right)\,\left(\frac{\Xi\,D}{2\,\left(\Pi_f\,+\,\Pi_l\right)}\,-\,\frac{\sigma_q}{T^2}\right)\,=\,\nonumber\\
    &\,=\,\frac{D}{\sigma}\,\left(\frac{s_f\,q_l\,-\,q_f\,s_l}{\Pi_f\,+\,\Pi_l}\,+\,\frac{\gamma}{T}\right)\,\left(\frac{s_f\,q_l\,-\,q_f\,s_l}{\Pi_f\,+\,\Pi_l}\,-\,\frac{\gamma'}{T}\right)\,.
\label{cc}
\end{align}

\begin{figure}[h]
     \centering
     \includegraphics[width=0.48 \linewidth]{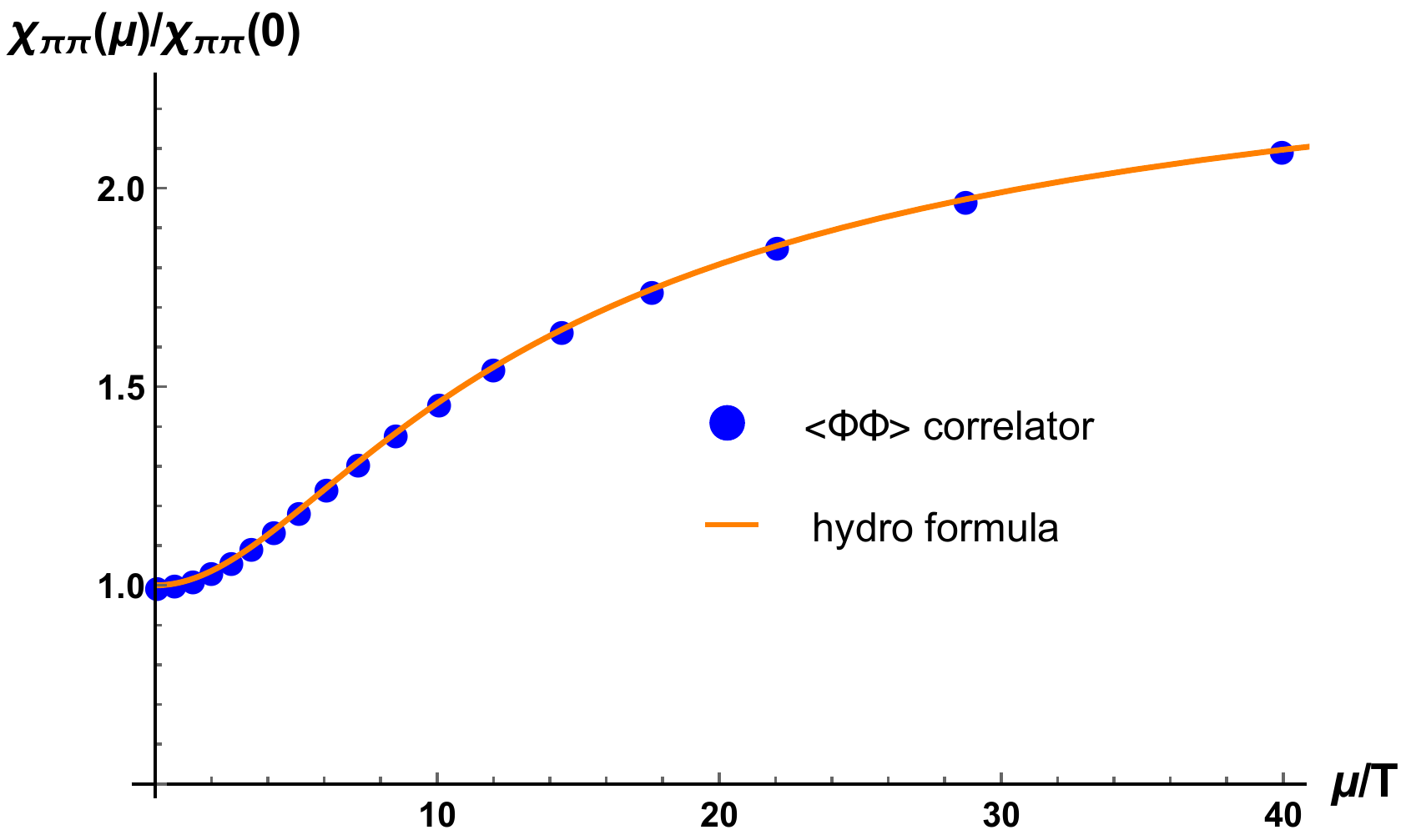} \quad 
     \includegraphics[width=0.45 \linewidth]{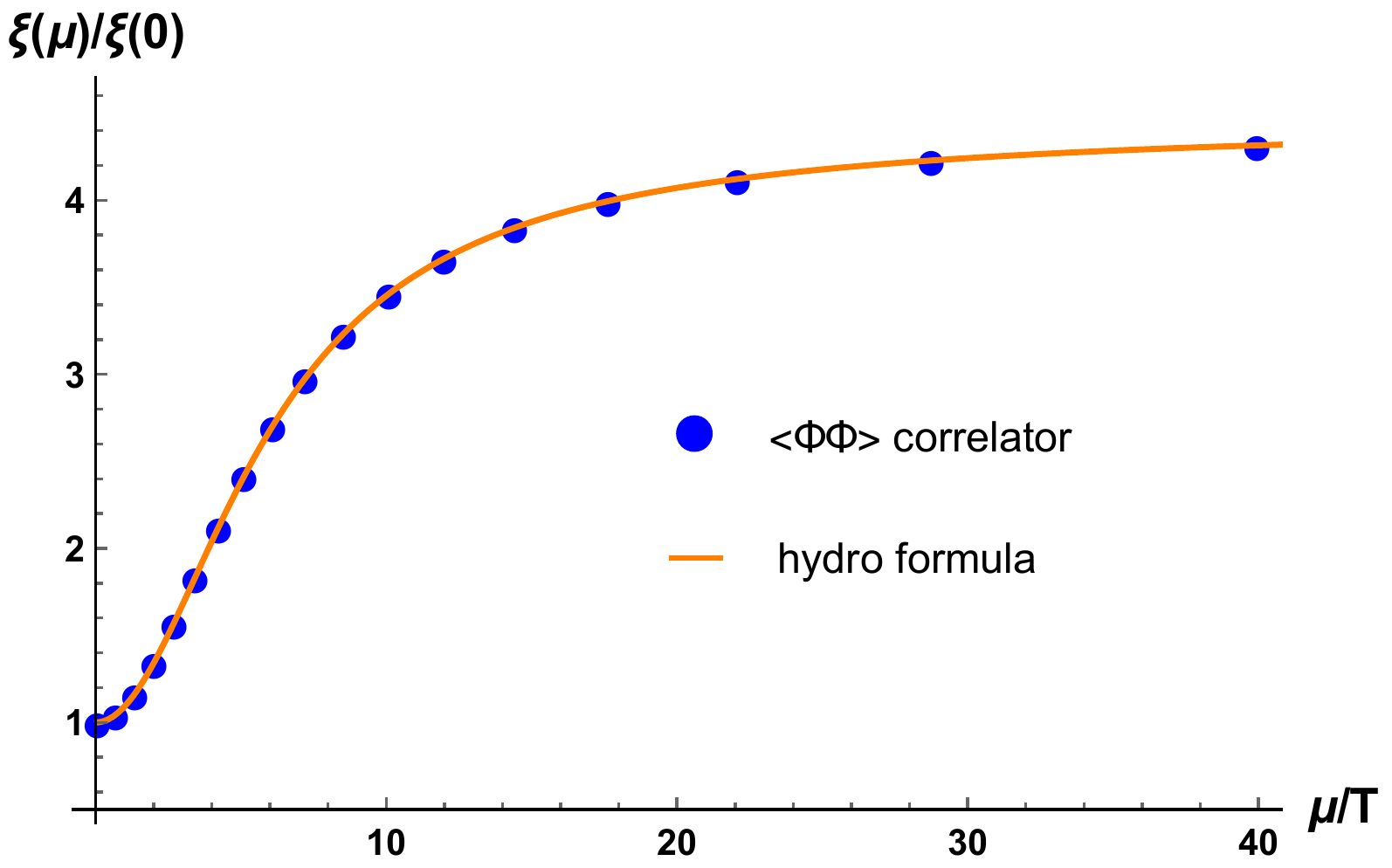}
     \caption{The analysis of the $\langle \Phi \Phi \rangle $ correlator for a benchmark potential $V(X)=m^2 X^3$ at zero magnetic field. The structure follows the hydrodynamic expectations shown in eq.\eqref{corr1}. For simplicity we show only one set of data where we keep $m=1,u_h=1$ fixed. \textbf{Left: }Comparison between the momentum susceptibility $\chi_{\pi\pi}$ extracted from the numerical correlator (blue dots) and the hydrodynamic formula (orange line). \textbf{Right: } Comparison between the dissipative coefficients $\xi$ (which relates to $\sigma$) extracted from the numerical correlator (blue dots) and the hydrodynamic formula (orange line).}
     \label{figphi}
 \end{figure}

Let us explain one by one all the terms entering into the equations above.
We define the hydrodynamic pressure $p$, the energy density $\epsilon$, the lattice pressure $\mathcal{P}$, and the momentum susceptibility $\chi_{\pi\pi}$ as:
\begin{equation}
    p = -\,\Omega\,, \quad
    \epsilon = \langle T^{tt} \rangle\,, \quad
    \mathcal{P} = \langle T^{xx} \rangle \,-\,p\,,\quad \chi_{\pi\pi}\,=\,\langle T^{tt} \rangle\,+\,\langle T^{xx} \rangle\,,
\end{equation}
where $\Omega$ is the free energy density, and $T^{\mu\nu}$ the stress-energy tensor.
The shear modulus $G$, the shear viscosity $\eta$, and the dissipative parameter $\sigma$ can be extracted using the Kubo formulas:
\begin{align}
&    G = \lim_{\omega\to0}\lim_{k\to0}
    \mathrm{Re}\,\mathcal{G}^R_{T^{xy}T^{xy}}\,,
   \quad   \eta = -\lim_{\omega\to0}\lim_{k\to0}
    \frac{1}{\omega}\mathrm{Im}\,\mathcal{G}^R_{T^{xy}T^{xy}}\,,
    \quad   \frac{(\epsilon+p)^2}{\sigma\chi_{\pi\pi}^2} \,=\,\xi
    = \lim_{\omega\to0}\lim_{k\to0}
    \omega\,\mathrm{Im}\,\mathcal{G}^R_{\Phi\Phi}\,,\label{corr1}
\end{align}
 where $\Phi$ is the Goldstone operator dual to the bulk scalars $\phi$\,\footnote{We are not indicating any $\parallel,\perp$ index because we are interested only at the limit $k\rightarrow 0$ where such distinction does not make any sense anymore.}. Moreover, we can use the results of~\cite{Amoretti:2017axe} to obtain a horizon formula for $\xi$ (and thus for $\sigma$), given by:
 \begin{equation}
     \xi\,=\,\frac{4\,\pi\,(\epsilon+p)^2}{2\,m^2\,N\,s\,\chi_{\pi\pi}^2}\,+\,\frac{\mu^2}{\chi_{\pi\pi}^2}\,, \label{fufu}
 \end{equation}
 which is valid for potentials of the form $V(X)=m^2X^N$, and correctly reproduces the results at zero chemical potential presented in~\cite{Ammon:2020xyv}. The expected behavior of the $\langle \Phi\Phi \rangle $ correlator is shown in figure~\ref{figphi} together with a numerical confirmation of formula \eqref{fufu}.
 \\
The bulk modulus $\mathfrak{B}$ can be derived using the following relation:
\begin{equation}
    T\,\frac{\partial \mathcal{P}}{\partial T}\,+\,\mu\,\frac{\partial \mathcal{P}}{\partial \mu}\,=\,3\,\mathcal{P}\,-\,2\,\mathfrak{B}\,,
\end{equation}
which holds due to the conformal symmetry.

In order to derive other parameters, we use the definitions of ~\cite{Armas:2020bmo}:
\begin{equation}
    s_f\,=\,\frac{\partial p}{\partial T}\,,\quad s_l\,=\,\frac{\partial \mathcal{P}}{\partial T}\,,\quad q_f\,=\,\frac{\partial p}{\partial \mu}\,,\quad q_l\,=\,\frac{\partial \mathcal{P}}{\partial \mu}\,,
\end{equation}
together with
\begin{equation}
    \Pi_f\,=\,\epsilon\,+\,p\,=\,s_f\,T\,+\,q_f\,\mu\,,\quad \Pi_l\,=\,\epsilon_l\,+\,\mathcal{P}\,=\,s_l\,T\,+\,q_l\,\mu\,.
\end{equation}
One can check explicitly that $s_f$ and $q_f$ are nothing but the entropy density $s$ and charge density $\rho$ of our system \eqref{backg}, respectively.
Additionally, because of the conformal invariance we have:
\begin{equation}
    \epsilon\,=\,2\,\left(p\,+\,\mathcal{P}\right)\,,\quad \epsilon_l\,=\,2\,\left(\mathcal{P}\,-\,\mathfrak{B}\right)\,.
\end{equation}
Interestingly, we find that $s_l<0$ for any choice of monomial potentials.\\
Finally, the remaining coefficients $\Xi,\sigma_q,\gamma,\gamma'$ can be obtained from
\begin{equation}
    \Xi\,=\,\frac{\partial s_f}{\partial T}\,\frac{\partial q_f}{\partial \mu}\,-\,\frac{\partial s_f}{\partial \mu}\,\frac{\partial q_f}{\partial T}\,,
\end{equation}
and from the low frequency expansion of the following Green's functions:
\begin{equation}
    \mathcal{G}_{J_xJ_x}^R\,=\,\frac{q_f^2}{\chi_{\pi\pi}}\,-\,i\,\omega\,\tilde{\sigma}_q\,,\quad \mathcal{G}_{J_x\phi_x}^R\,=\,-\,\frac{q_f}{i\,\omega\,\chi_{\pi\pi}}\,+\,\tilde \gamma\,,\quad \mathcal{G}_{\phi_xJ_x}^R\,=\,\frac{q_f}{i\,\omega\,\chi_{\pi\pi}}\,+\,\tilde \gamma'\,,\label{corr2}
\end{equation}
where
\begin{align}
    &\tilde \sigma_q\,=\,\sigma_q\,+\,\frac{1}{\sigma}\,\left(\frac{q_f\,\mathcal{P}}{\chi_{\pi\pi}}\,-\,\gamma\right)\,\left(\frac{q_f\,\mathcal{P}}{\chi_{\pi\pi}}\,+\,\gamma'\right)\,,\\
    &\tilde{\gamma}\,=\,\frac{\Pi_f}{\sigma}\,\left(\frac{\gamma}{\chi_{\pi\pi}}-\,\frac{q_f\,\mathcal{P}}{\chi_{\pi\pi}^2}\right)\,,\\
    &\tilde{\gamma}'\,=\,\frac{\Pi_f}{\sigma}\,\left(\frac{\gamma'}{\chi_{\pi\pi}}+\,\frac{q_f\,\mathcal{P}}{\chi_{\pi\pi}^2}\right)\,.
\end{align}
We have checked numerically the structure of the $JJ,J\Phi,\Phi J$ correlators and found perfect agreement with the expressions in eq.\eqref{corr2}. Moreover, we have verified numerically that 
\begin{equation}
    \tilde{\gamma}\,=\,-\,\tilde{\gamma}'\quad \rightarrow\,\quad \gamma\,=\,-\,\gamma'\,,
\end{equation}
which comes simply from the fact that our system is invariant under time-reversal  and it is imposed by so-called Onsager constraints.

\begin{figure}
    \centering
    \includegraphics[width=0.45\linewidth]{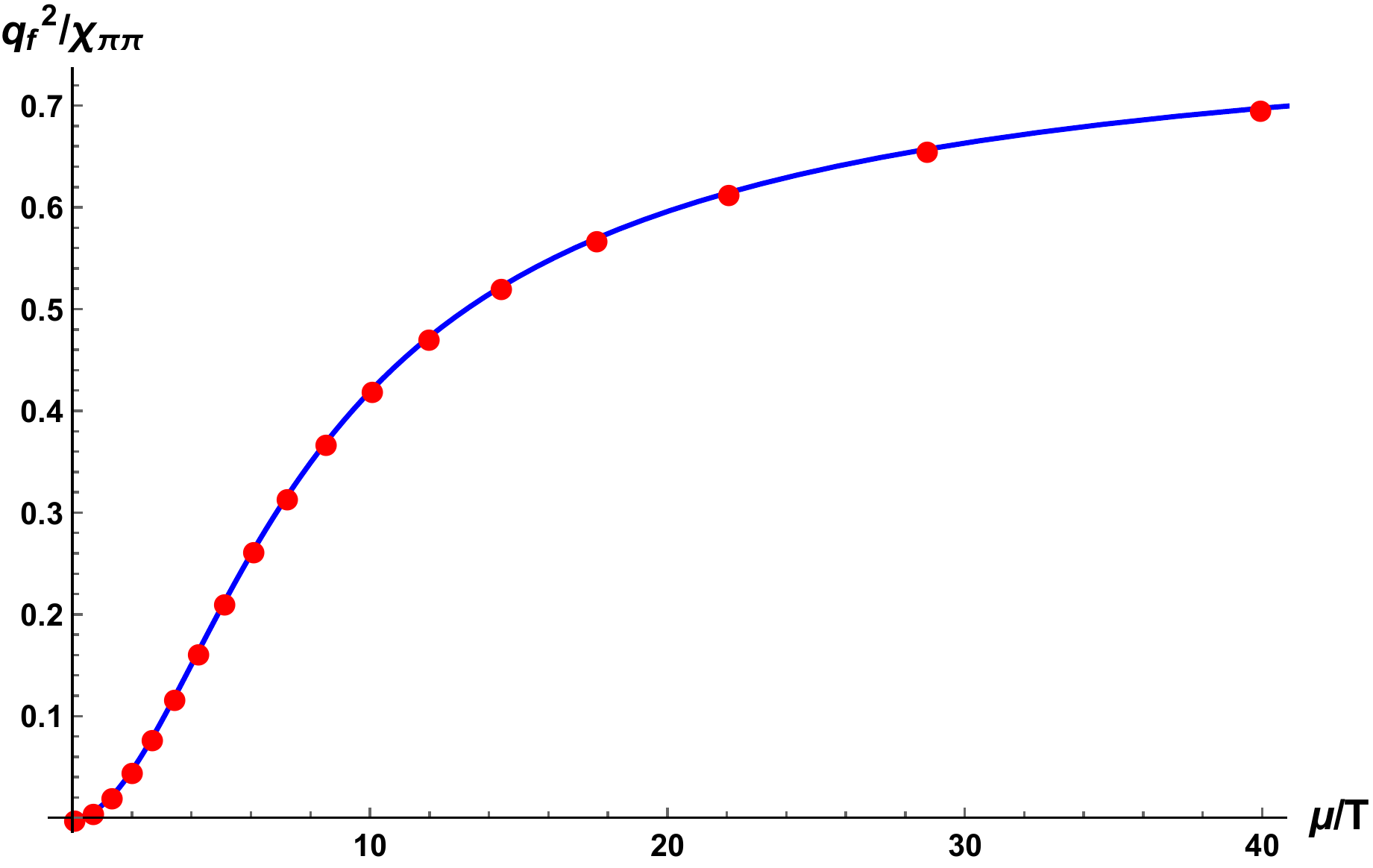}\quad 
     \includegraphics[width=0.45\linewidth]{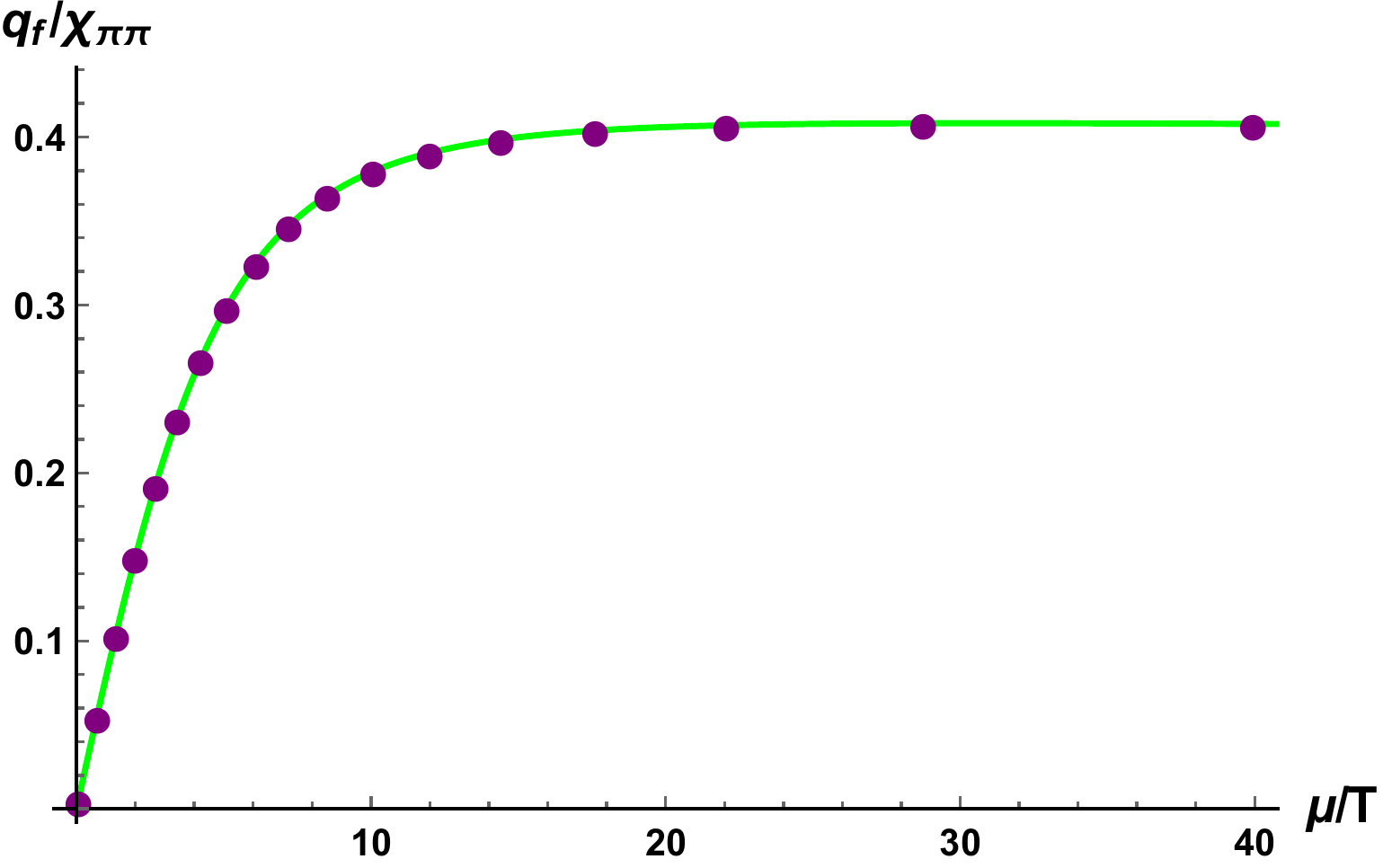}\\
     
     \vspace{0.3cm}
     
     \includegraphics[width=0.45\linewidth]{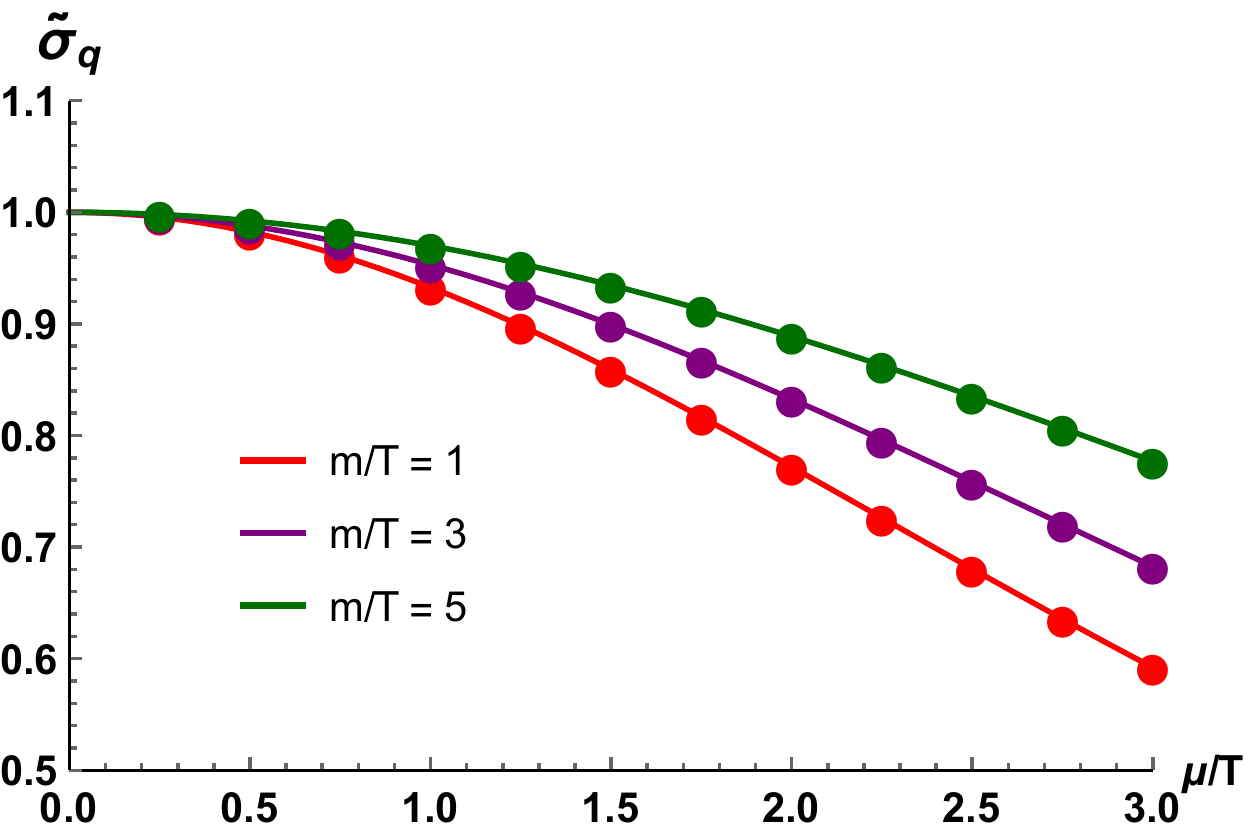}\quad 
     \includegraphics[width=0.45\linewidth]{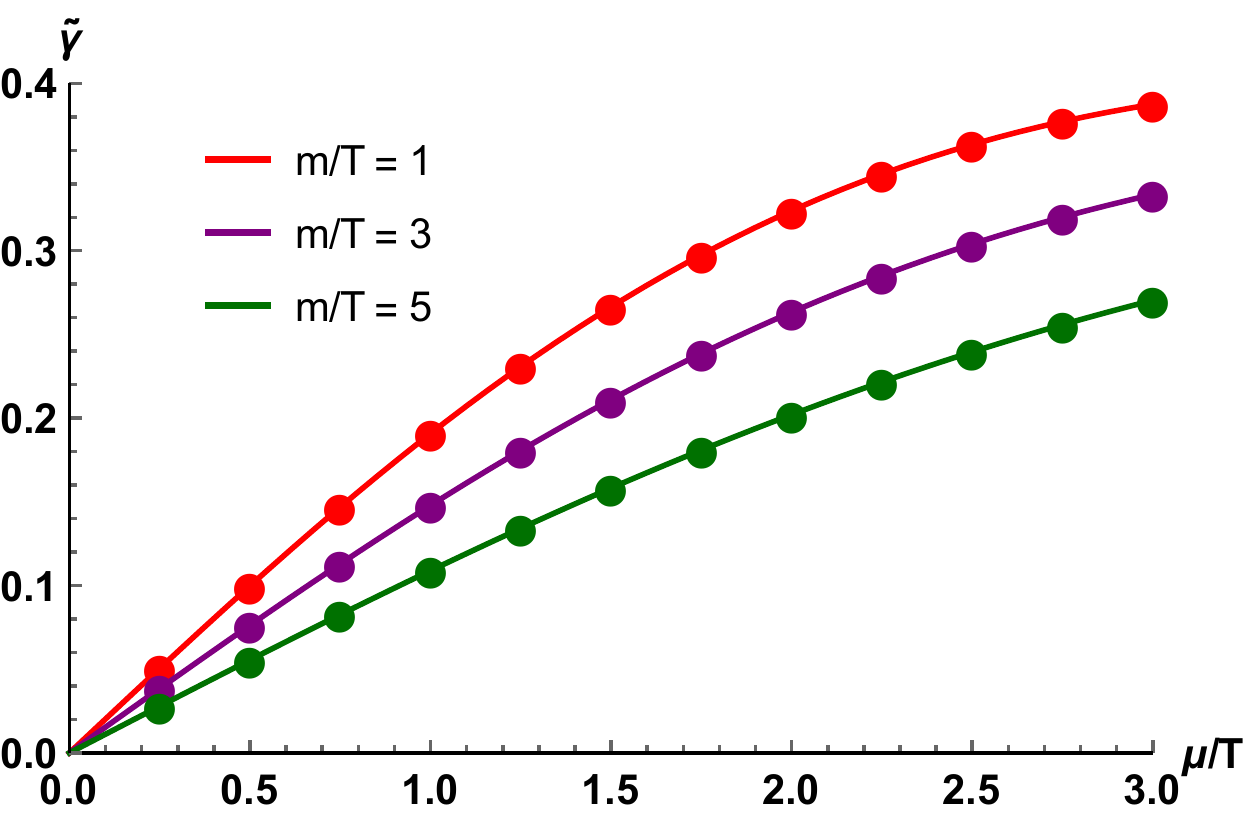}
    \caption{\textbf{Top Left: } The real part of the $J J$ correlator (red dots) compared to the hydrodynamic formula for $q_f^2/\chi_{\pi\pi}$ (blue line). We fixed $m=u_h=1$. \textbf{Top Right: } The imaginary part of the mixed $J \Phi$ correlator (purple dots) compared to the hydrodynamic formula for $q_f/\chi_{\pi\pi}$ (orange line). We fixed $m=u_h=1$. \textbf{Bottom Left :} The parameter $\tilde \sigma_q$ as a function of $\mu/T$ for several values of $m/T$. The lines are the analytic formula \eqref{ff}, while the dots are the numerical values extracted from the $JJ$ correlator. \textbf{Bottom Right :} The parameter $\tilde \gamma$ as a function of $\mu/T$ for several values of $m/T$. The lines are the analytic formula \eqref{ff} and the dots the numerical values extracted from the $J\Phi$ correlator.}
    \label{testfig}
\end{figure}

Following~\cite{Amoretti:2018tzw}, one can derive simple formulas for the parameters $\tilde{\gamma}$ and $\tilde \sigma_q$:
\begin{align}
    &\tilde{\gamma}\,=\,\frac{\mu  \,(m^2+s\, T)}{\chi _{\pi\pi}^2}+\frac{2 \,\pi\,  \rho\,  (\mu \, \rho +s \,T)}{\chi_{\pi\pi} ^2 \,N\, s}\,,\label{ff0}\\
    &\tilde \sigma_q\,=\,\frac{(\chi_{\pi\pi} -\mu\,  \rho )^2}{\chi_{\pi\pi} ^2}+\frac{2\, \pi\,  \rho ^2 \,(\chi_{\pi\pi} -\mu \, \rho -s\, T)^2}{\chi_{\pi\pi} ^2\,m^2\,s \,N}\,, 
\label{ff}
\end{align}
which are valid for potentials $V(X)=m^2 X^N$ assuming the scalars profile to be simply $\phi^I=x^I$ \footnote{Taking the more generic profile $\phi^I=\kappa x^I$, it is easy to realize that for monomial potentials of the form $V(X)=m^2 X^N$ the $\kappa$ parameter is redundant and it can be re-asborbed in the definition of  $m^2$.}.
Our results are shown in figure~\ref{testfig}, where we verify the expressions presented in \eqref{corr2}, and plot the behavior of the parameters $\tilde \sigma_q, \tilde{\gamma}$ as a function of $\mu/T$ and $m/T$. Finally, we check numerically that the analytic formulas \eqref{ff} are indeed correct.

After discussing the various Green's functions and hydrodynamic predictions, we move to a concrete and complete check of the hydrodynamic framework. To simplify the discussion, we will focus on a specific potential:
\begin{equation}
    V(X)\,=\,m^2\,X^3\,,\quad \text{with}\quad \phi^I\,=\,x^I\,,
\end{equation}
which we take as a benchmark model. 
%%%%%%%%%%%%%%%%%%%%%
%%%%%%%%%%%%%%%%%%%%
\begin{figure}
    \centering
    \includegraphics[width=0.45\linewidth]{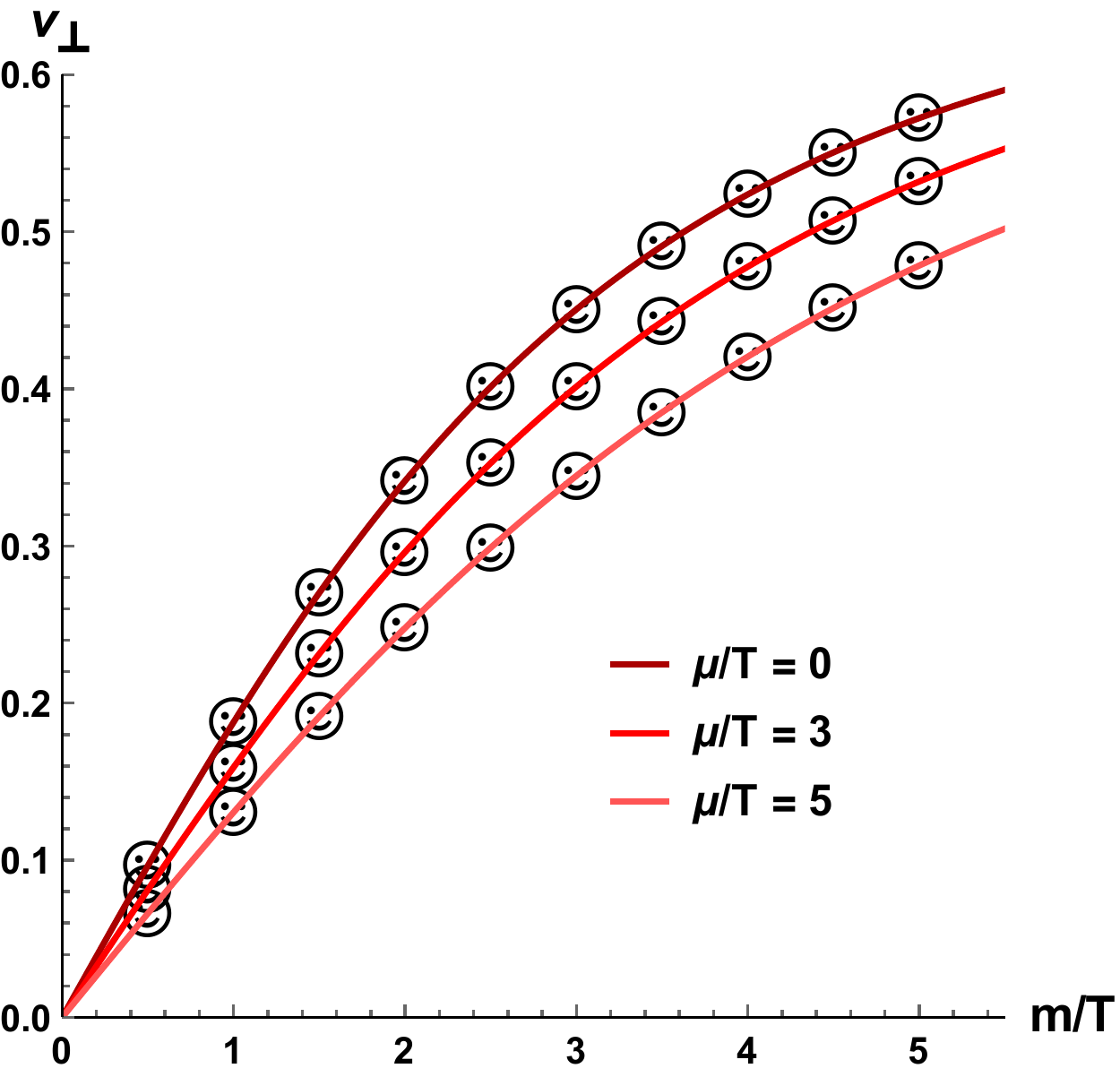} \quad
    \includegraphics[width=0.45 \linewidth]{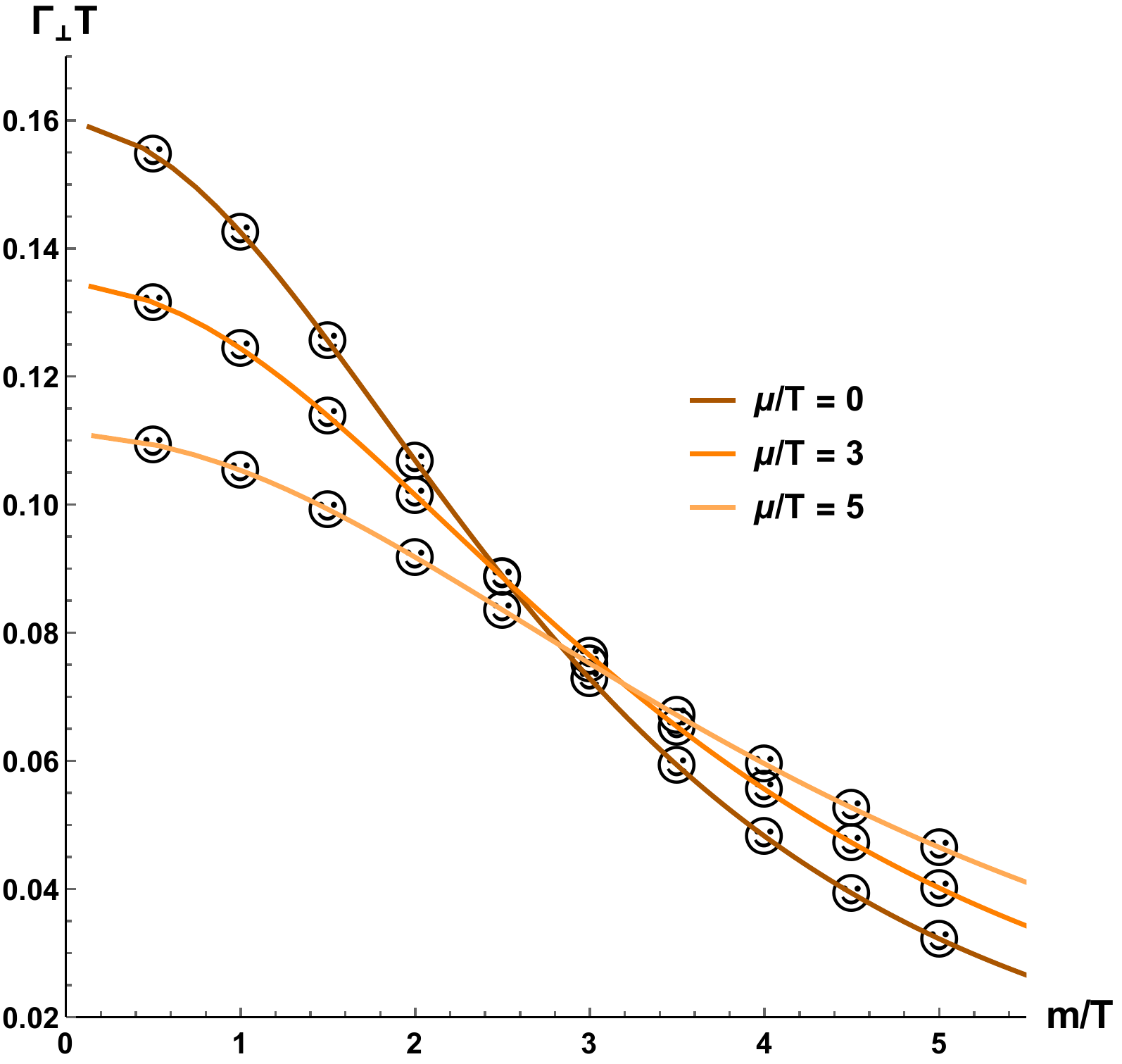}
    \caption{The predictions from hydrodynamics for the benchmark potential $V(X)=m^2 X^3$ and various values of $\mu/T$. \textbf{Left: }The speed of transverse sound. The smileys indicate the numerical values obtained by fitting the real part of the quasinormal modes $\mathrm{Re}[\omega]\,=\,v_\perp\,k$. \textbf{Right: }The transverse sound attenuation constant. The smileys indicate the numerical values obtained by fitting the real part of the quasinormal modes $\mathrm{Im}[\omega]\,=\,-\,\frac{1}{2}\Gamma_\perp\,k^2$.}
    \label{fig:hydro1}
\end{figure}
We start in figure~\ref{fig:hydro1} by comparing the hydrodynamic formulas with the numerical data regarding the propagating transverse sound. The numerical values extracted from the quasinormal modes (QNMs) are in good agreement with hydrodynamics for all values of $m/T$ and $\mu/T$. The presence of a finite charge density always decreases the speed of propagation of transverse sound. This can be easily understood from the fact that the momentum susceptibility $\chi_{\pi\pi}$ grows with the chemical potential $\sim\mu^2$. In other words, sound is slower because the ``mass density'' (the non-relativistic analogous of the momentum susceptibility) is larger. The dynamics of transverse sound attenuation is more elaborated. Even though the finite density decreases the sound attenuation constant $\Gamma_\perp$ at small $m/T$, the effect is reversed at large $m/T$, where the finite charge density decreases the lifetime of transverse sound.
For the longitudinal sound mode, we show our results in figure~\ref{fig:hydro2}. The results from hydrodynamics and holography match perfectly.
\begin{figure}
    \centering
    \includegraphics[width=0.45\linewidth]{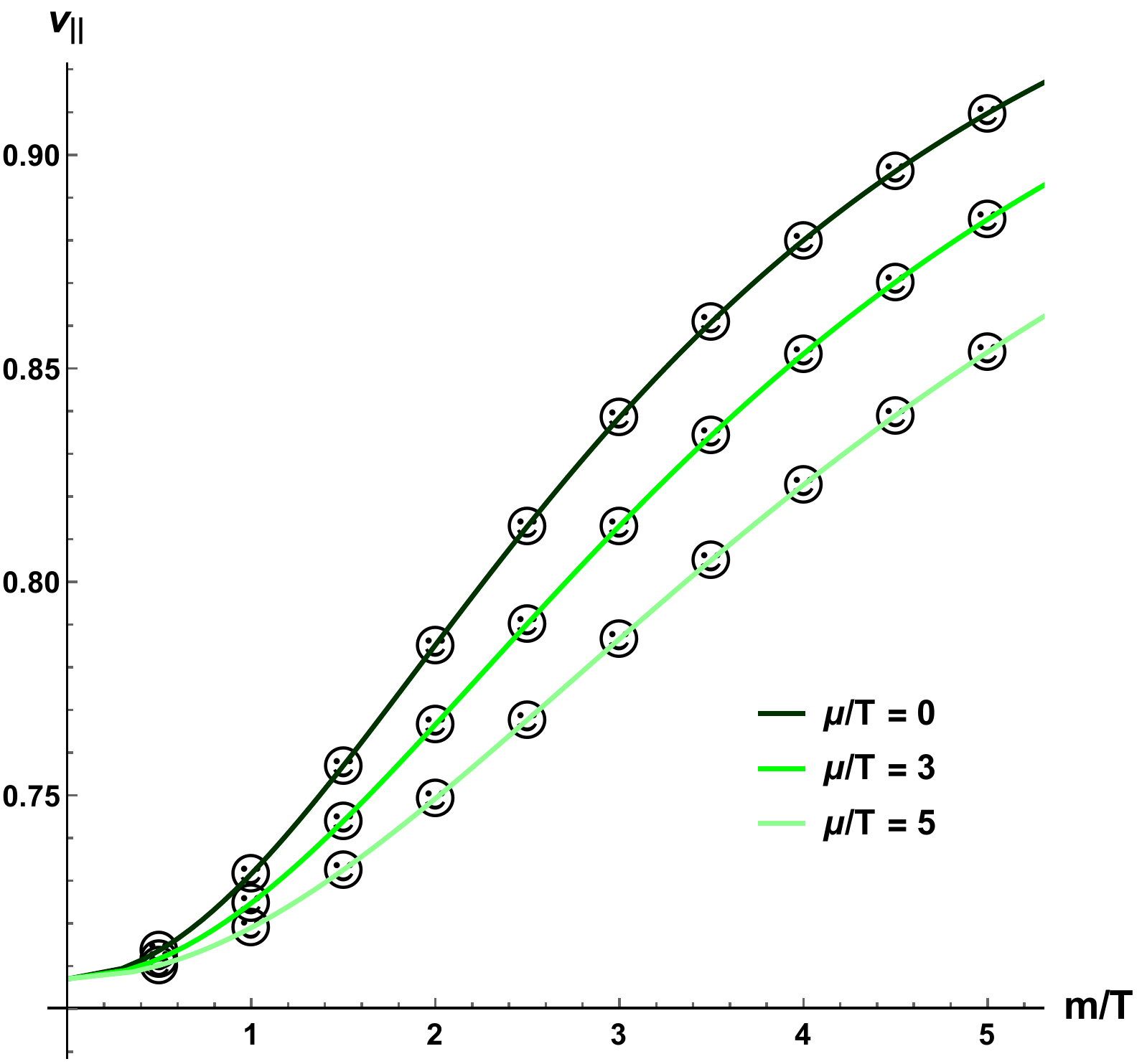} \quad
    \includegraphics[width=0.45 \linewidth]{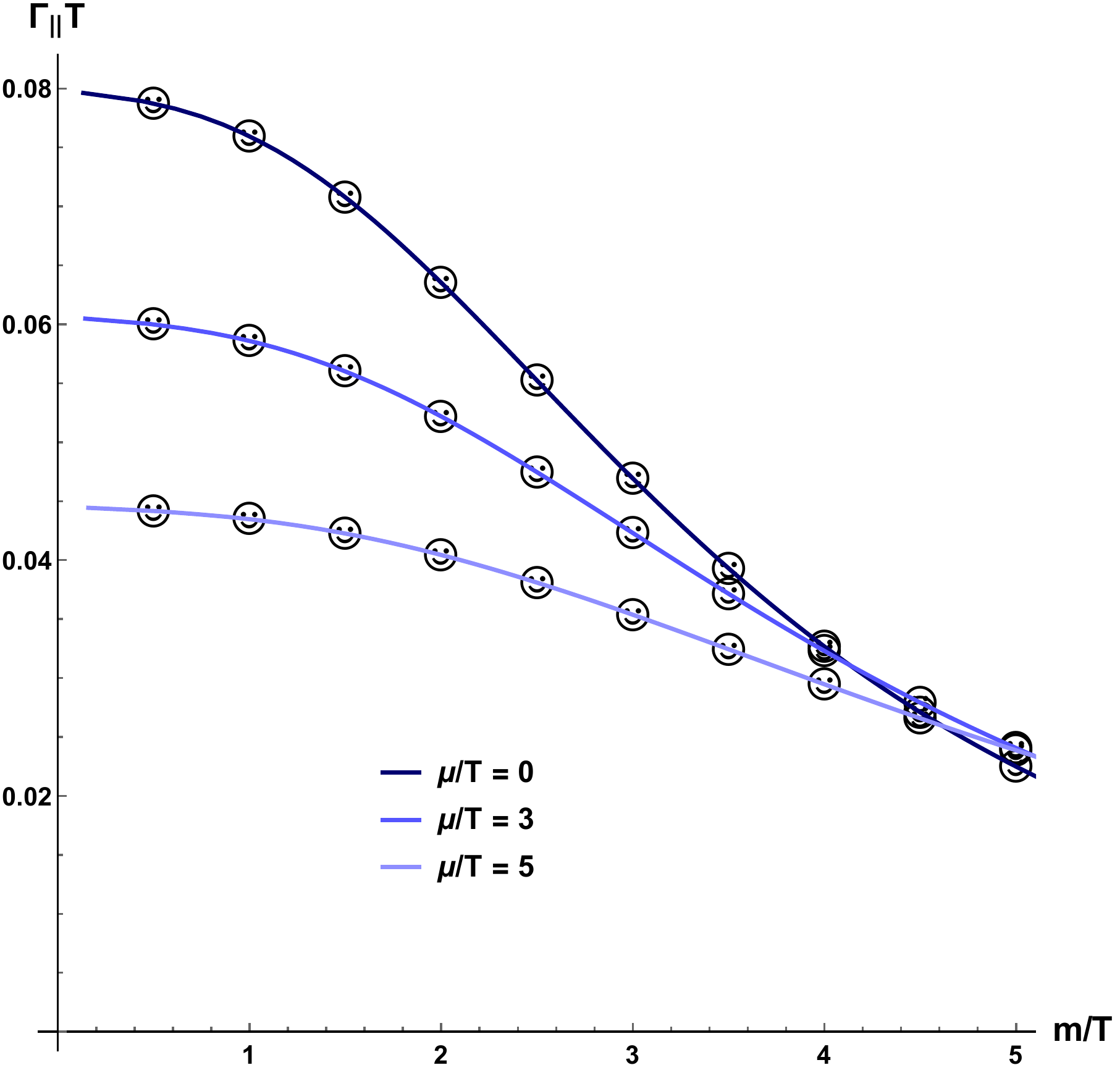}
    \caption{The predictions from hydrodynamics for the benchmark potential $V(X)=m^2 X^3$ and various values of $\mu/T$. \textbf{Left: }The speed of longitudinal sound. \textbf{Right: }The longitudinal sound attenuation constant.}
    \label{fig:hydro2}
\end{figure}
\begin{figure}
    \centering
    \includegraphics[width=0.45\linewidth]{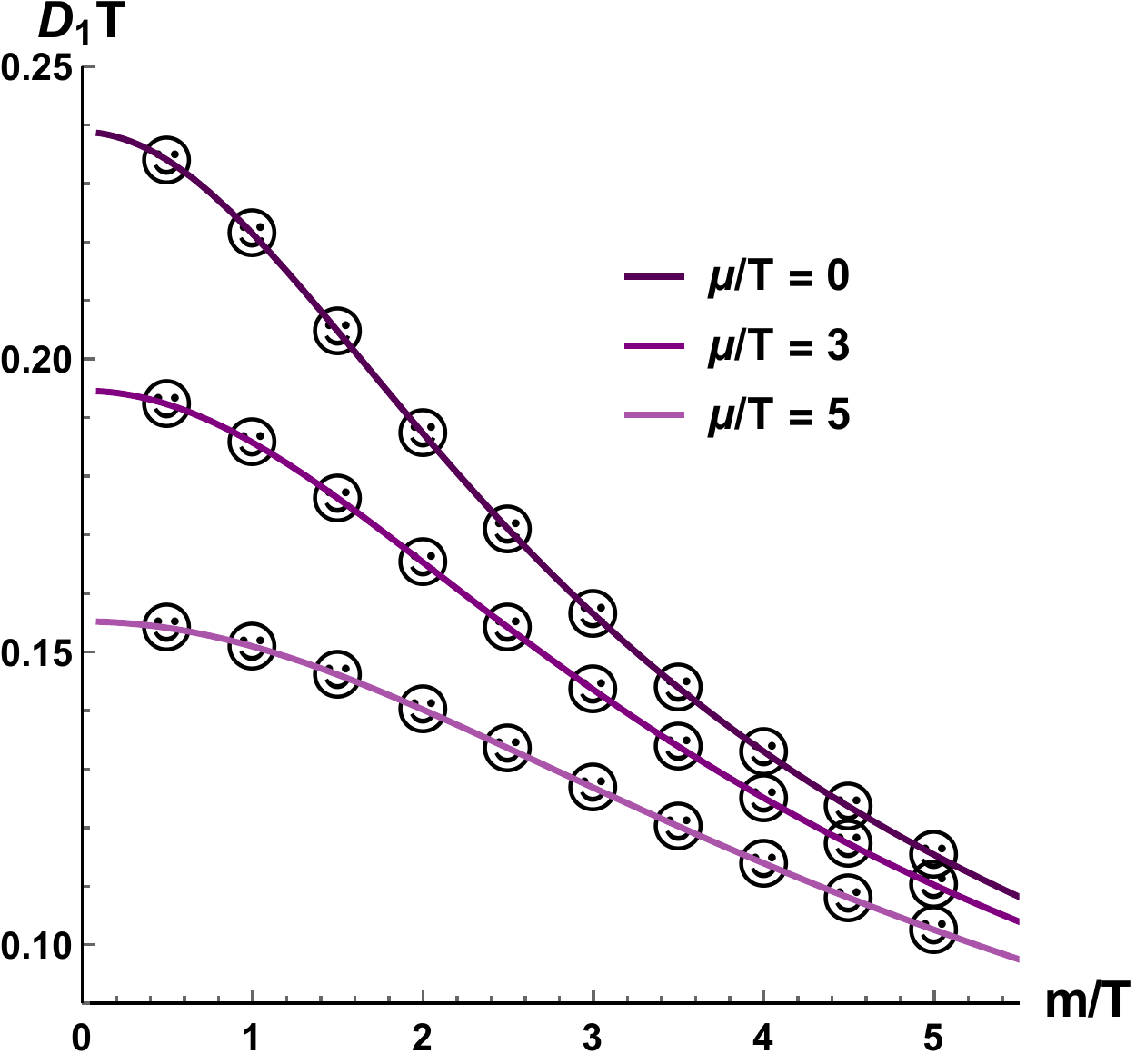} \quad
    \includegraphics[width=0.45 \linewidth]{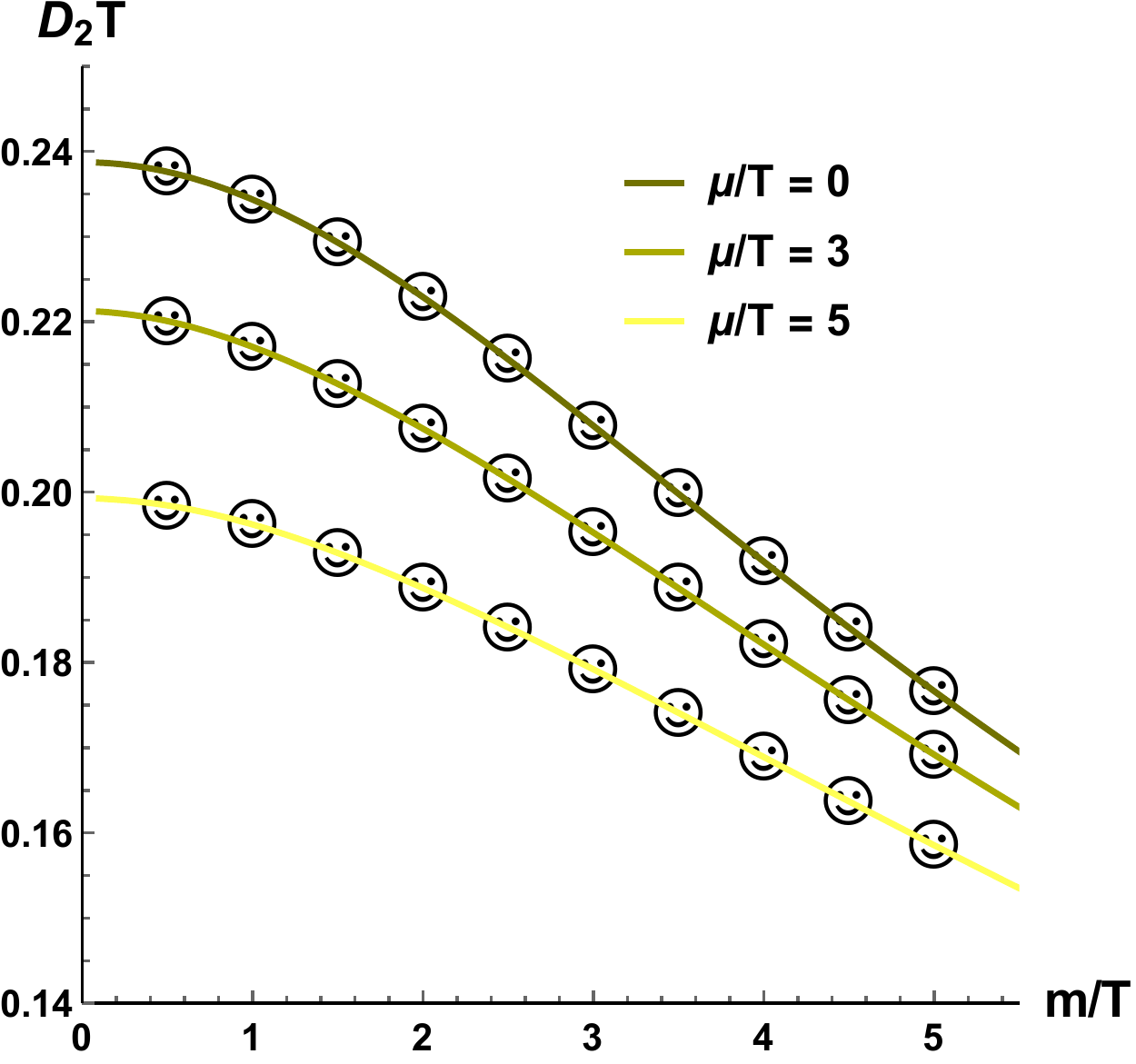}
    \caption{The predictions from hydrodynamics for the benchmark potential $V(X)=m^2 X^3$ and various values of $\mu/T$. The diffusion constants are obtained by solving the coupled equation \eqref{cc}. \textbf{Left: }The diffusion constant $D_1$ which in the limit $\mu \rightarrow 0$ corresponds to the crystal diffusion. \textbf{Right: }The diffusion constant $D_2$ which in the limit $\mu \rightarrow 0$ corresponds to the charge diffusion.}
    \label{fig:hydro3}
\end{figure}

We now discuss the two diffusive modes present in the longitudinal sector. We compare hydrodynamics and the holographic results in figure~\ref{fig:hydro3}. The hydrodynamic predictions are satisfied in our holographic model. Both diffusion constants decrease with the dimensionless chemical potential $\mu/T$.

\begin{figure}
    \centering
    \includegraphics[width=0.6\linewidth]{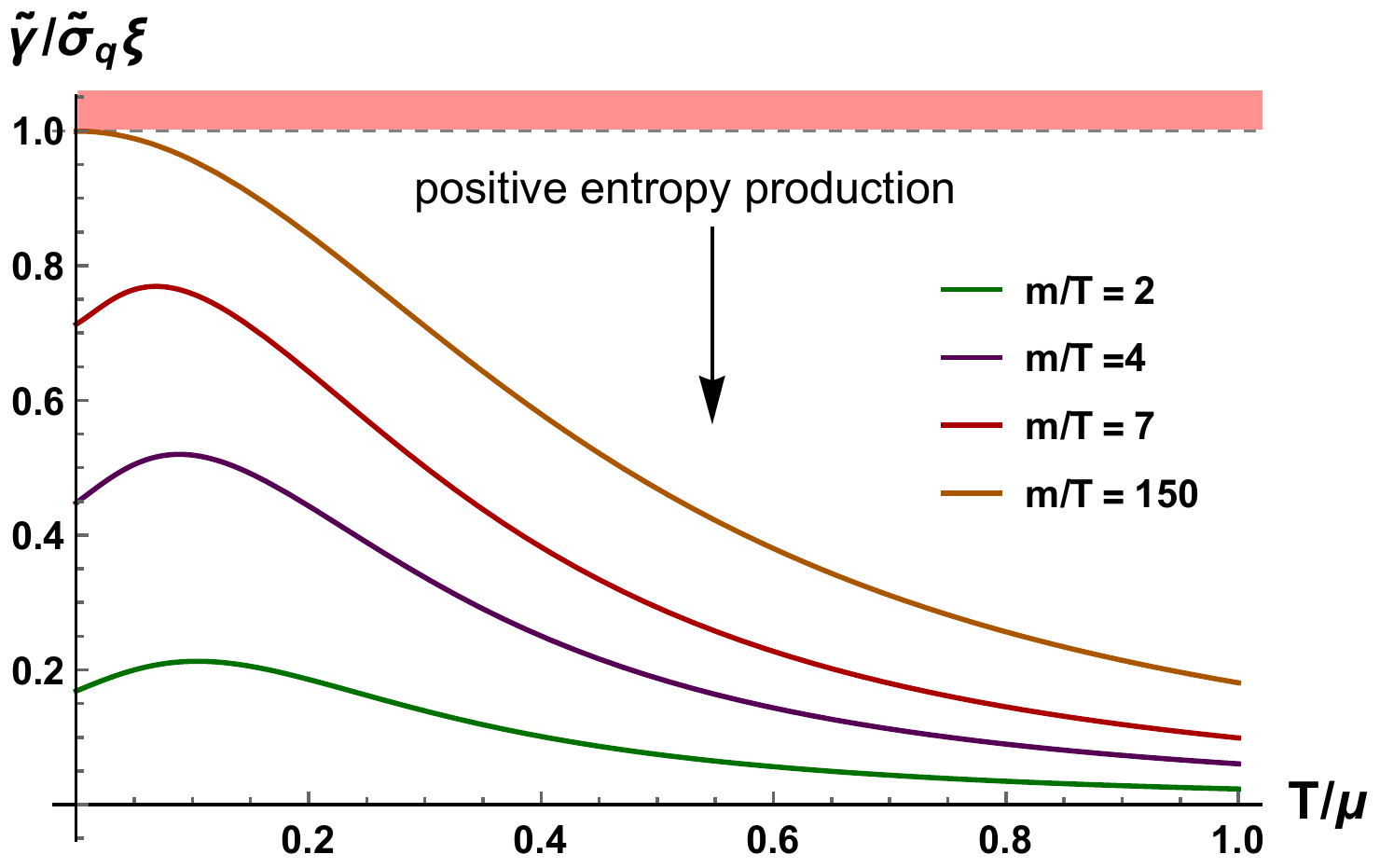}
    \caption{The combination $\tilde \gamma^2/\sigma_q \xi$ as a function of $T/\mu$ for various values of $m/T$. This figure has to be compared with figure~4 of~\cite{Amoretti:2018tzw}.}
    \label{figbound}
\end{figure}

Finally, we discuss the bound~\cite{Amoretti:2018tzw}
\begin{equation}
    \tilde \gamma^2\,\leq\,\sigma_q\,\xi\,, \label{bb}
\end{equation}
which follows from the positivity of the entropy production in the hydrodynamic theory of~\cite{Delacretaz:2017zxd}. %We focus on potentials of the type $V(X)=m^2 X^N$ with $N>5/2$ for which momentum is a conserved quantity, and therefore hydrodynamic applies at all values of $m/T$. 
All the quantities involved in this bound can be obtained analytically in our holographic model. Hence, it is immediate to verify the inequality \eqref{bb} as a function of the parameters $\mu/T$ and $m/T$.
We plot our results in figure~\ref{figbound}. The bound in eq.\eqref{bb} holds at any temperature. Moreover, it is saturated exactly in the limit $m/T \rightarrow \infty$, $T/\mu \rightarrow 0$.

In summary, we conclude this section by celebrating a perfect agreement between the hydrodynamic framework of~\cite{Armas:2020bmo} and the results obtained from our holographic theory.

\section{The magnetophonon as a type-B Goldstone}\label{sec:typeB}
In this section, we add a finite magnetic field to the setup of the last section and continue to consider the case in which translations are spontaneously, but not explicitly broken. In terms of the parameters of our system \eqref{param}, this corresponds to the choice $\alpha=0$. In absence of magnetic field, the presence of the following hydrodynamic modes has been verified:
\begin{align}
    & \text{transverse sector:}\quad \omega_\perp\,=\,v_\perp\,k\,-\,\frac{i}{2}\,\Gamma_\perp\,k^2\,,\\
    & \text{longitudinal sector:}\quad \omega_\parallel\,=\,v_\parallel\,k\,-\,\frac{i}{2}\,\Gamma_\parallel\,k^2\,,\quad \quad \omega_{1,2}\,=\,-\,i\,D^\parallel_{1,2}\,k^2\,,
\end{align}
where the two modes $\omega_{\perp,\parallel}$ are the expected phononic vibrational modes whose speeds are set by the shear and bulk elastic moduli \cite{Ammon:2020xyv}. The additional longitudinal diffusive modes are a combination of the crystal diffusive mode, emerging because of the spontaneous breaking of the global symmetry $\phi \rightarrow \phi+b$~\cite{Donos:2019txg,Baggioli:2020nay}, and the standard charge diffusion mode (see previous section \ref{sec:hydro}).

At zero magnetic field, the two sectors are decoupled and the longitudinal/transverse phonons represent indeed a pair of linearly propagating (type A) Goldstone bosons, corresponding to the breaking of translations in the $x,y$ spatial directions\,\footnote{As expected, no additional Goldstone modes for broken rotations appear due to the Inverse Higgs mechanism.}. As explained in section \ref{magn}, at finite magnetic field the two sectors couple, and one expects the presence of a type-B mode and a gapped partner:
\begin{align}
   & \mathrm{Re}[\omega]\,=\,\mathcal{C}\,+\,\mathcal{B}\,k^2\,,\quad \quad \text{magnetoplasmon}\,,\nonumber\\
   & \mathrm{Re}[\omega]\,=\,\mathcal{A}\,k^2\,,\quad \quad \quad \quad \text{magnetophonon}\,, \label{vv}
\end{align}
where these dispersion relation can be derived formally using hydrodynamics~\cite{PhysRevB.100.085140}.

Before continuing, let us remind the reader which are the expectations from hydrodynamics and field theory. The real part of the two modes, at small momentum, should follow:
\begin{equation}\label{vv2b}
    \mathrm{Re}\,[\omega_+]\,=\,\omega_c\,+\,\frac{(v_\parallel^2+v_\perp^2)}{2\,\omega_c}\,k^2\,+\,\dots\,,\quad \quad \mathrm{Re}\,[\omega_-]\,=\,\frac{v_\perp\,v_\parallel}{\omega_c}\,k^2\,+\,\dots \,,
\end{equation}
where the cyclotron frequency $\omega_c$~\cite{Hartnoll:2007ip} is defined as
\begin{equation}
    \omega_c\,=\,\frac{\rho\,B}{\chi_{\pi\pi}} \label{cyclo}\,.
\end{equation}
Additionally, the magnetoplasmon mode $\omega_+$ displays a damping term:
\begin{equation}
    \mathrm{Im}\,[\omega_+]\,=\,-\,\gamma_B\,,\quad \quad \textit{with}\quad \gamma_B\,=\,\frac{\tilde{\sigma}_q\,B^2}{\chi_{\pi\pi}}\,,\label{vv3}
\end{equation}
which is valid in the limit of small magnetic field $B/T^2 \ll 1$~\cite{Hartnoll:2007ip}. A full analysis of our numerical results would necessitate a complete hydrodynamic theory in the presence of lattice pressure $\mathcal{P}$, spontaneously broken translations, and finite charge density and magnetic field. To the best of our knowledge, such theory has not been built yet.

\begin{figure}
    \centering
    \includegraphics[width=0.45\linewidth]{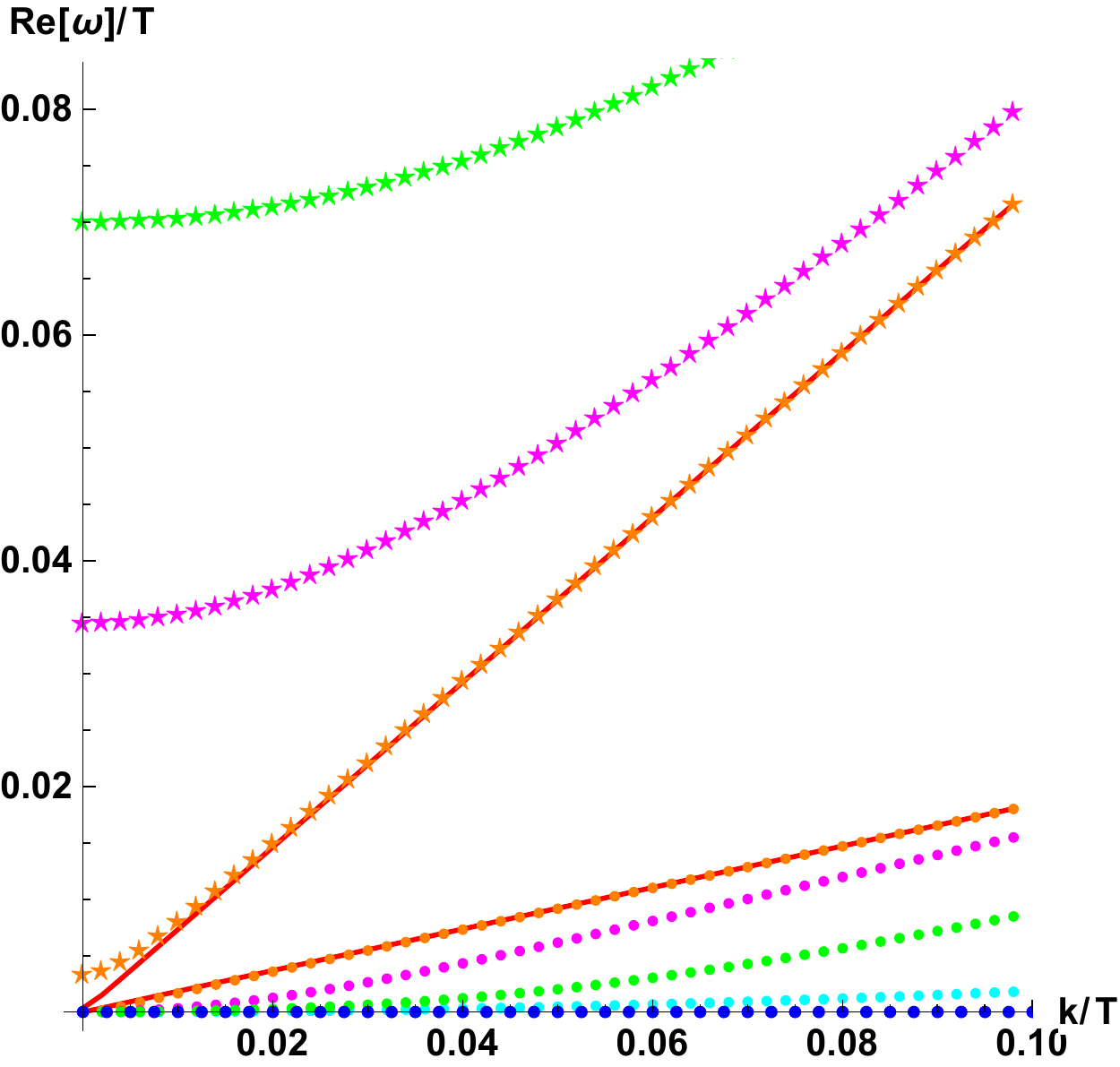}
    \quad 
    \includegraphics[width=0.45\linewidth]{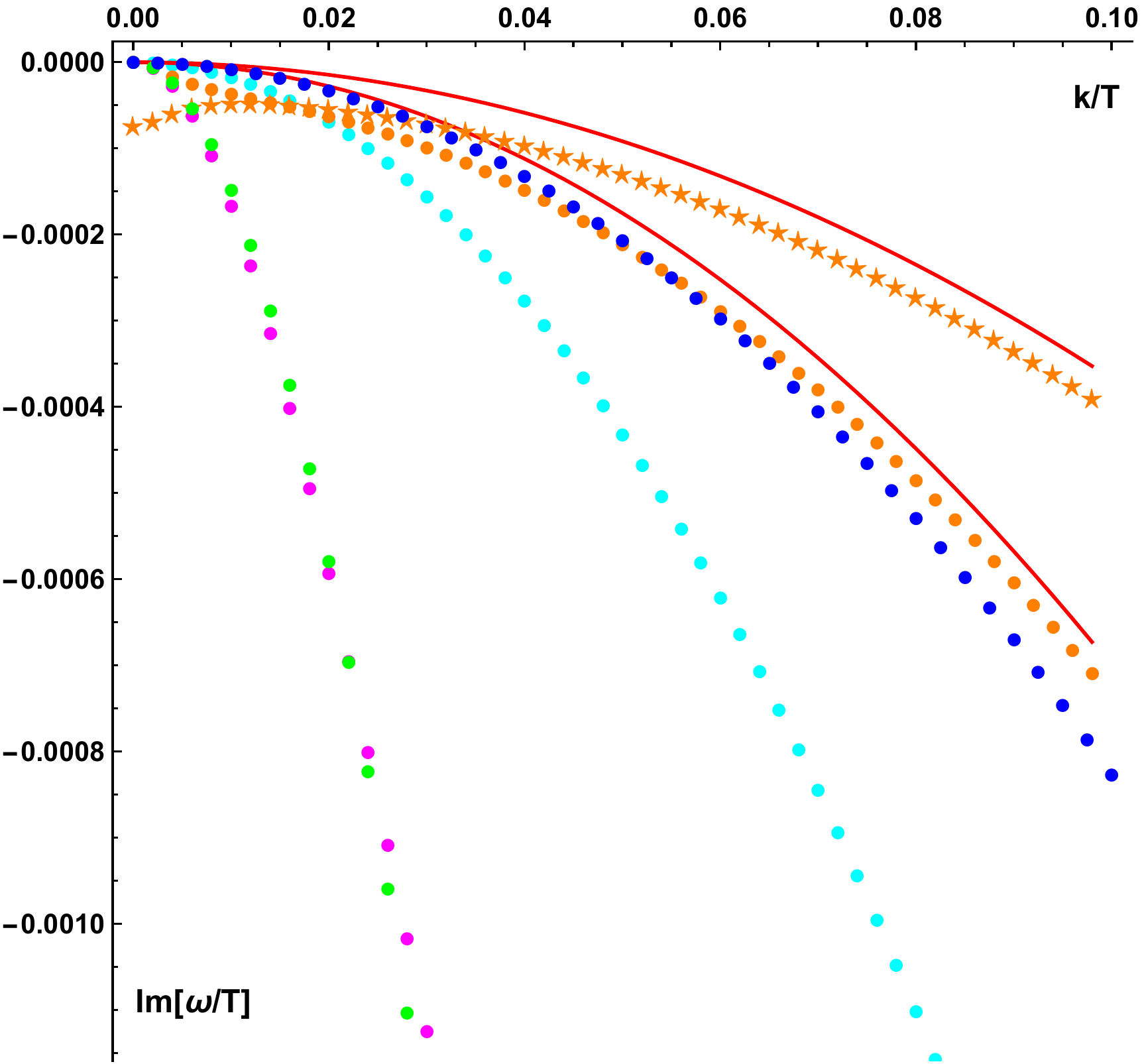}
    
     \vspace{0.3cm}
    
    \includegraphics[width=0.6\linewidth]{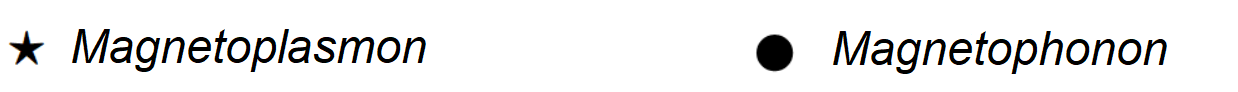}
    
    \caption{The fate of the two propagating hydrodynamic sound modes under increasing the magnetic field. The red lines show the results at $B=0$ and the two linear sound modes. We fix $\mu/T=1$ and $m/T=1$ moving $B/T^2=$ 0.01 (orange), 0.1 (pink), 1 (green), 2 (cyan) and 4 (blue). The circles indicate the magnetophonon mode while the stars the magnetoplasmon mode. For simplicity we do not show the behavior of the diffusive modes. \textbf{Left:} The real part of the dispersion relations. Cyan and blue stars are not shown as the gap becomes very large. \textbf{Right:} The imaginary part. The rest of the magnetoplasmons data (corresponding to green, cyan and blue stars) are not shown since the damping becomes very large.}
    \label{m1}
\end{figure}

The aim of this section is to verify the dispersion relation in eq.~\eqref{vv} and to compare them with the hydrodynamic formulas in eq.~\eqref{vv2}. For simplicity, we focus on a single case:
\begin{equation}
    V(X)\,=\,m^2\,X^3\,,\quad \phi^I\,=\,x^I\,,
\end{equation}
as a prototype for spontaneous symmetry breaking of translations.
First, we show in figure~\ref{m1} the dynamics of the sound modes by increasing the magnetic field. At $B=0$, we can identify two linear propagating sound modes (solid lines). After turning on the magnetic field, the modes combine forming the gapped magnetoplasmon (stars) and the quadratic type-B magnetophonon (circles). As evident from figure~\ref{m1}, the gap of the magnetoplasmon grows with $B$, while the coefficient of the $k^2$ scaling of the magnetophonon decreases with it.

\begin{figure}
    \centering
    \includegraphics[width=0.496\linewidth]{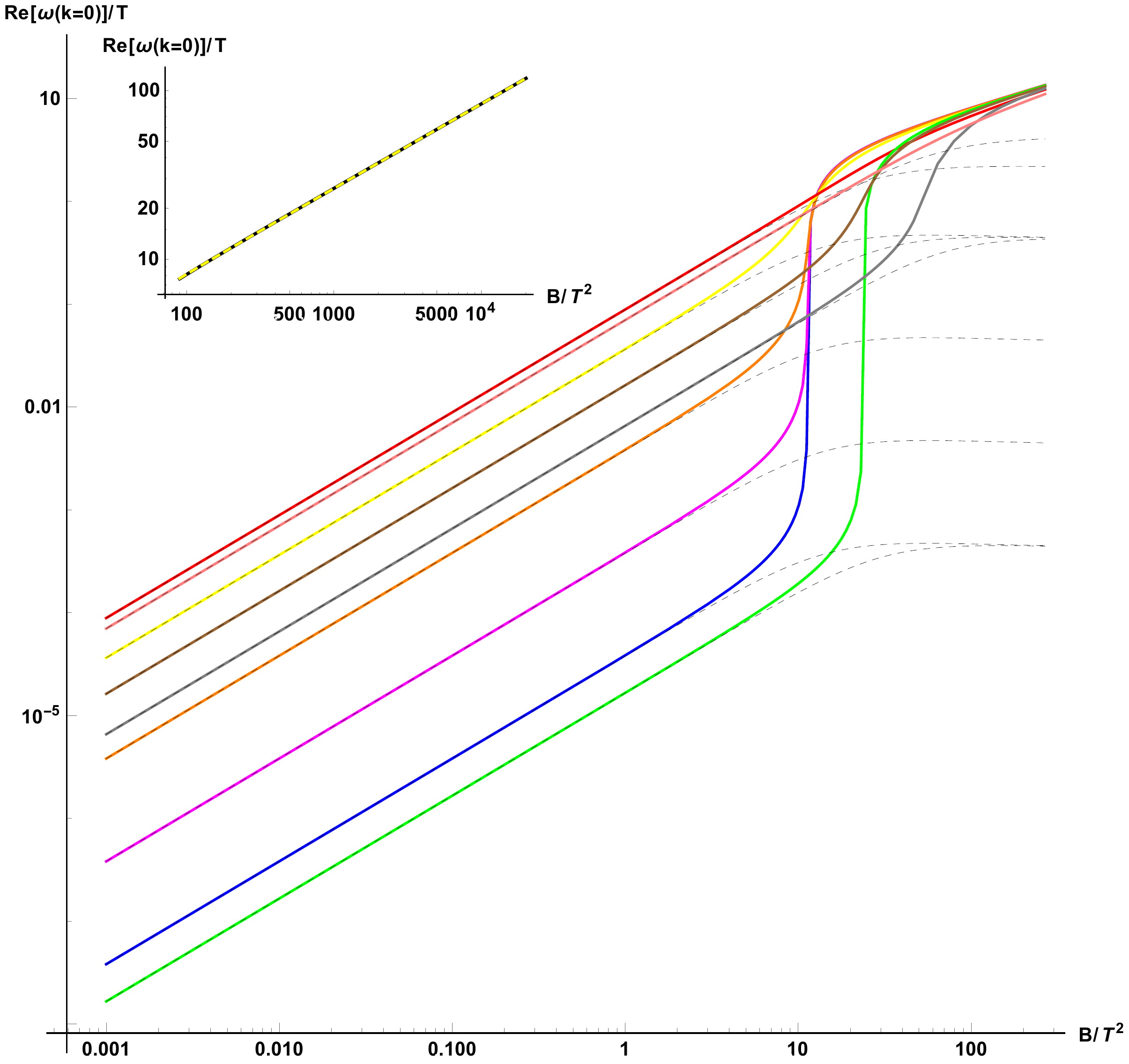}
  %  \quad 
    \includegraphics[width=0.496\linewidth]{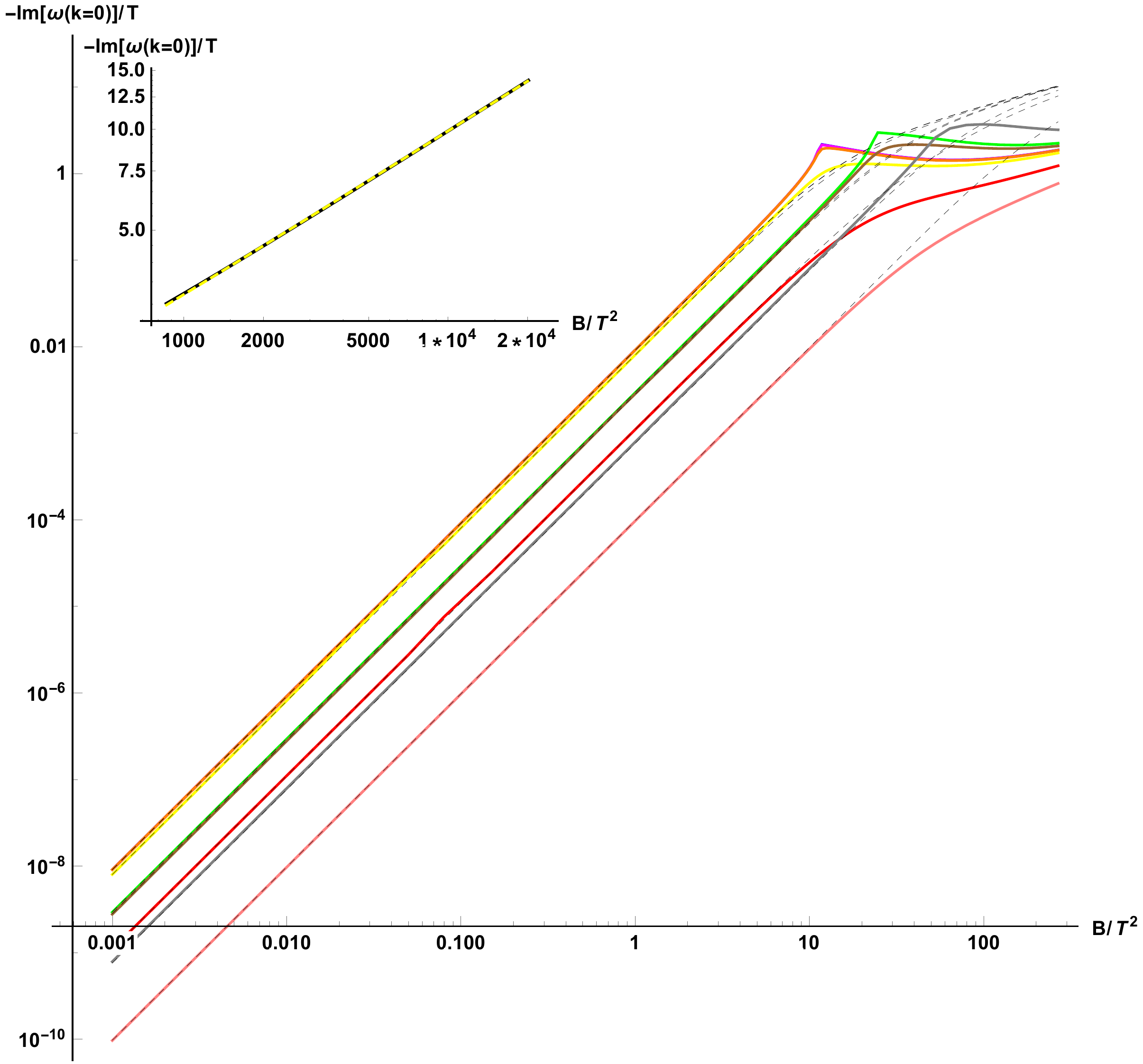}
    \caption{The cyclotron frequency $\omega_c$ and the cyclotron damping $\gamma_B$ extracted from the dispersion relation of the magnetoplasmon mode at zero momentum. The parameters are $\{\mu/T,\,m/T\}=\{\{0.001,0.001\},\{0.01, 0.001\}, \{0.1, 0.001\}, \{1, 0.001\},  \{5, 0.001\},\{0.001, 5\},\{1, 5\}, \{1, 10\},\{10,1\}\}$ (blue, magenta, orange, yellow, red, green, brown, gray, pink). \textbf{Left:} $\omega_c$ as a function of $B/T^2$ varying other parameters. The dashed lines are the formula for  $\omega_c$ \eqref{cyclo} valid at small magnetic field. The solid lines are the numerical data. The inset shows the large $B$ behavior consistent with $\omega/T\,\sim\,0.84\,\sqrt{B/T^2}$. \textbf{Right:} The damping $\gamma_B$. The dashed lines are the small $B$ analytic formula \eqref{vv3}. The inset is the large $B$ limit.}
    \label{m2}
\end{figure}

Next, we focus more in detail on the dispersion relation of the two modes. First, in figure~\ref{m2}, we consider the magnetoplasmon gapped mode. We plot both the real part (the gap) and the imaginary part (the damping) of its dispersion relation, and we verify that the hydrodynamic formulas \eqref{vv2b} and \eqref{vv3} are valid. We find that at small magnetic field, $B/T^2 \ll 1$, the numerical data are in very good agreement with those formulas. At large magnetic field, the real part clearly shows a different scaling, $\mathrm{Re}\,[\omega]\,\sim\,B^{1/2}$. Interestingly, the numerical prefactor is completely independent of $\mu/T$ and $m$. It is tempting to connect this behavior to the fact that at very large magnetic field the physics is purely quantum and is controlled by the lowest Landau level:
\begin{equation}
    \omega_{LLL}\,\sim\,\sqrt{B}\,.
\end{equation}
In any case, we are able to recover the large $B$ behavior using simple arguments. Let us start with the definition of the magnetic length:
\begin{equation}
    l\,=\,\sqrt{\frac{\hbar\,v}{e\,B}}\,,
\end{equation}
which can be converted into a frequency using the relation $l\,=\,v/\omega\,;$
in units $\hbar=e=1$, we obtain $\omega\,=\,\sqrt{v}\,\sqrt{B}.$
Now, in the large $B$ limit, $B/T^2 \ll 1$, we approach the conformal UV fixed point for which the characteristic speed is given by the conformal value $v=1/\sqrt{2}$. This means that the resulting frequency is given by
\begin{equation}
    \omega\,=\,\frac{1}{2^{1/4}}\,\sqrt{B}\,\sim\,0.84\,\sqrt{B}\,.\label{boh}
\end{equation}
This estimate is in perfect agreement with the numerical data shown in figure~\ref{m2}. Increasing the mass of the graviton $m/T$, the universal behavior is reached at larger values of the magnetic field. Note that a $\sqrt{B}$-like behavior of the real part of the QNMs for large magnetic fields was also observed in \cite{Ammon:2016fru,Grieninger:2017jxz,Ammon:2017ded}.

We proceed with discussing the dynamics of the magnetophonon mode -- our type-B Goldstone. The real part of the dispersion relation is consistent with a quadratic scaling $\omega\,=\,\mathcal{A}\, k^2$. At the same time, we observe that the imaginary part is compatible with a quadratic diffusive behavior. In summary, we observe that the dispersion relation of the type-B Goldstone mode at small momentum is of the type:
\begin{equation}\label{eqtypeB}
    \omega_{TYPE-B}\,=\,\mathcal{A}\,k^2\,-\,i\,\mathcal{D}\,k^2\,+\,\dots\,.
\end{equation}
These results are in agreement with what observed in the context of SU(2)$ \rightarrow $U(1) symmetry breaking in~\cite{Amado:2013xya}. Interestingly, a diffusive damping for type-B Goldstone is not envisaged from EFT methods~\cite{Hayata:2014yga}. Field theory approaches predict a $\sim k^4$ imaginary part for quadratic type-B Goldstone modes manifesting the quasiparticle nature of the excitation. To ensure a quasiparticle excitation in our present case~\eqref{eqtypeB}, we have to require $\mathcal{D}<\mathcal{A}$ which, in general, is not guaranteed in our holographic theory. The presence of a particle-hole continuum in the holographic model -- the so called incoherent sector -- can be a possible mechanism behind the $\sim k^2$ imaginary term observed\,\footnote{We thank C.Hoyos for suggesting this point to us.}. The continuum is not taken into account in the EFT description and it is known to have already important consequences in other situations, such as Fermi liquid theory~\cite{valentinis2020optical} and holographic models~\cite{HoyosBadajoz:2010kd,Krikun:2018agd,Baggioli:2019sio}.
\begin{figure}
    \centering
    \includegraphics[width=0.3\linewidth]{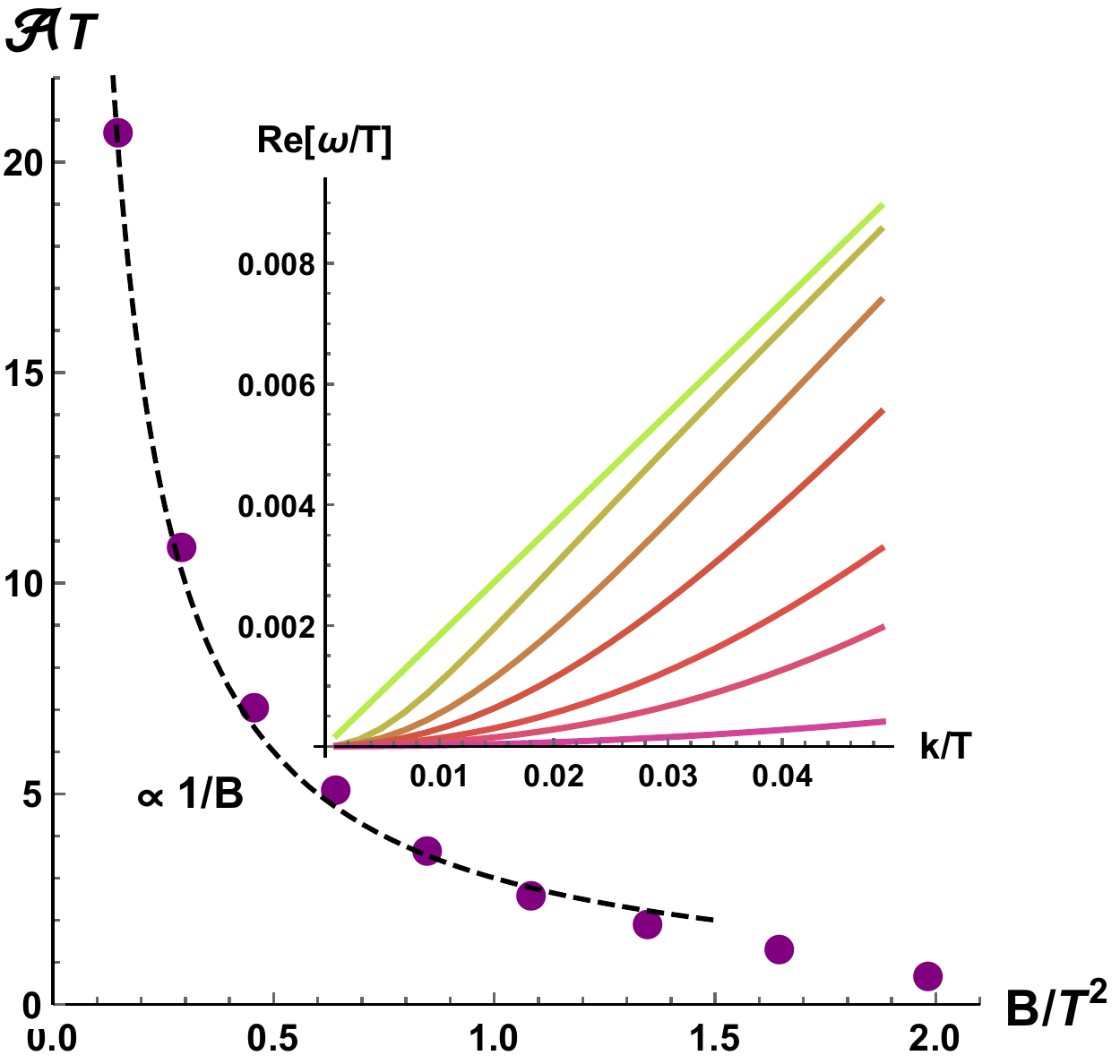}
    \quad 
    \includegraphics[width=0.3\linewidth]{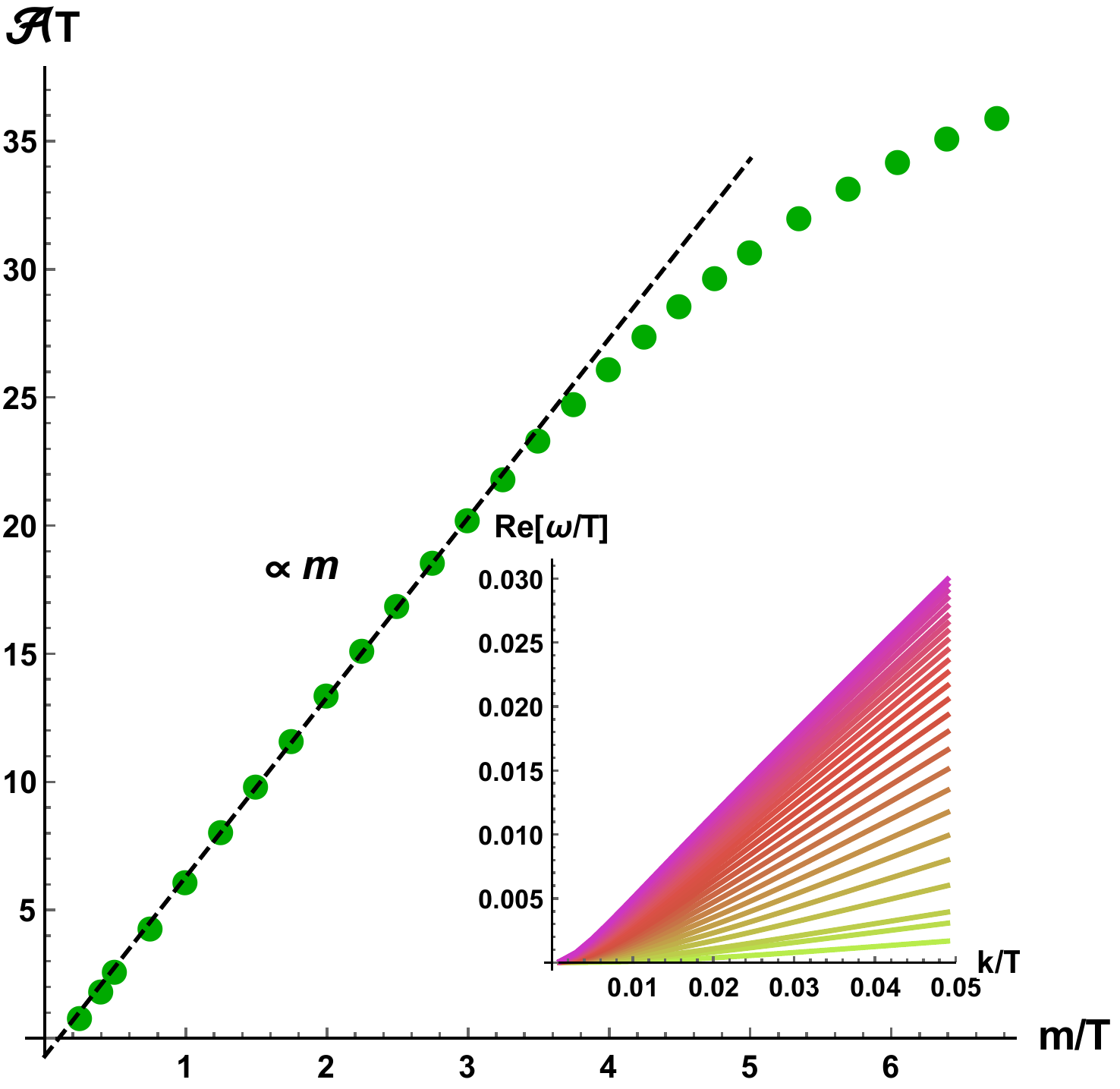}
    \quad 
    \includegraphics[width=0.3\linewidth]{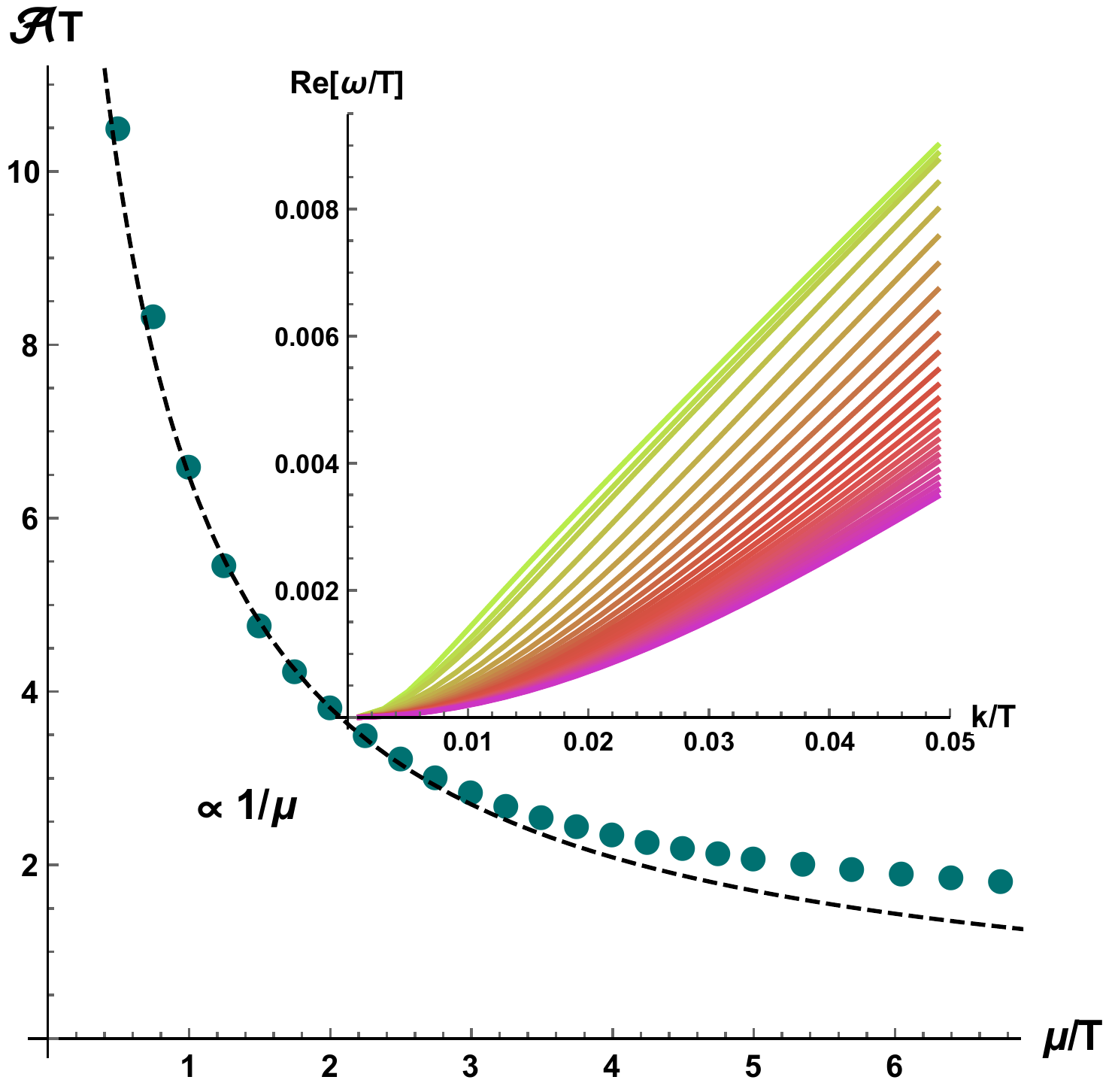}
    \caption{The coefficient $\mathcal{A}$ extracted from the dispersion relation of the type-B magnetophonon $\mathrm{Re}\,[\omega]\sim \mathcal{A}\,k^2$ as a function of the parameters of the system. \textbf{Left:} As a function of the dimensionless magnetic field $B/T^2$ fixing $m/T=\mu/T=1$. The inset shows the dispersion relation of the magnetoplasmon increasing the magnetic field (from green to pink). \textbf{Center:} As a function of the graviton mass $m/T$ fixing $\mu/T=1$ and $B/T^2=0.5$. The inset shows the dispersion relation of the magnetoplasmon increasing $m/T$ (from green to pink). The dashed line shows the low $m$ linear scaling compatible with the hydrodynamic formula. \textbf{Right:} As a function of the dimensionless chemical potential $\mu/T$ fixing $m/T=1$ and $B/T^2=0.5$. The inset shows the dispersion relation of the magnetoplasmon increasing the chemical potential (from green to pink). The dashed line shows the $\sim 1/\mu$ scaling compatible with the hydrodynamic formula.}
    \label{m3}
\end{figure}

\begin{figure}
    \centering
    \includegraphics[width=0.3\linewidth]{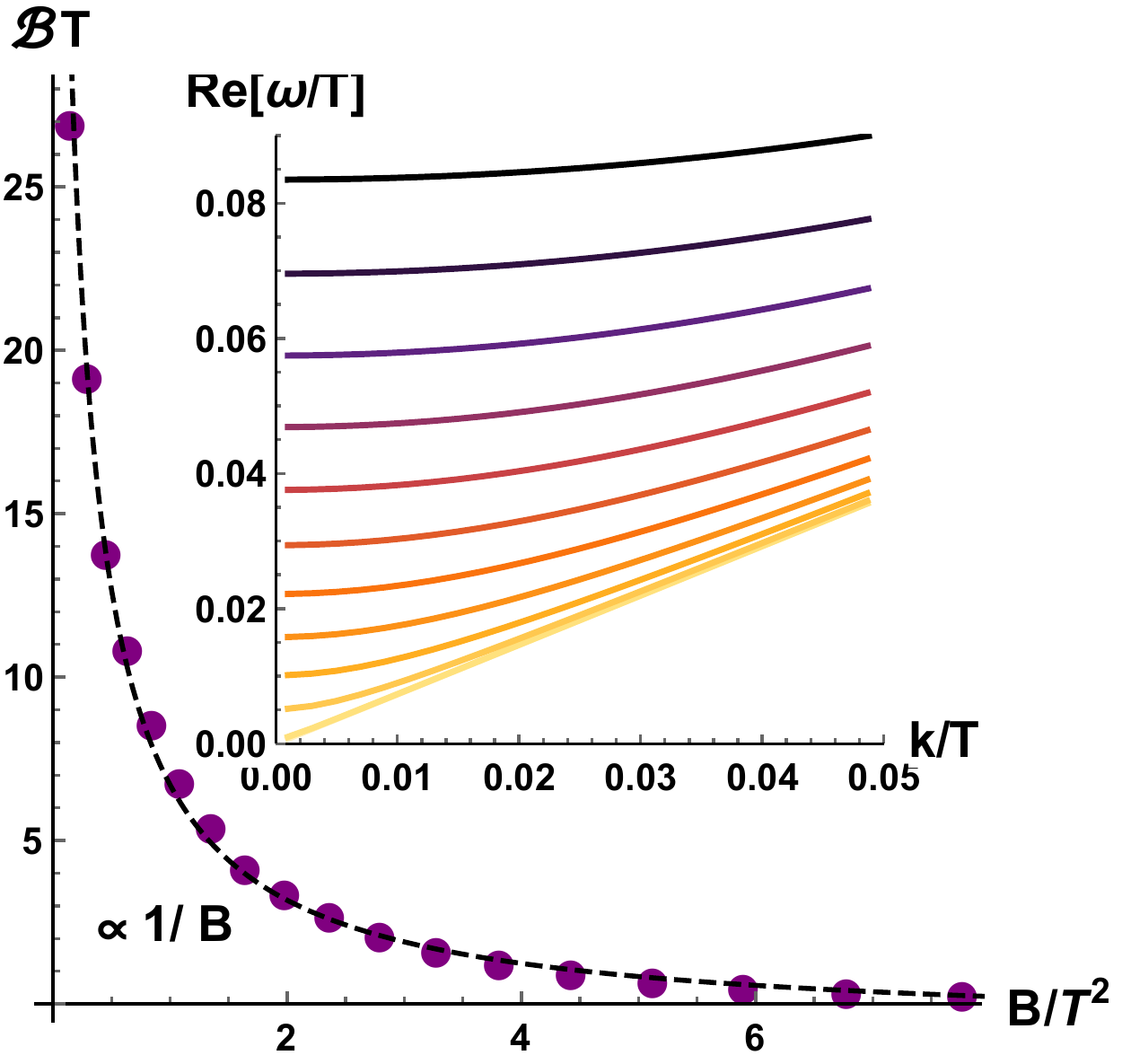}
    \quad 
    \includegraphics[width=0.3\linewidth]{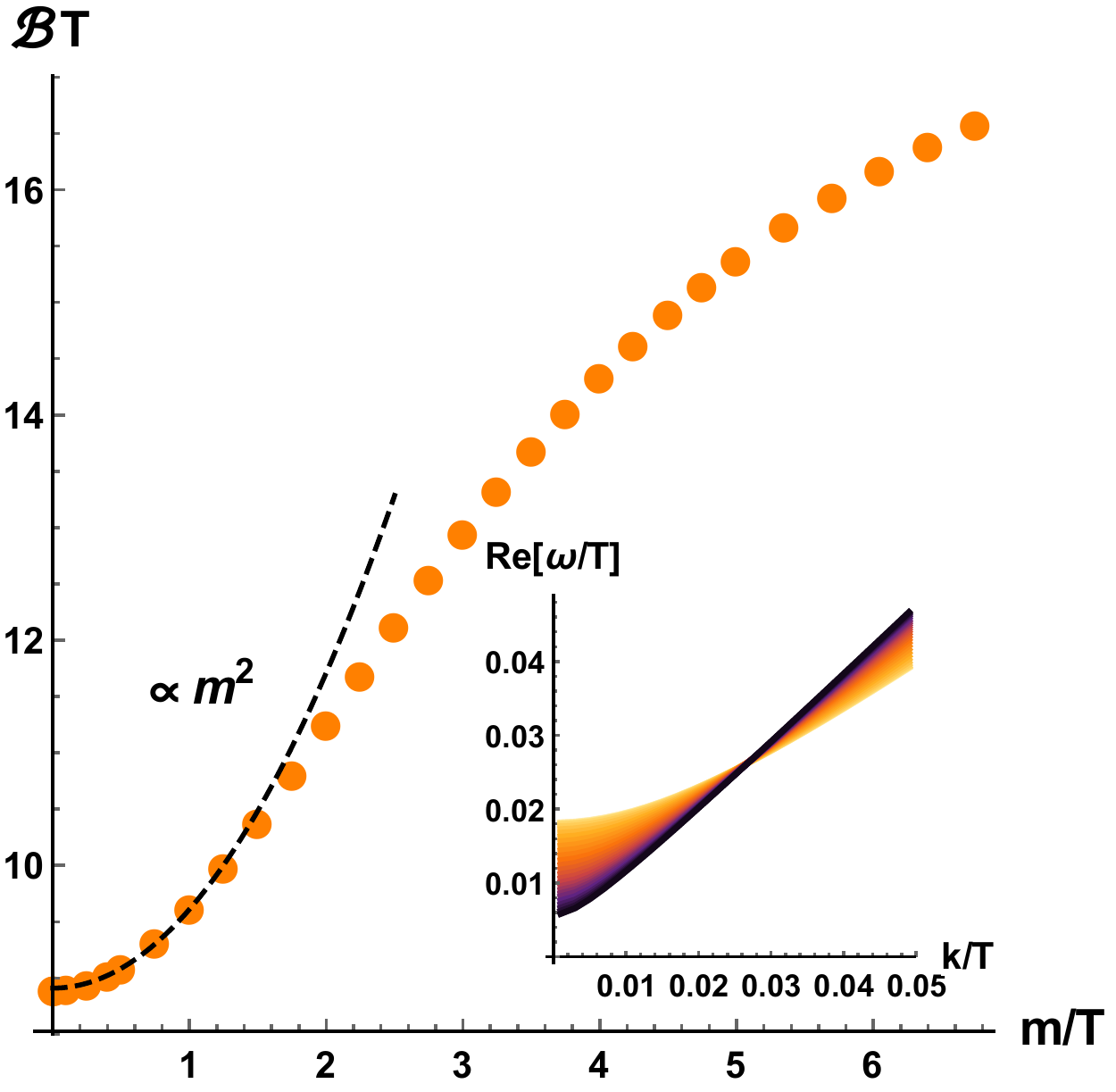}
    \quad 
    \includegraphics[width=0.3\linewidth]{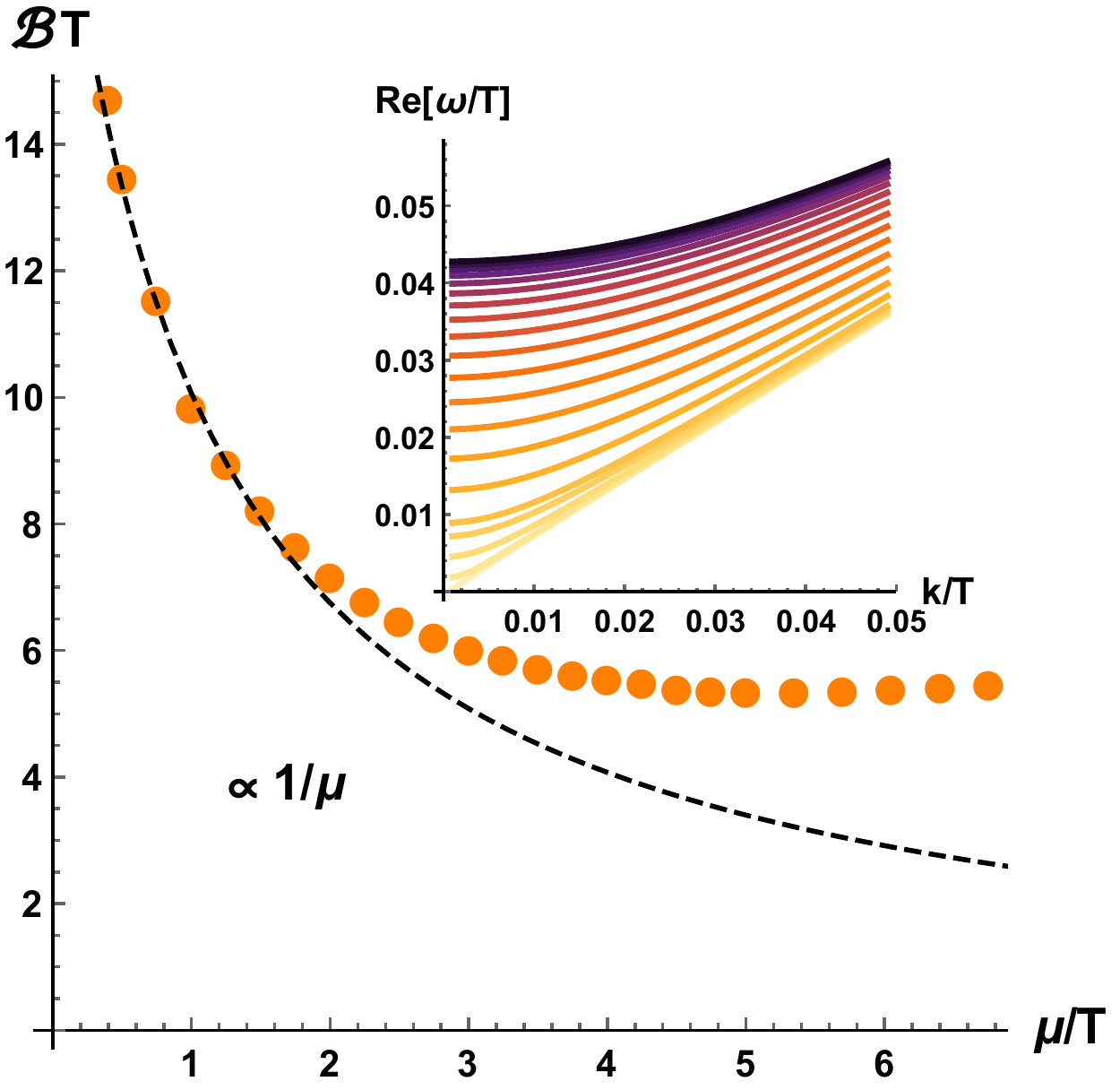}
    \caption{The coefficient $\mathcal{B}$ extracted from the dispersion relation of the type-B magnetoplasmon $\mathrm{Re}\,[\omega]\sim \mathcal{C}\,+\,\mathcal{B}\,k^2$ as a function of the parameters of the system. \textbf{Left: }As a function of the dimensionless magnetic field $B/T^2$ fixing $m/T=\mu/T=1$. The inset shows the dispersion relation of the magnetoplasmon increasing the magnetic field (from orange to black). \textbf{Center: }As a function of the graviton mass fixing $\mu/T=1$ and $B/T^2=0.5$. The inset shows the dispersion relation of the magnetoplasmon increasing $m/T$ (from orange to black). The dashed line shows the low $m$ quadratic scaling compatible with the hydrodynamic formula. \textbf{Right: }As a function of the dimensionless chemical potential $\mu/T$ fixing $m/T=1$ and $B/T^2=0.5$. The inset shows the dispersion relation of the magnetoplasmon increasing the chemical potential (from orange to black). The dashed line shows the $\sim 1/\mu$ scaling compatible with the hydrodynamic formula.}
    \label{m4}
\end{figure}

We examine the behavior of the coefficients $\mathcal{A},\mathcal{B}$ appearing in the dispersion relations in eq.\eqref{vv}. We have confirmed numerically that the hydrodynamic formula:
\begin{equation}
    \mathcal{B}\,=\,\frac{(v_\parallel^2+v_\perp^2)}{2\,\omega_c}\,\,,\quad \quad \mathcal{A}\,=\,\frac{v_\perp\,v_\parallel}{\omega_c}\,,\quad \quad v_\parallel^2\,=\,\frac{1}{2}\,+\,v_\perp^2\,,\quad \quad \mathcal{C}\,=\,\omega_c\,,
\end{equation}
fit very well the scalings obtained from the numerical data.
In absence of a complete hydrodynamic framework, we plot our results from the numerical data. More precisely, in figure~\ref{m3} and figure~\ref{m4}, we show the behavior of these the coefficients appearing in the dispersion relations as a function of the parameters of our system $\mu/T$, $m/T$ and $B/T^2$.

The results are compatible with hydrodynamics. In particular:
\begin{itemize}
    \item  Both parameters $\mathcal{A}$ and $\mathcal{B}$ decrease with the magnetic field. This is due to the fact that both the parameters are inversely proportional to the cyclotron frequency $\omega_c$:
    \begin{equation}
        \mathcal{A}\,\sim\,\omega_c^{-1}\,,\quad \quad \mathcal{B}\,\sim\,\omega_c^{-1}\,,
    \end{equation}
    and $\omega_c$ grows linearly with the magnetic field at small $B$. Indeed, our numerical results are compatible with a $\sim 1/B$ decay of both the parameters at small magnetic field.
    \item Both the parameters grow with $m/T$. This is due to the fact that both are proportional to the speed  of transverse of sound. Let us remind that
    \begin{equation}
        v_\perp\,\sim\,m\,,\quad \quad \mathcal{A}\,\sim\,v_\perp\,,\quad \quad \mathcal{B}\,\sim\,v_\perp^2\,.
    \end{equation}
    As expected, we observe that $\mathcal{A} \,T\sim m/T $ and $\mathcal{B}\,T\,\sim m^2/T^2$ at small $m/T$. This is another proof that the hydrodynamic formulas are correct.
    \item Both parameters decrease by increasing the chemical potential (or equivalently the charge density). This is expected from hydrodynamics since:
    \begin{equation}
        \mathcal{A}\,\sim\,\omega_c^{-1}\,,\quad \quad \mathcal{B}\,\sim\,\omega_c^{-1}\,,\quad \quad \omega_c\,\sim\,\rho\,\sim\,\mu\,.
    \end{equation}
    Our numerical results fully support this scaling.
\end{itemize}
In summary, the results shown in figure~\ref{m3} and figure~\ref{m4} confirm the hydrodynamic behavior:
\begin{equation}
    \mathrm{Re}\,[\omega_+]\,=\,\omega_c\,+\,\frac{(v_\parallel^2+v_\perp^2)}{2\,\omega_c}\,k^2\,+\,\dots\,,\quad \quad \mathrm{Re}\,[\omega_-]\,=\,\frac{v_\perp\,v_\parallel}{\omega_c}\,k^2\,+\,\dots\,,
\end{equation}
and suggest that the physics of our holographic model is indeed captured by such effective description.
\begin{figure}
    \centering
    \includegraphics[width=0.4\linewidth]{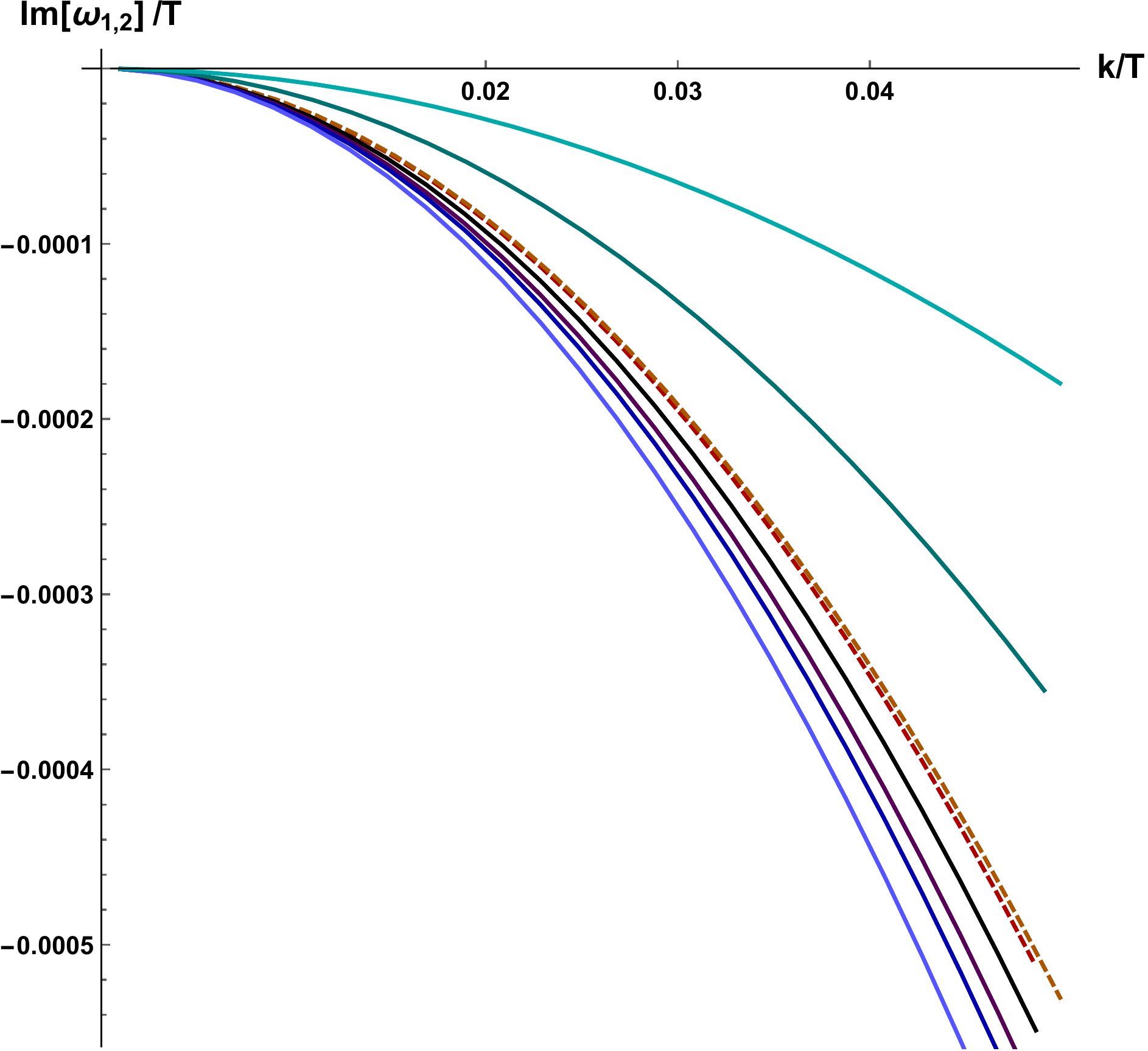}
    \quad 
    \includegraphics[width=0.4\linewidth]{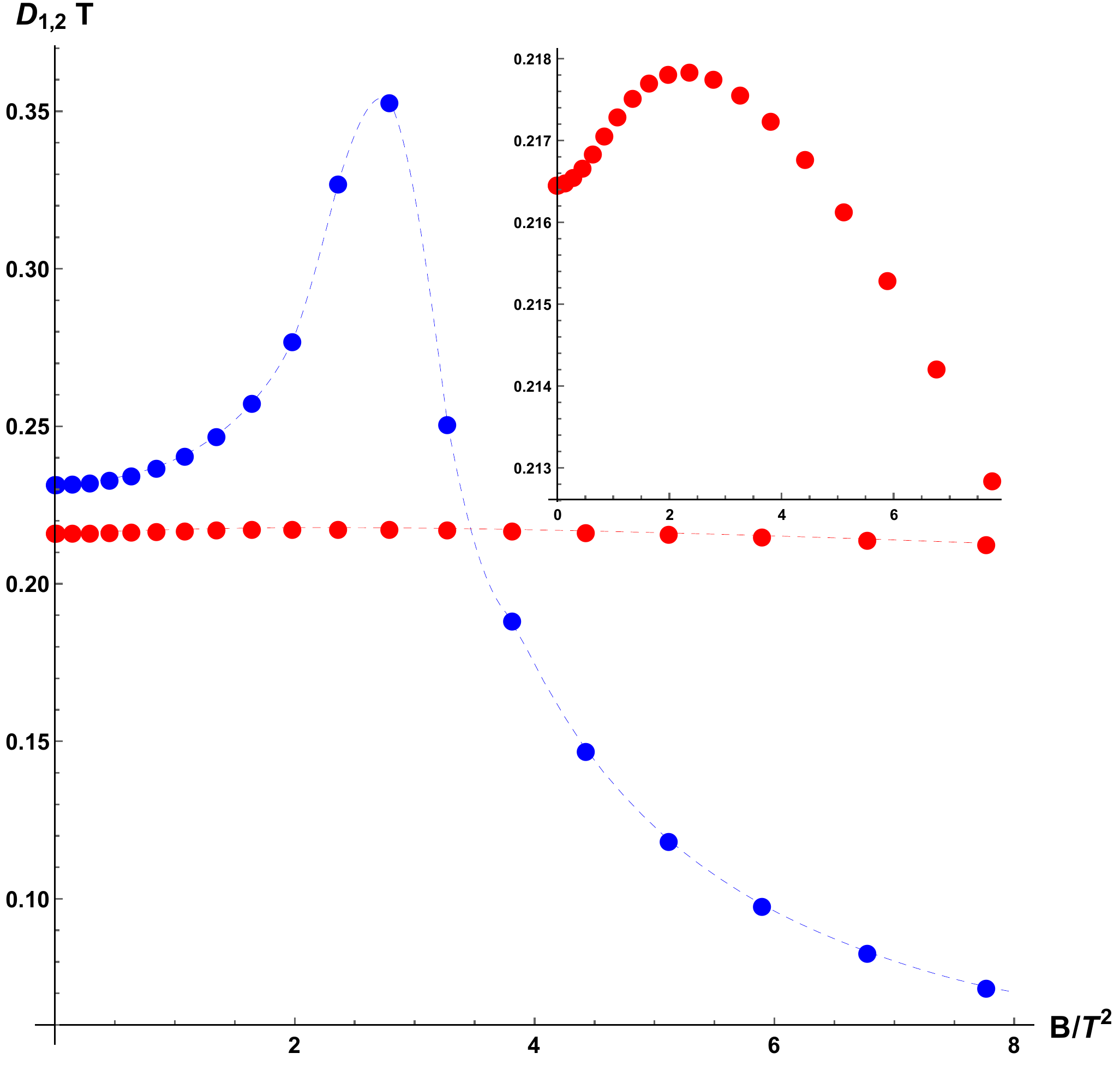}
    \caption{The dynamics of the coupled charge and crystal diffusive modes increasing the magnetic field $B$. We fix $\mu/T=1$ and $m/T=1$ moving $B/T^2 \in [0, 8]$. \textbf{Left: } The dispersion relation increasing $B/T^2$. The first mode is displayed with dashed lines and it moves from red $\rightarrow$ orange. The second mode is shown with filled lines and it moves from black $\rightarrow$ blue $\rightarrow$ light blue. \textbf{Right: } The diffusion constants as a function of the magnetic field $B$. The red dots correspond to the mode displayed with dashed lines in the left panel, and the blue dots to the filled lines.}
    \label{m6}
\end{figure}

Finally, we show in figure~\ref{m6} the dynamics of the coupled charge and crystal diffusive modes by increasing the magnetic field $B$. Interestingly, even in presence of a strong magnetic field both the modes remain gapless and diffusive; no damping appears. The diffusion constant of the modes which connecting to crystal diffusion (red dots in the right panel of figure~\ref{m6}) is almost unaffected by the presence of the small magnetic field. Its diffusion constant grows very slowly by increasing $B/T^2$ (see inset in the right panel of figure~\ref{m6}). It reaches a maximum around $B/T^2 \sim 2.7$ and then slowly decreases. On the contrary, the diffusion constant of the mode related to charge diffusion (blue dots in the right panel of figure~\ref{m6}) is strongly modified by the magnetic field, and displays a similar, but much more pronounced, non-monotonic behavior. It shows a maximum around the same position $B/T^2 \sim 2.7$, and then decreases to zero at large $B$.\\

\section{The magnetophonon peak and the effects of the magnetic field}\label{sec:peak}
In this section, we introduce a small external source of explicit breaking parametrized by the dimensionless parameter $\alpha$. We will always consider the pseudo-spontaneous regime defined in \eqref{ps}.

We focus our analysis on the electric conductivities:
\begin{equation}
    \mathcal{J}_i\,=\,\sigma_{ij}\,E_j\,,
\end{equation}
and~\cite{Baumgartner:2017kme}, in particular, on the longitudinal conductivity $\sigma_{xx}(\omega)$ and the Hall conductivity $\sigma_{xy}(\omega)$, defined via the following Kubo formulas:
\begin{align}
    \sigma_{xx}(\omega)\,=\,\frac{i}{\omega}\,\langle \mathcal{J}_x\,\mathcal{J}_x\,\rangle\,,\quad \sigma_{xy}(\omega)\,=\,\frac{i}{\omega}\,\langle \mathcal{J}_x\,\mathcal{J}_y\,\rangle\,.
\end{align}
Both correlators above may be computed using the standard holographic techniques (see appendix \ref{app1} for details).

\begin{figure}
    \centering
    \includegraphics[width=0.3\linewidth]{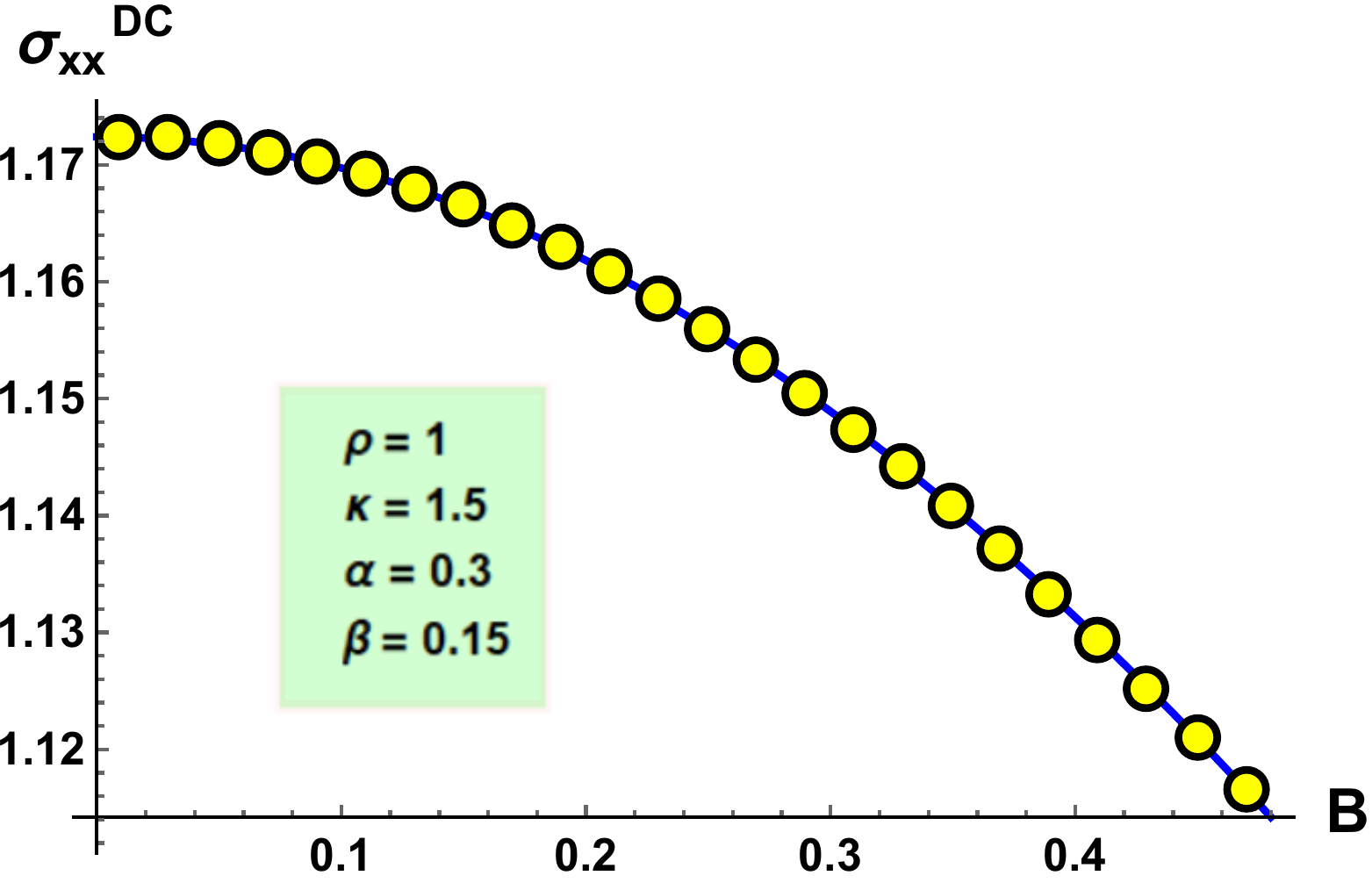}
    \quad
    \includegraphics[width=0.3\linewidth]{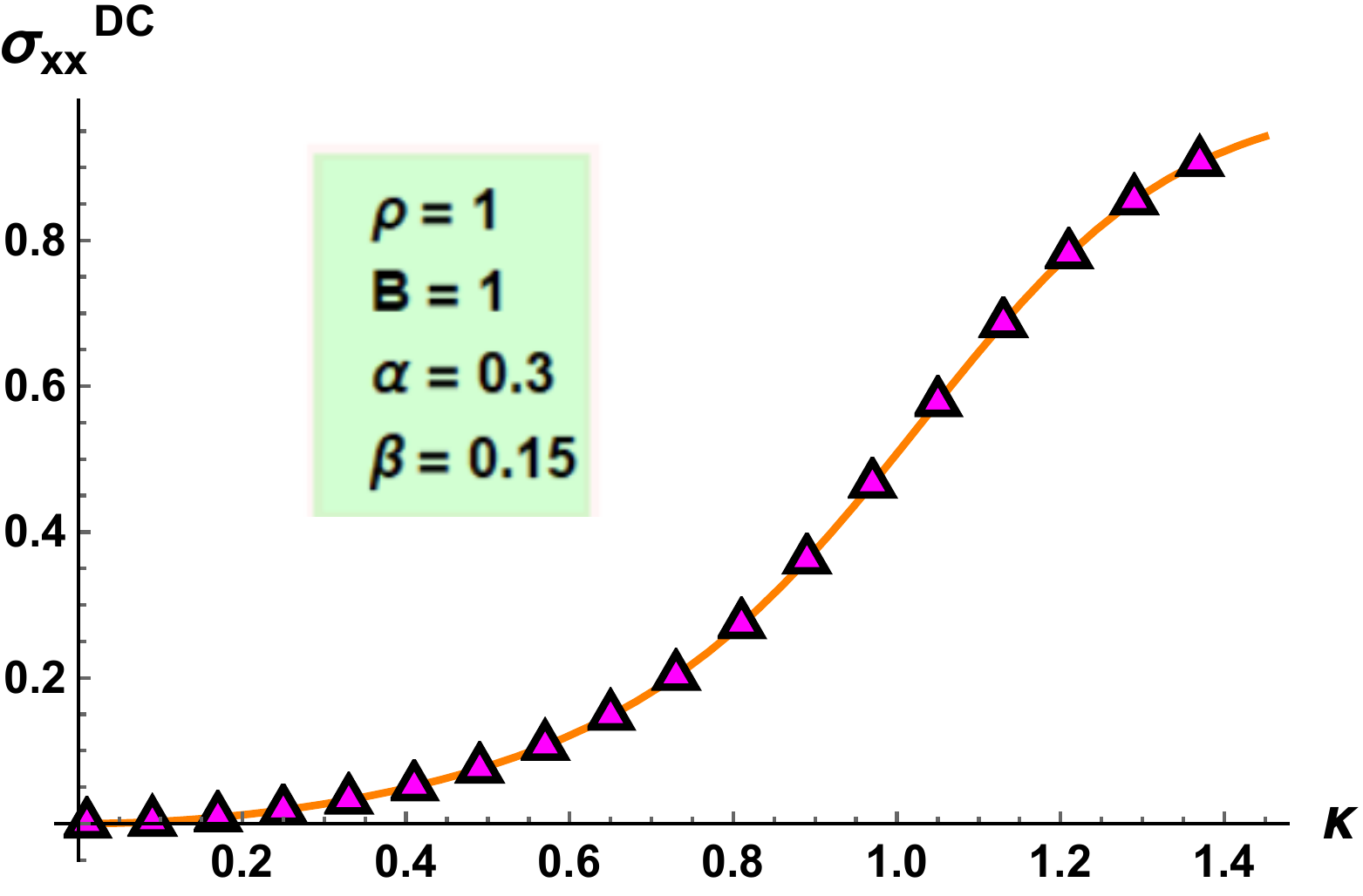}
    \quad
    \includegraphics[width=0.3\linewidth]{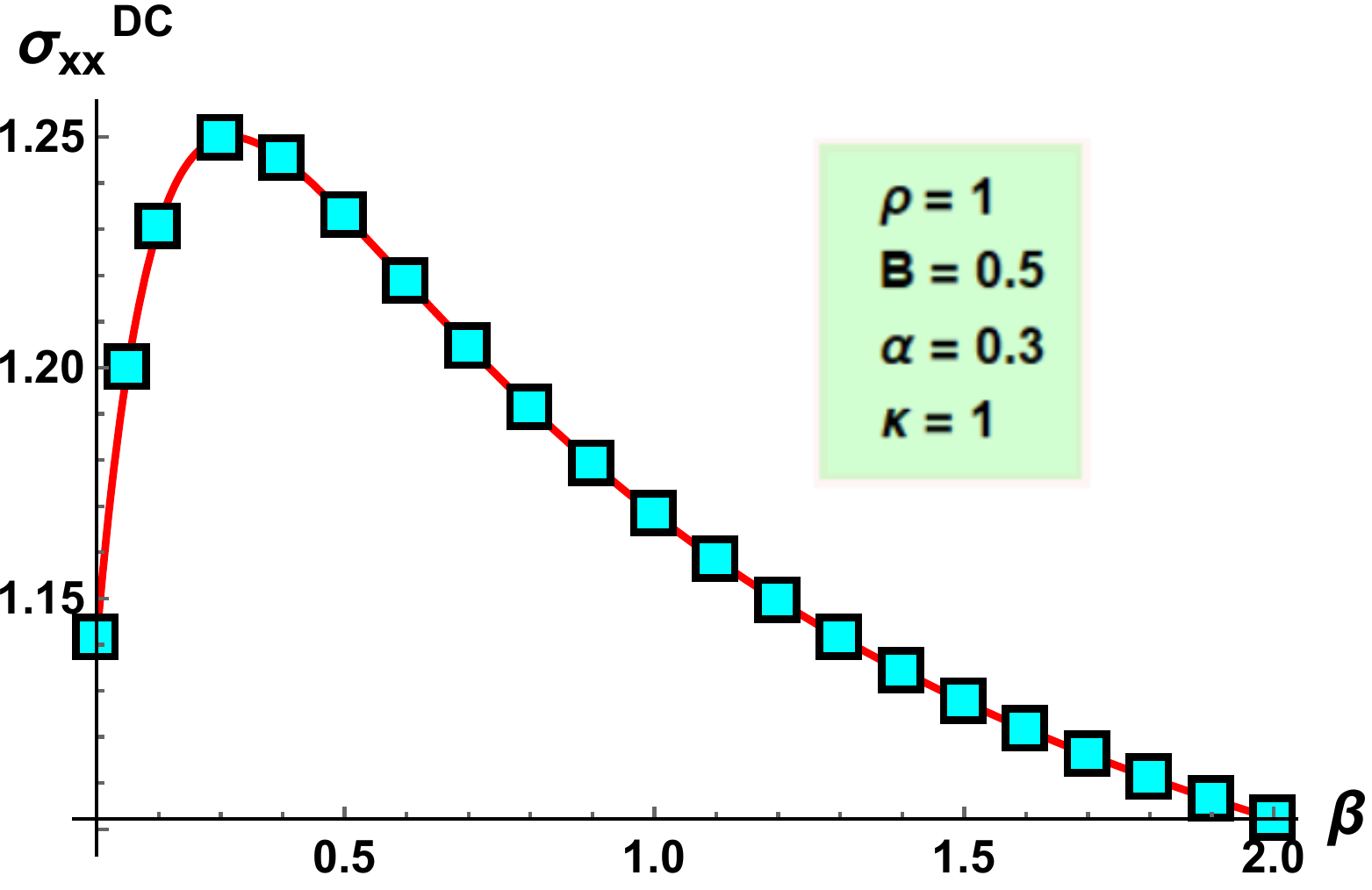}
    
    \vspace{0.3cm}
    
    \includegraphics[width=0.3\linewidth]{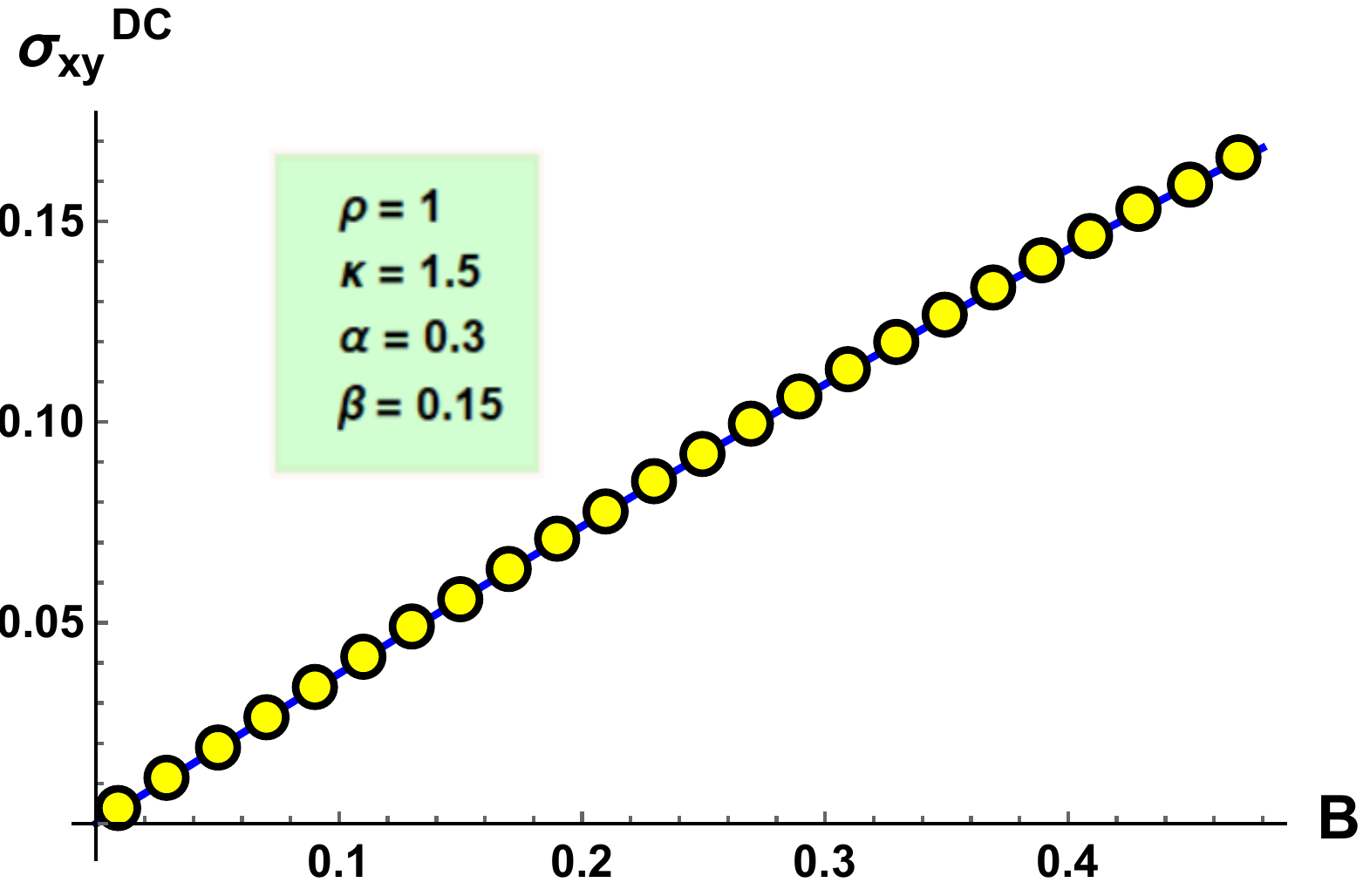}
    \quad
    \includegraphics[width=0.3\linewidth]{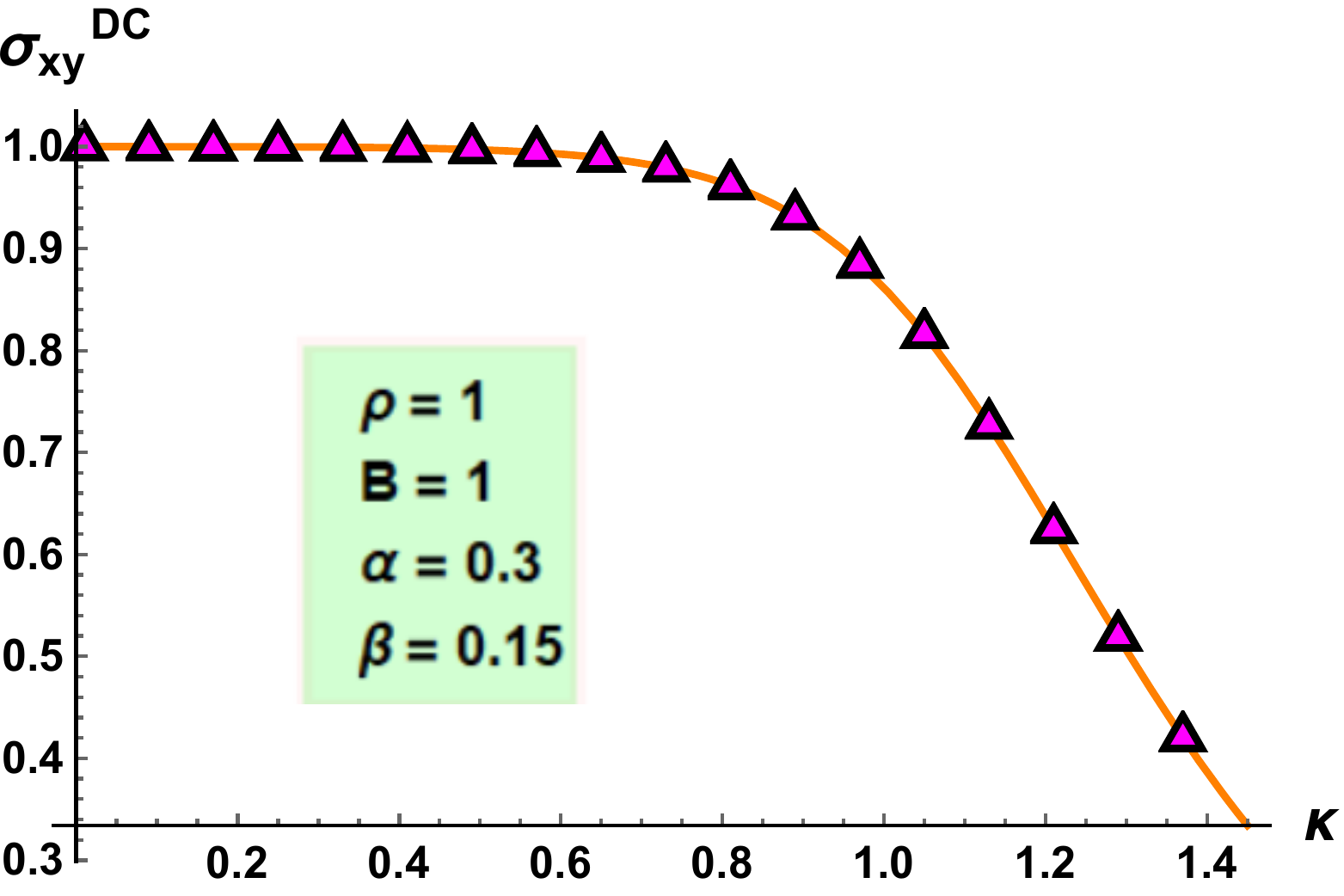}
    \quad
    \includegraphics[width=0.3\linewidth]{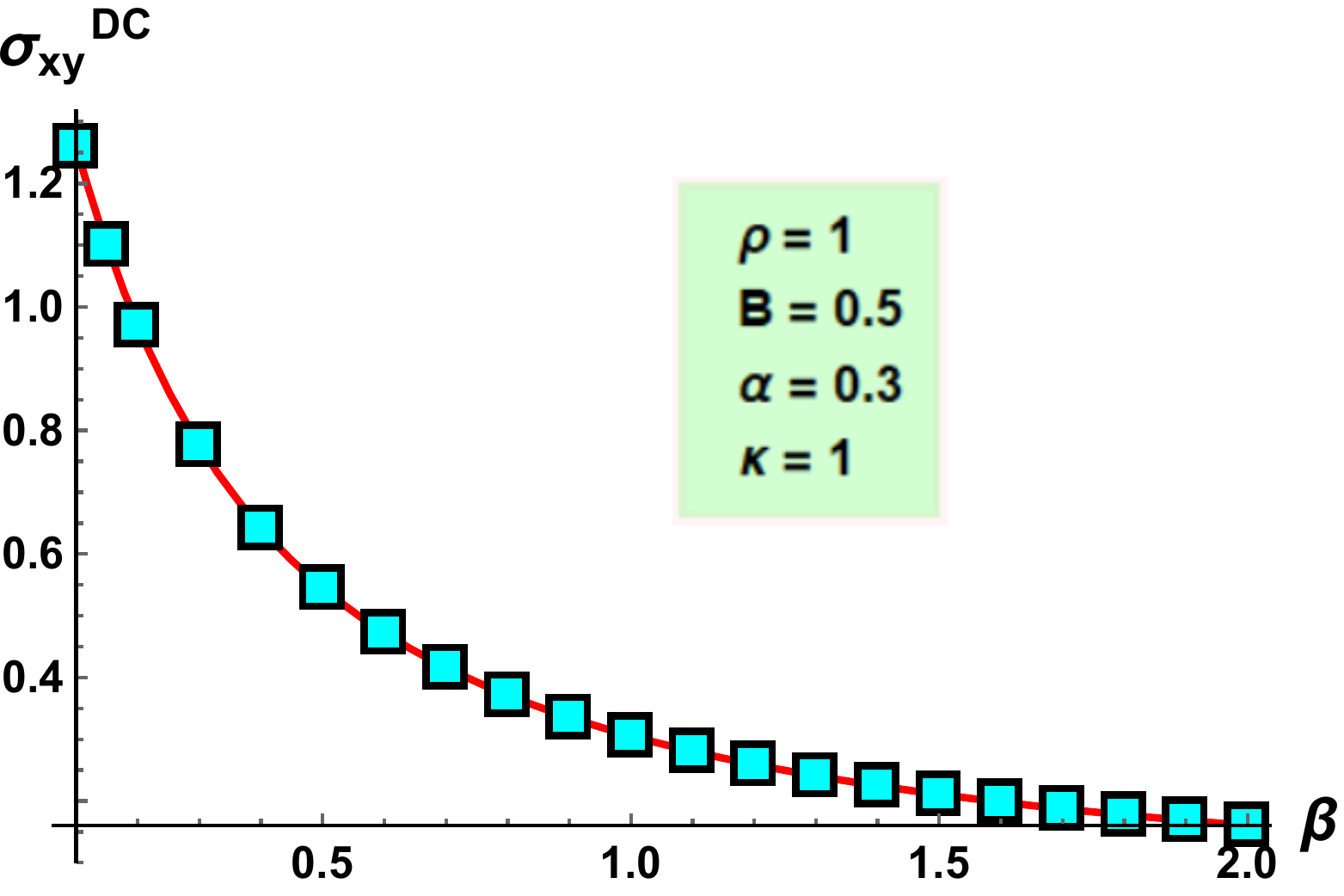}
    \caption{Comparison between the numerical conductivities at $\omega=0$ (markers) and the DC formulas in equations \eqref{DCformula} (lines). For simplicity we fixed $u_h=1$. The agreement is excellent.}
    \label{fig1b}
\end{figure}

The DC ($\omega=0$) values of such conductivities can be obtained using the methods of \cite{Donos:2014cya} and they read
\begin{align}
   & \sigma_{xx}^{DC}\,=\,\frac{\kappa^2\,V'\,g_{xx}\,\left(B^2\,+\,\kappa^2\,V'\,g_{xx}\,+\,\rho^2\right)}{B^2\,\rho^2\,\,+\,\left(B^2\,+\,\kappa^2\,g_{xx}\,V'\,\right)^2}\,\big|_{u_h},\quad \sigma_{xy}^{DC}\,=\,B\,\rho\,\frac{\left(B^2\,+\,2\,\kappa^2\,V'\,g_{xx}\,+\,\rho^2\right)}{B^2\,\rho^2\,\,+\,\left(B^2\,+\,\kappa^2\,g_{xx}\,V'\,\right)^2}\,\big|_{u_h}\,,
    \label{DCformula}
\end{align}
generalizing the expressions in \cite{Blake:2015ina,Amoretti:2016cad}. Recall that $\kappa$ determines the coupling of the linear axions to gravity~\eqref{eq:bulkscalar} and that we fixed $m=1$ throughout this section. In the limit where translations are not broken (i.e. $\kappa\to0$), we recover the standard result
\begin{equation}
    \sigma_{xx}\,=\,0\,,\quad \sigma_{xy}\,=\,\frac{\rho}{B}\,\equiv \nu\,,
\end{equation}
which was already found in~\cite{Hartnoll:2007ih}, with $\nu \equiv \rho/B$ being the filling fraction. As a warm-up, we check our numerical results with the DC formulas provided in \eqref{DCformula}. The comparison is shown in figure~\ref{fig1b} for a random sample of data. The agreement is very good, confirming that our numerics are reliable.

Next, we turn into studying the effects of the magnetic field on the pseudo-phonon peak manifesting in the electric conductivities. At zero magnetic field, the structure of the low-lying transverse excitations in the system is well described by the hydrodynamic expression
\begin{equation}
    (i\,\omega\,+\,\Gamma)\,\left(i\,\omega\,+\,\Omega\right)\,+\,\omega_0^2\,=\,0\,,
\end{equation}
as confirmed numerically in~\cite{Baggioli:2019abx,Ammon:2019wci}. $\Gamma$ is the momentum relaxation rate, corresponding to the explicit breaking of spacetime translations $x \rightarrow x+a$. Analogously, $\Omega$ is the ``phase relaxation'' rate, associated to the explicit breaking of the internal shit symmetry $\phi \rightarrow \phi+b$. $\omega_0$ is the already mentioned pinning frequency which is responsible for the off-axes peak in the electric conductivity.

\begin{figure}
    \centering
    \includegraphics[width=0.45\linewidth]{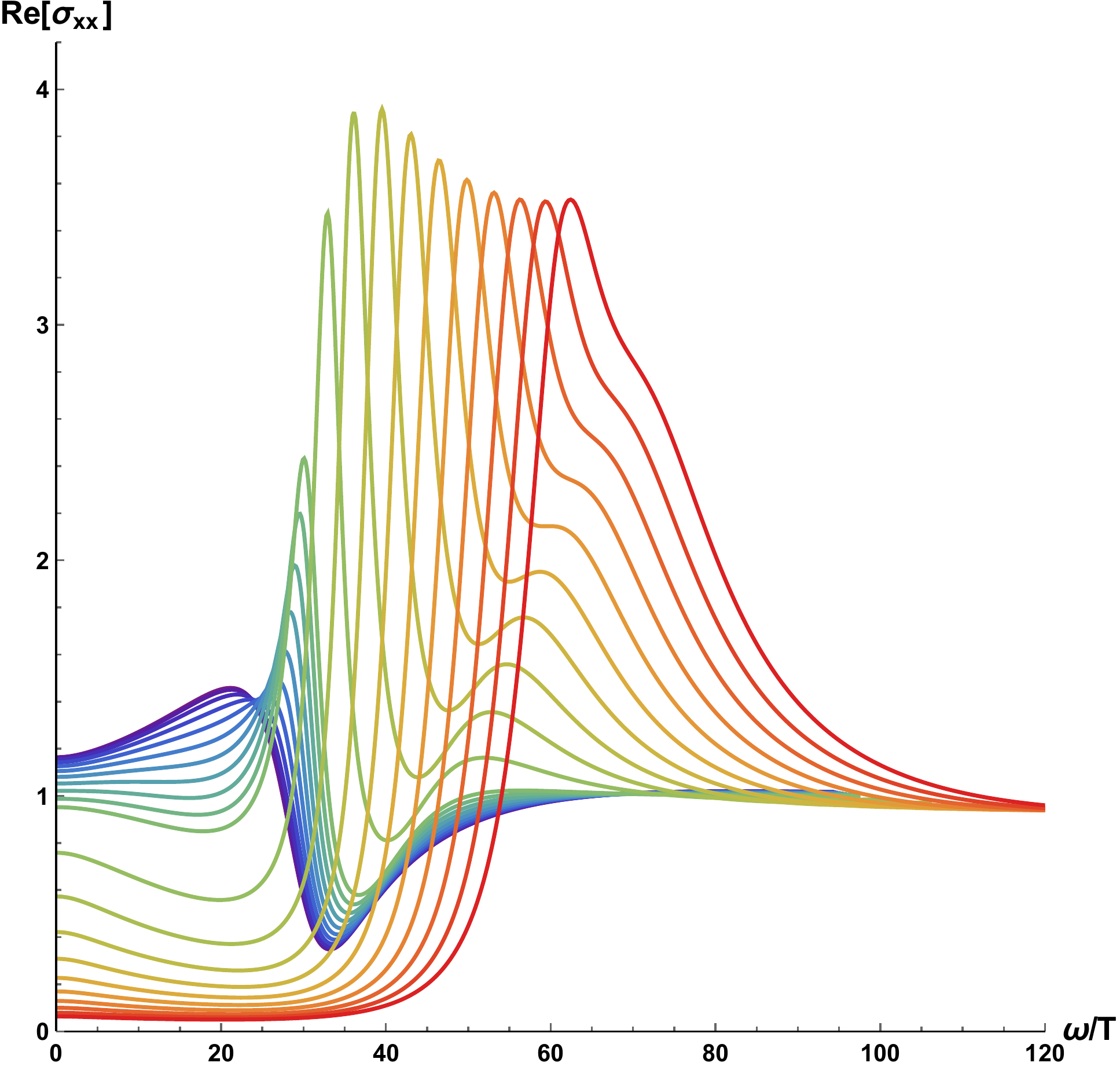}
\quad  \includegraphics[width=0.45\linewidth]{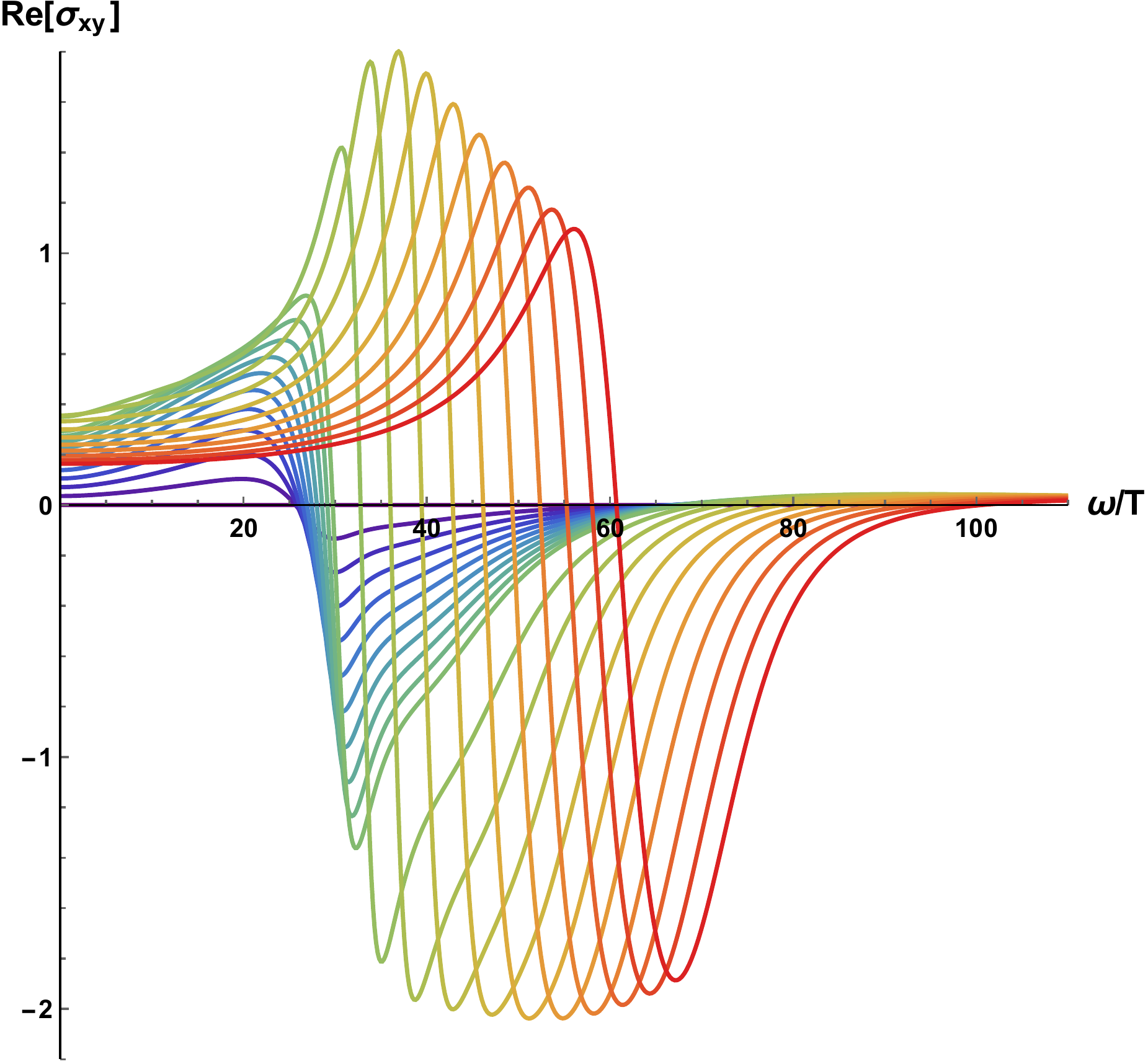}
    \caption{The evolution of the pseudo-phonon peak increasing the magnetic field. We move the magnetic field $B/T^2 \in [0,5\times10^3]$ (from blue to red). The other parameters $\rho/T^2=987,\kappa/T=31.4,\alpha=0.1, \beta=2$ are kept fixed.}
    \label{fig2b}
\end{figure}

We present the effects of the magnetic field $B$ on the electric conductivities in figure~\ref{fig2b}. We start from a choice of parameters which exhibit a clear pseudo-phonon peak at $B=0$, visible as the blue line of figure~\ref{fig2b}, and then increase the value of the dimensionless magnetic field $B/T^2$ until very large values. We observe that the position of the peak, expressed as the maximum of the longitudinal conductivity $\mathrm{Re}[\sigma_{xx}]$ \footnote{Note that the maximum in the conductivities does not coincide with the value of the pinning frequency $\omega_0$ but it is rather a result of a complicated interplay between the various parameters. However, the measured peak is not the pinning frequency itself but the maximum of the conductivity \cite{Chen2005QuantumSO}.}, increases monotonically with the magnetic field $B$. On the contrary, the width of the peak, which determines the lifetime of the associated resonance, becomes first sharper and then starts increasing again at very large magnetic fields.

\begin{figure}
    \centering
    \includegraphics[height=5cm]{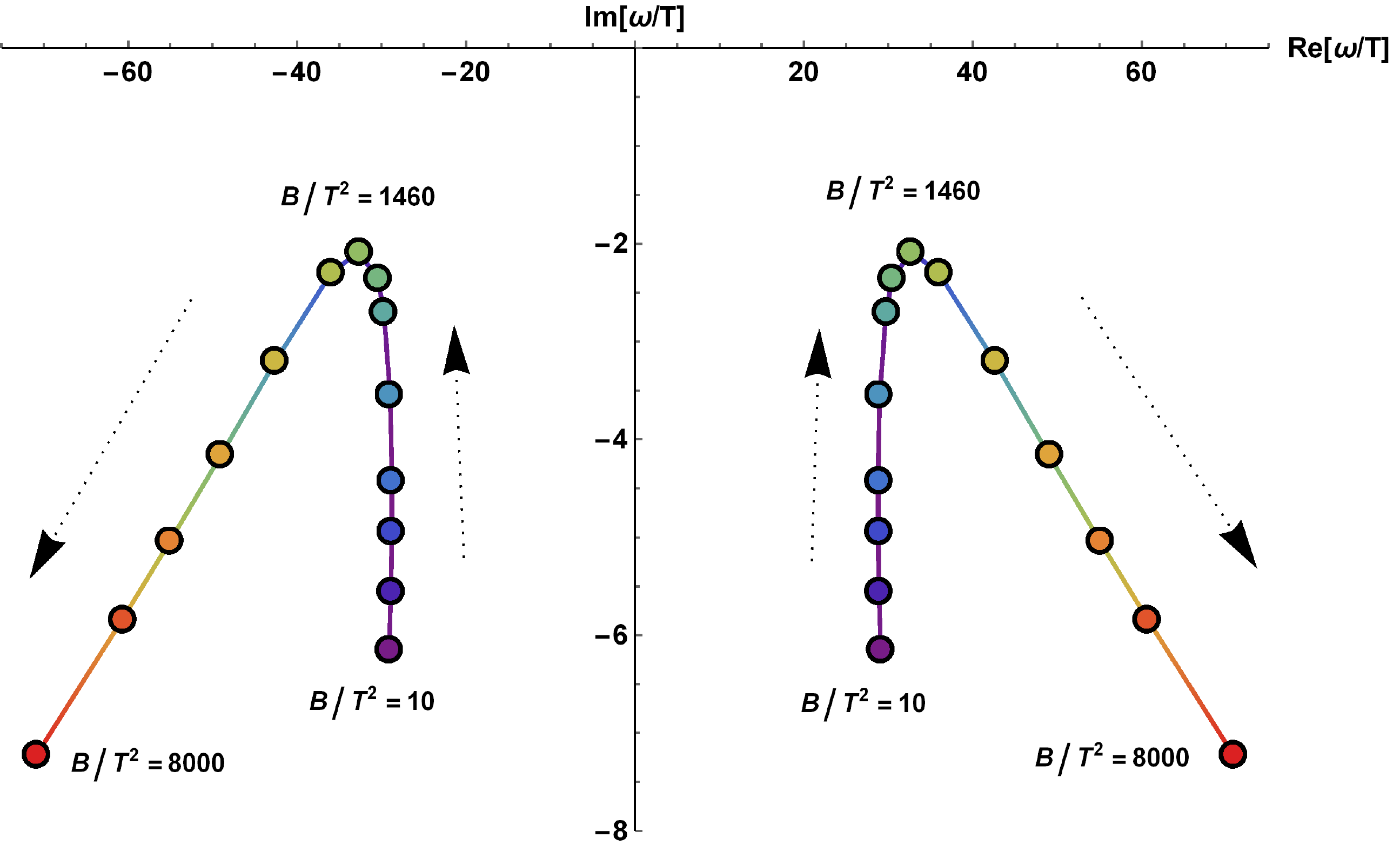}\qquad
    \includegraphics[height=5cm]{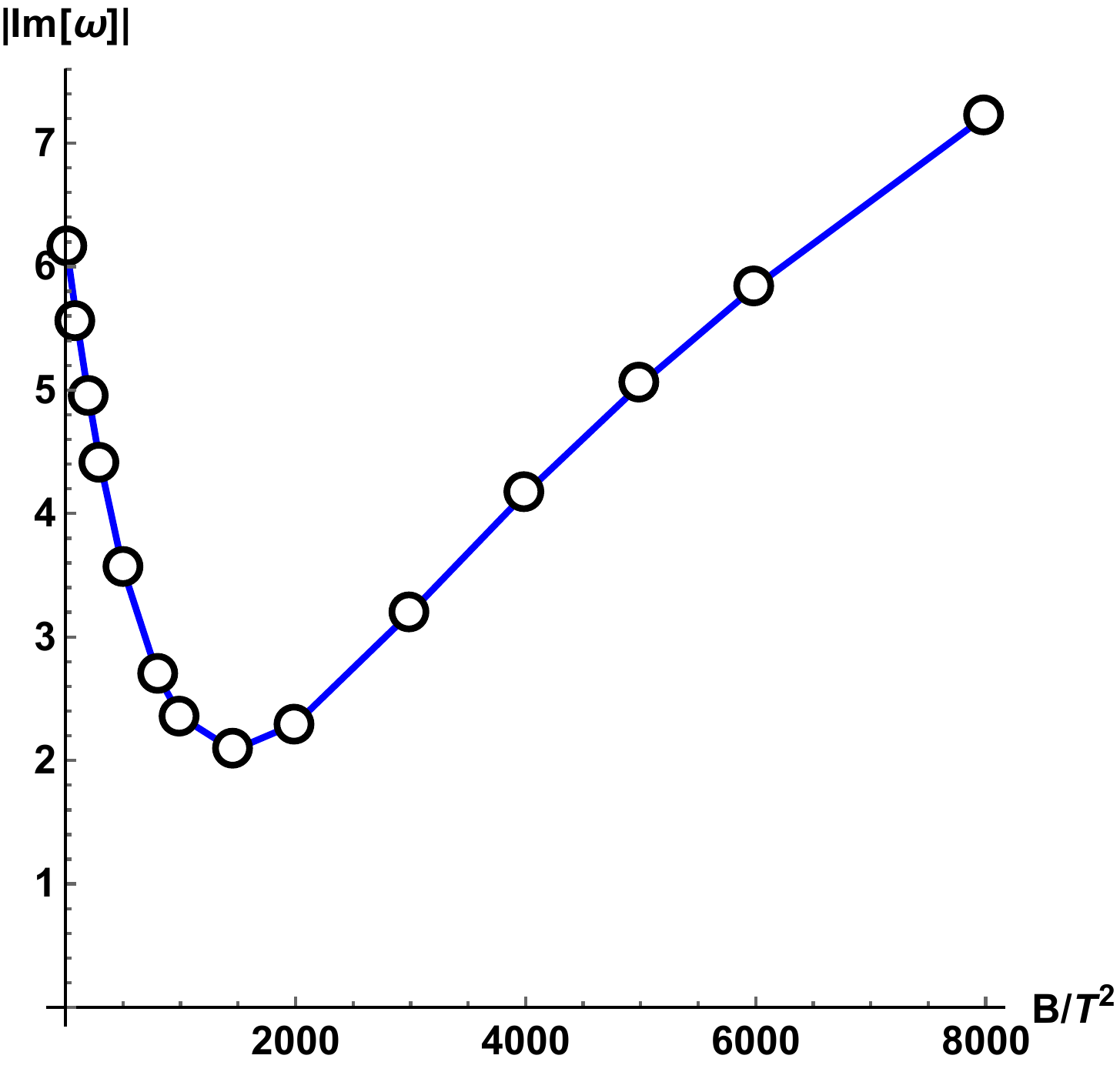}
    \caption{QNMs motion in the complex plane for the same parameters of figure~\ref{fig2b}. The dynamics is produced by dialing the dimensionless magnetic field $B/T^2$, while keeping other parameters fixed.}
    \label{figqnm}
\end{figure}

In order to confirm our results, we compute the quasinormal modes of the system at finite charge density and magnetic field. Our results are displayed in figure~\ref{figqnm}. The motion of the QNMs is consistent with what already found in the electric conductivities. The real part of the lowest QNM increases monotonically with the strength of the magnetic field, corresponding to the peak moving continuously to higher frequency in the electric conductivities. More interestingly, the imaginary part of the lowest pole is non-monotonic. The lifetime of the QNMs first becomes longer as a function of $B/T^2$, and then decreases at larger values of the magnetic field. This is again consistent with the width of the peak in the conductivity first becoming smaller and then increasing with $B$. Notice that this non-monotonic behavior of the imaginary part as a function of the magnetic field is quite typical, and has been observed also in the absence of translations breaking~\cite{Hartnoll:2007ip}.
\begin{figure}
    \centering
    \includegraphics[width=0.45\linewidth]{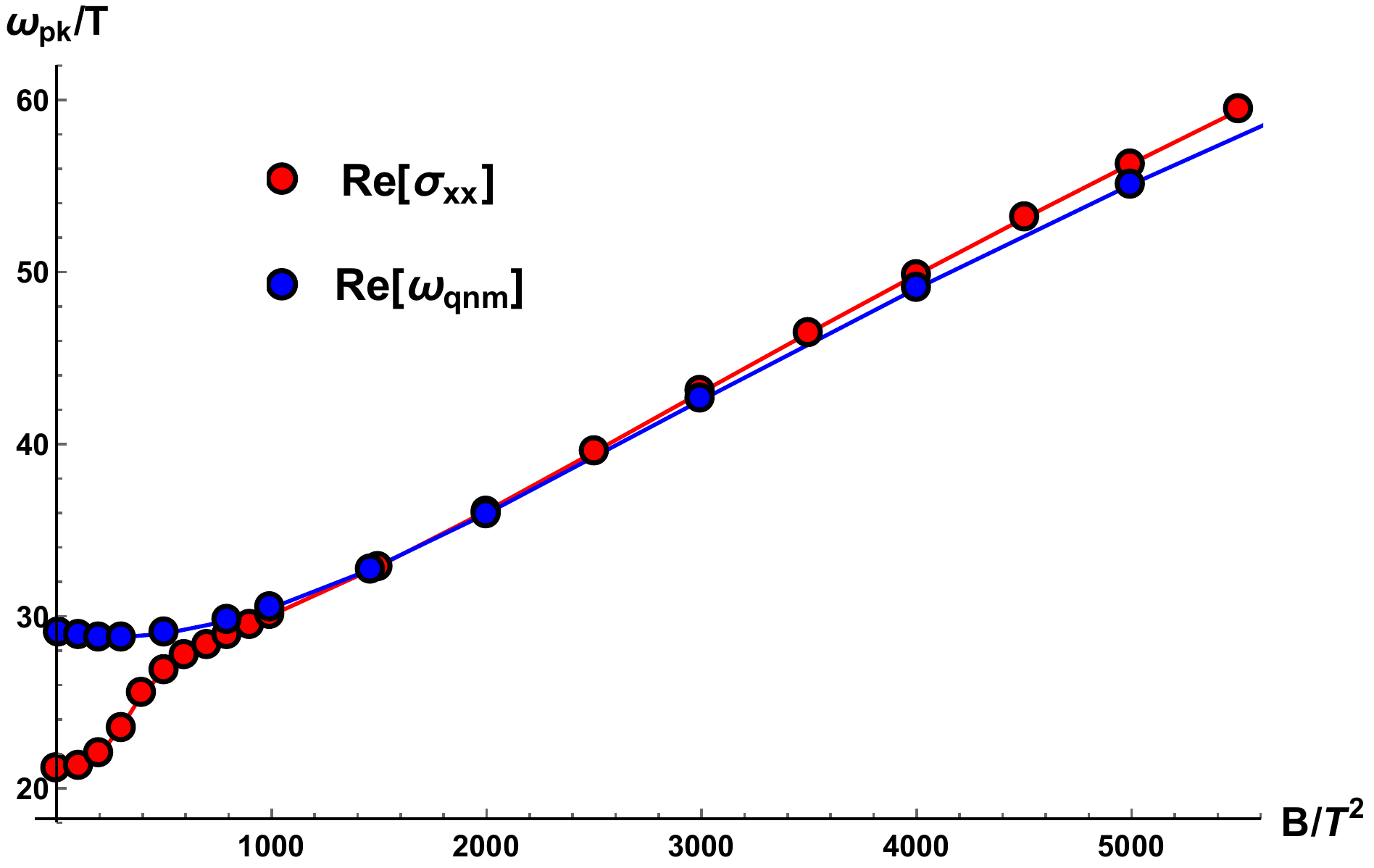}
    \quad \quad 
    \includegraphics[width=0.45\linewidth]{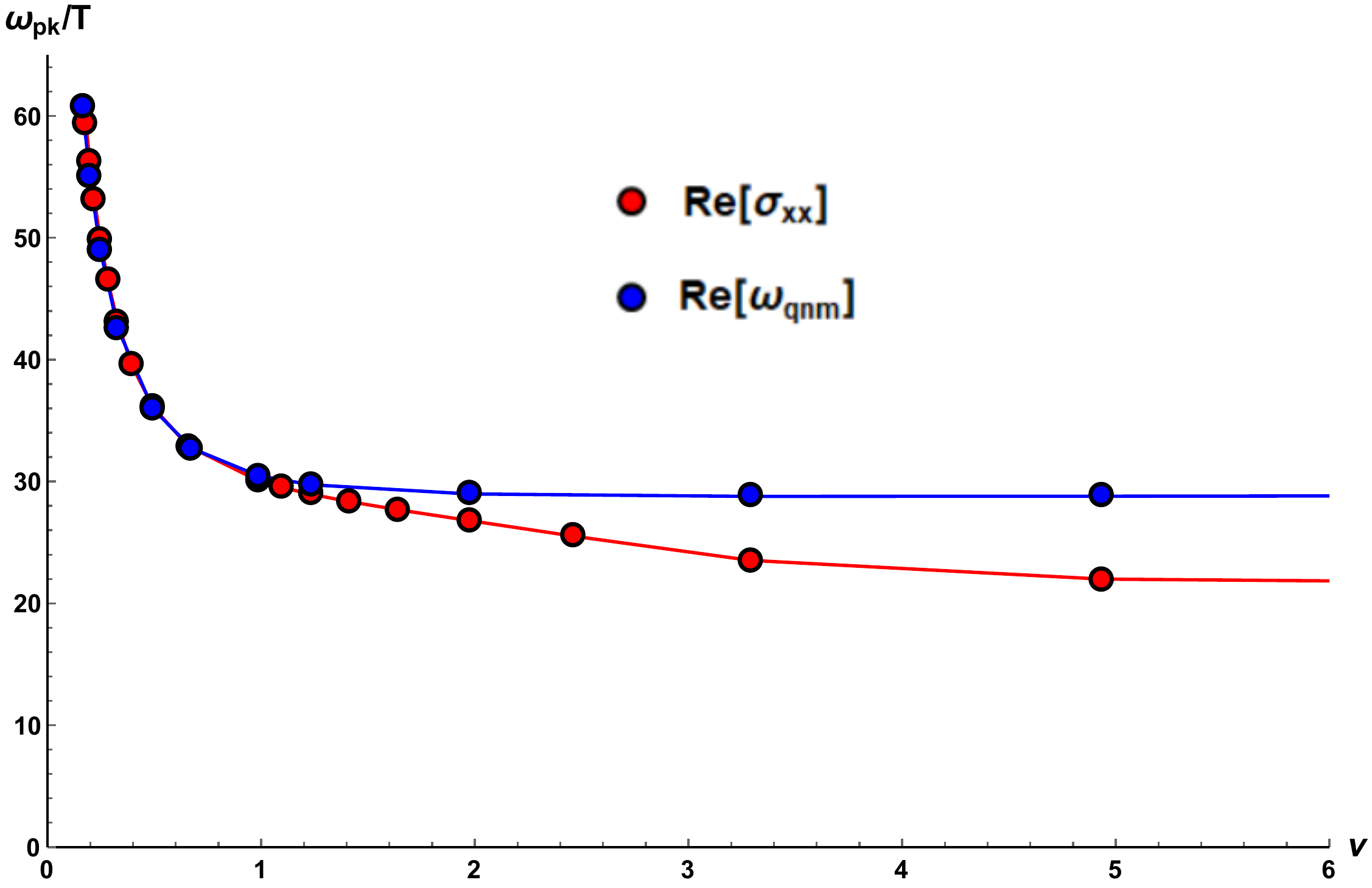}
    \caption{Dependence of the peak position $\omega_{pk}$ as a function of the magnetic field $B/T^2$ and the filling fraction $\nu\equiv \rho/B$. The peak is extracted both from the real part of the longitudinal conductivity (red dots) and from the quasinormal modes spectrum (blue dots).}
    \label{fig3}
\end{figure}

In figure~\ref{fig3}, we plot the position of the peaks $\omega_{pk}$ as a function of the magnetic field $B$ and the filling fraction $\nu$.  First, we compare the position of the peak extracted from the maximum of the longitudinal conductivity with the real part of the lowest QNM. The precise numerical values do not match as expected since generically
\begin{equation}
    \omega_{pk}\,\neq\,\mathrm{Re}[\omega_{qnm}]\,.
\end{equation}
Nevertheless, the qualitative trend as a function of the magnetic field $B$ is very similar. Moreover, at large magnetic field, $B/T^2 \gg 1$, the two almost coincide, meaning that the contribution of $\mathrm{Im}[\omega_{qnm}]$ to $\omega_{pk}$ becomes negligible. Interestingly, the position of the peak seems to saturate to a constant value for large values of the filling fraction, while it rapidly decreases at small values of $\nu$. Qualitatively, our results are in agreement with the experimental fits obtained in~\cite{Delacretaz:2019wzh} (see the right panel of figure~2 therein).

\begin{figure}
    \centering
   % \includegraphics[width=0.45\linewidth]{largeB1.pdf}
    %\quad \quad
    \includegraphics[width=0.45\linewidth]{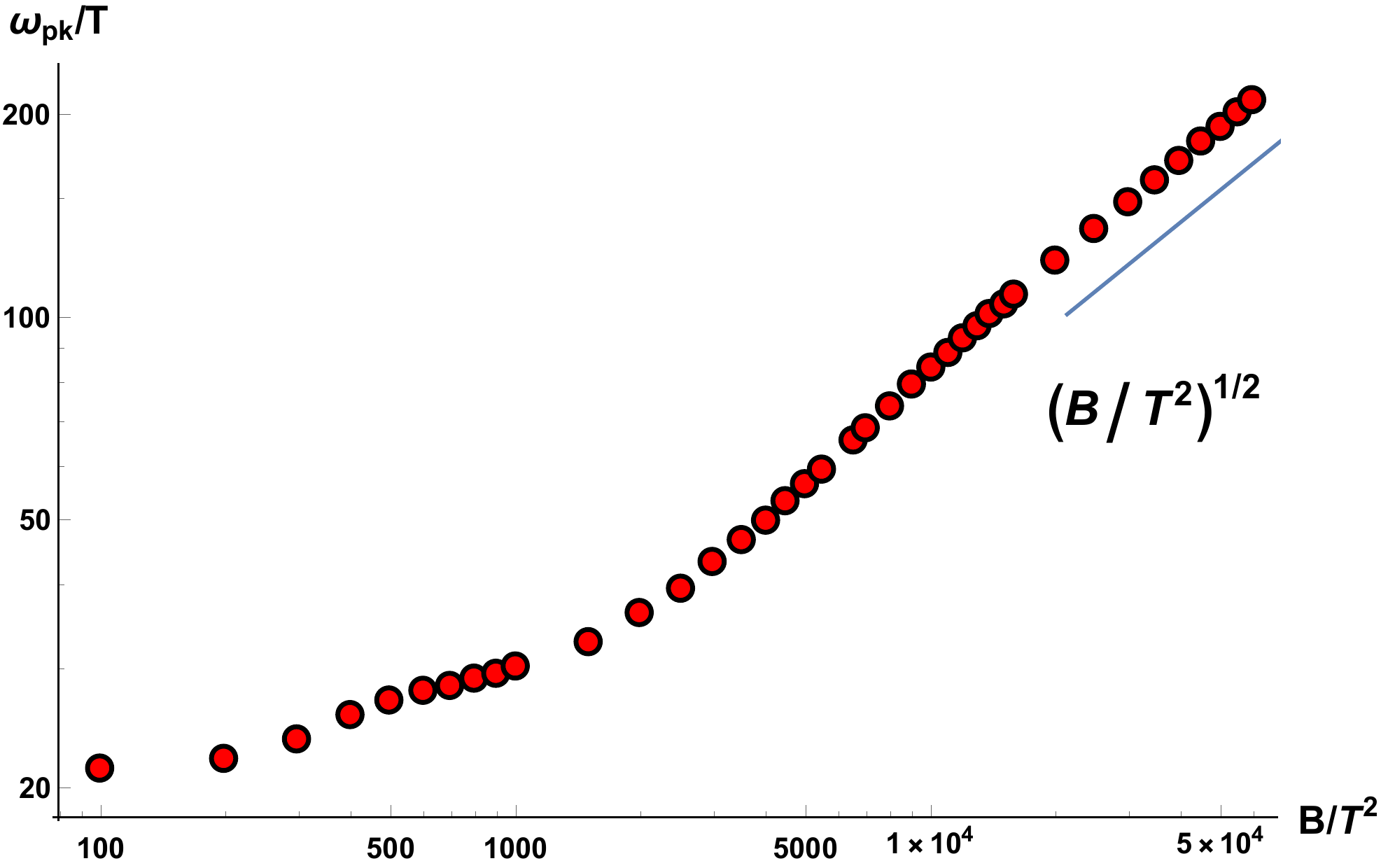}\quad\quad \includegraphics[width=0.45\linewidth]{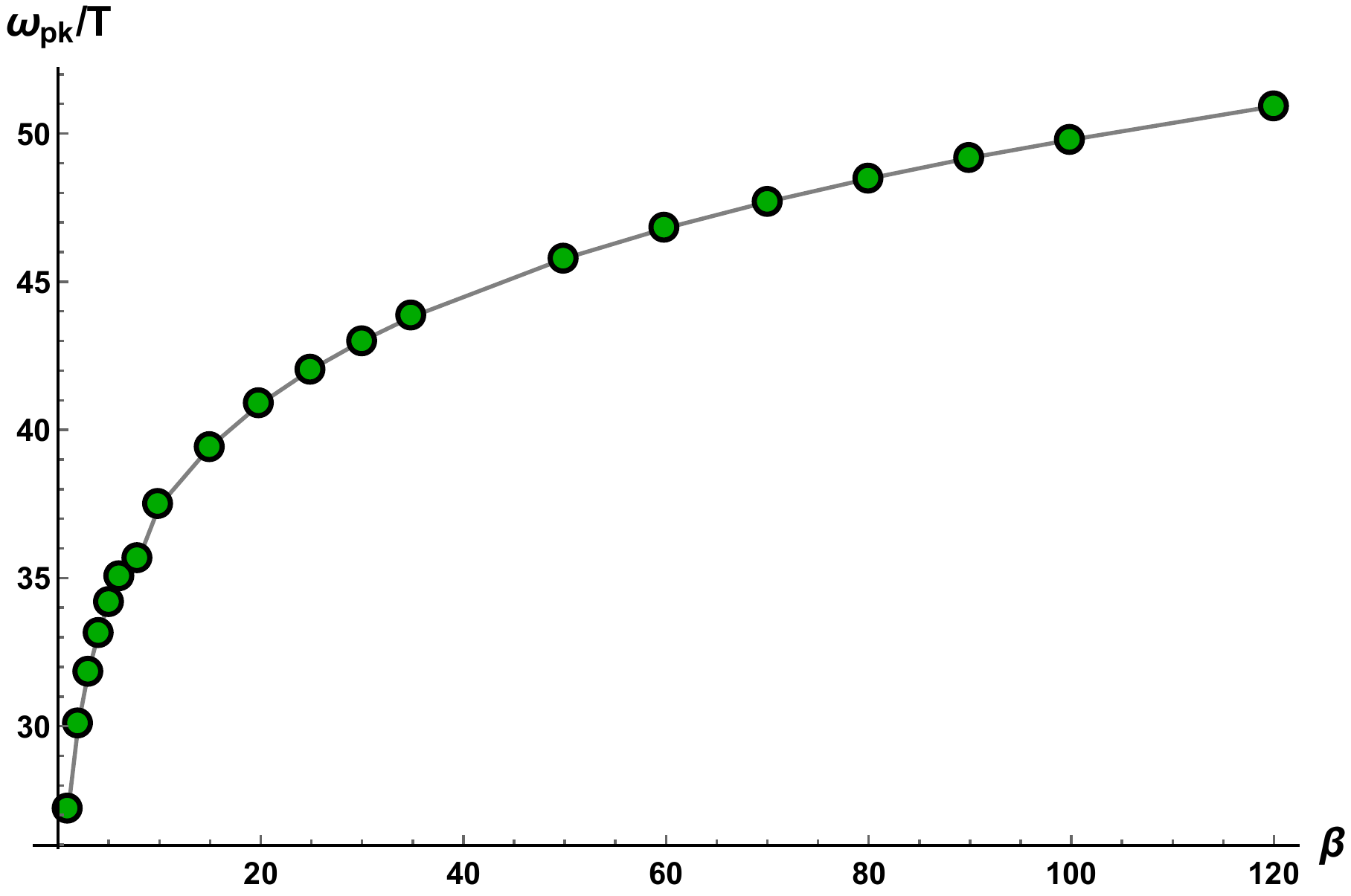}
    \caption{\textbf{Left: }Behavior of the peak in the longitudinal conductivity $Re[\sigma_{xx}]$ at large $B/T^2$. The asymptotic behavior is consistent with a scaling $\omega_{pk}\sim B^{1/2}$. \textbf{Right: }Dependence of the peak position $\omega_{pk}$ as a function of the parameter $\beta$, which determines the strength of the SSB -- the shear elastic modulus. The peak is extracted both from the real part of the longitudinal conductivity.}
    \label{fig3b}
\end{figure}
To make this analysis more precise, we follow the position of the magnetophonon peak $\omega_{pk}$ until very large values of $B$ in the left side of figure~\ref{fig3b}. The position of the peak is extracted as the maximum of $\mathrm{Re}[\sigma_{xx}(\omega)]$. At large magnetic field, a quite robust scaling
\begin{equation}
    \omega_{pk}\,\sim B^{1/2}\,,
\end{equation}
is identified, which is compatible with dimensional analysis. More precisely, for $B \gg \mu^2$ we can neglect the effects of the chemical potential and the only dimensionless constant we can construct is exactly $\omega/B^{1/2}$.
Notice that this scaling is incompatible with the results of ~\cite{Delacretaz:2019wzh}, and the idea that at large magnetic field the magnetophonon resonance becomes light, despite the presence of strong explicit breaking. As we will analyze further in the conclusions, our results are in agreement with certain experimental results, and suggest a precise interpretation of the nature of the ``disorder'' mimicked by these homogeneous holographic models.
\begin{figure}[h!]
    \centering
     \includegraphics[width=0.45\linewidth]{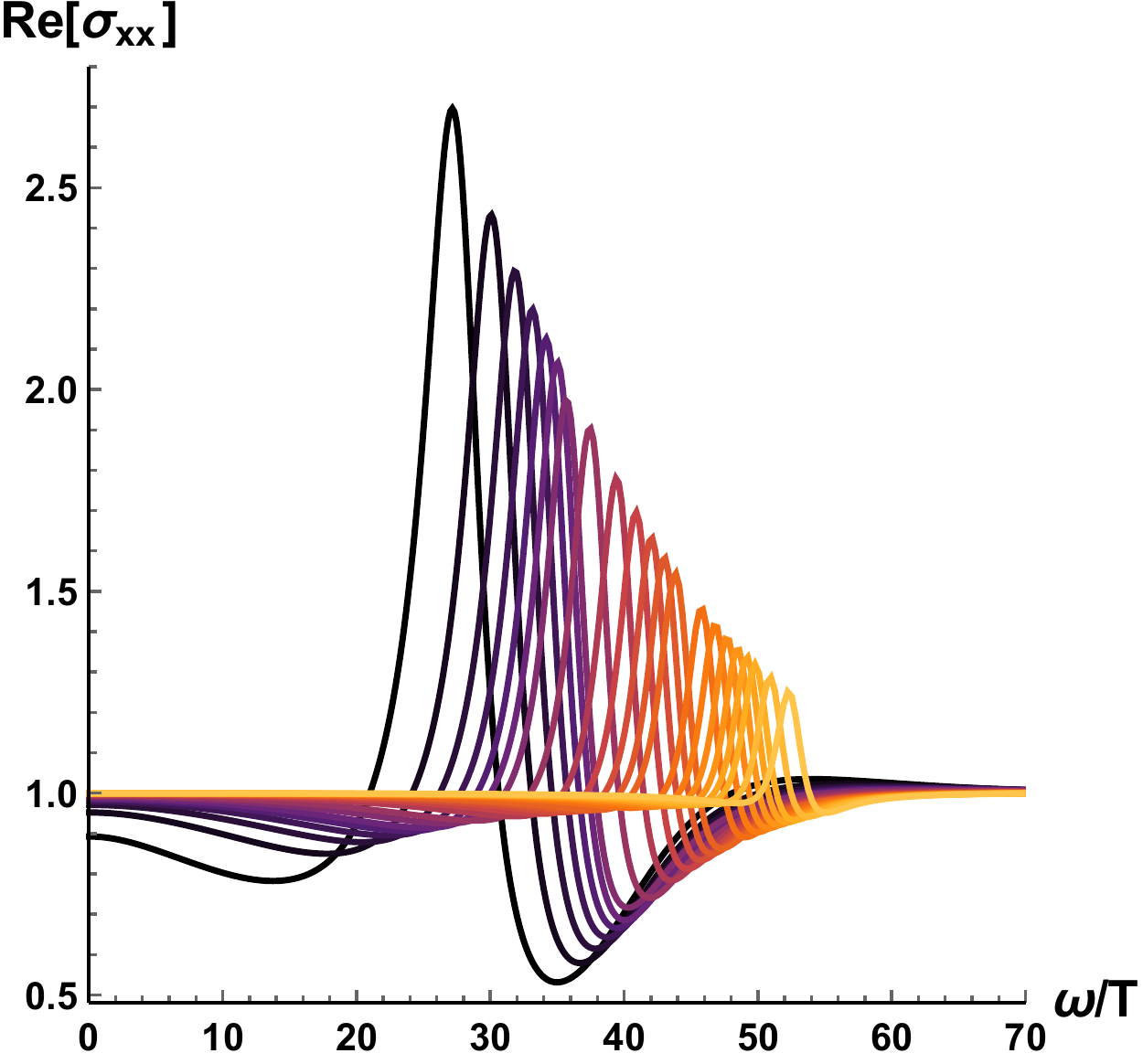}\quad \quad
    \includegraphics[width=0.45\linewidth]{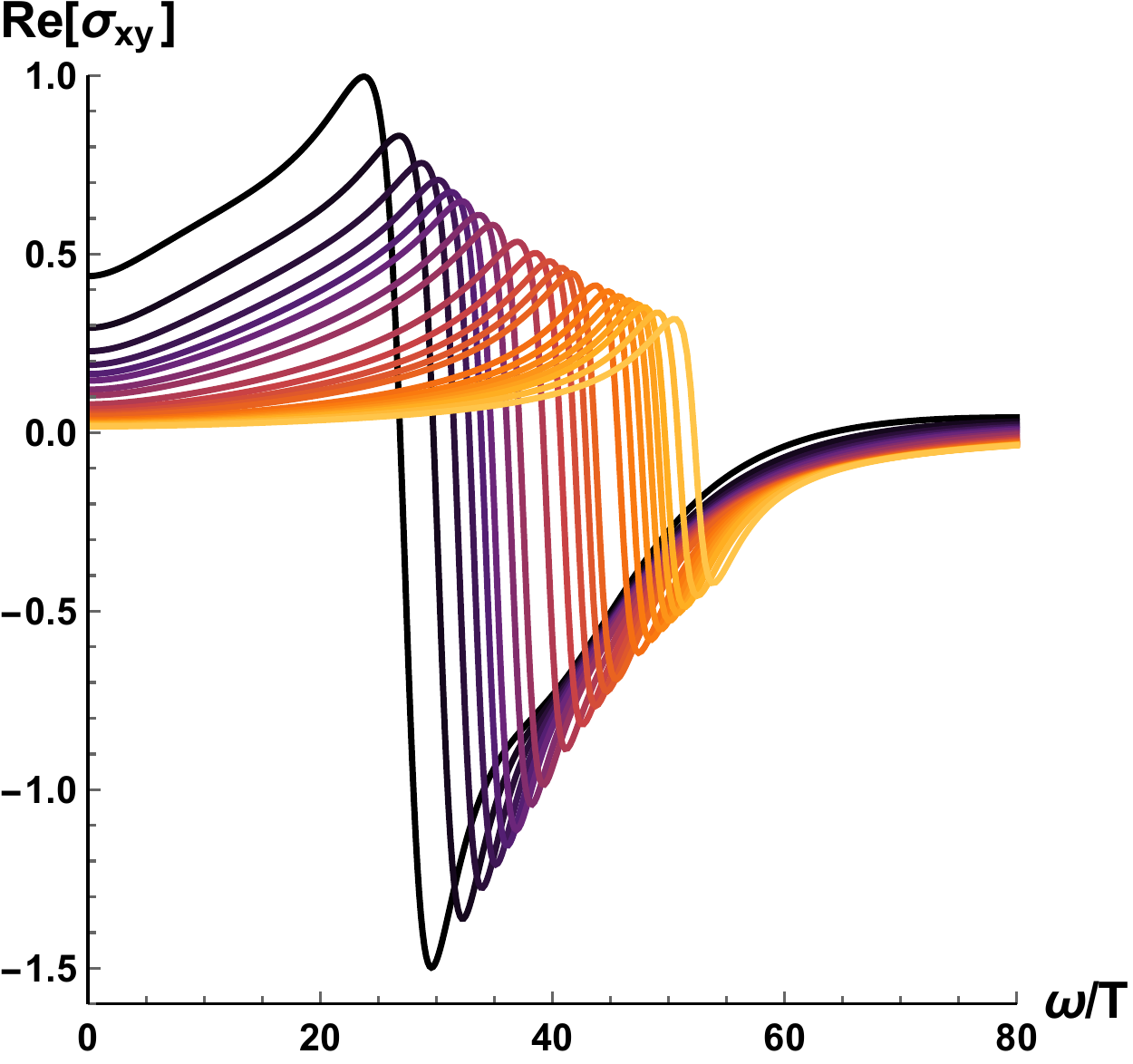}
    \caption{The evolution of the pseudo-phonon peak increasing the shear modulus. We move the parameter $\beta \in [1,150]$ (from black to yellow). Other parameters $\rho/T^2=987,\kappa/T=31.4,\alpha=0.1, B/T^2=10^3$ are kept fixed.}
    \label{fig4}
\end{figure}

We continue our analysis by studying the effects of the parameter $\beta$ on the magnetophonon peak. The parameter $\beta$ is dimensionless, and it parametrizes the strength of spontaneous symmetry breaking, or in other words, the rigidity of the dual system~\cite{Ammon:2019wci}.
In figure~\ref{fig4} we show the dynamics of both the longitudinal and Hall electric conductivities by moving the dimensionless parameter $\beta$ from small to large values. The position of the peak, given as the maximum of $\mathrm{Re}[\sigma_{xx}]$, moves towards higher frequency. The precise motion is depicted in the right side of figure~\ref{fig3b}. The scaling is consistent with
\begin{equation}
    \omega_{pk}\,\sim\,\sqrt{\beta}\,.
\end{equation}
In our setup, the spontaneous breaking scale $\langle SSB \rangle$ is indeed proportional to $\sqrt{\beta}$~\cite{Baggioli:2019abx,Ammon:2019wci}. Using the GMOR relation~\cite{PhysRev.175.2195}, we also know that $\omega_0^2 \sim \langle EXB \rangle\, \langle SSB \rangle \sim\sqrt{\beta}$, with $\langle EXB \rangle$ the explicit breaking scale that is independent of $\beta$. All in all, our numerical data supports the hydrodynamic prediction that ~\cite{PhysRevB.18.6245}
\begin{equation}
    \omega_{pk}\,\sim\,\frac{\omega_0^2}{\omega_c} \quad (\sim \sqrt{\beta})\,,
\end{equation}
and reproduces the correct scaling with the rigidity of the system.

\begin{figure}
    \centering
     \includegraphics[width=0.45\linewidth]{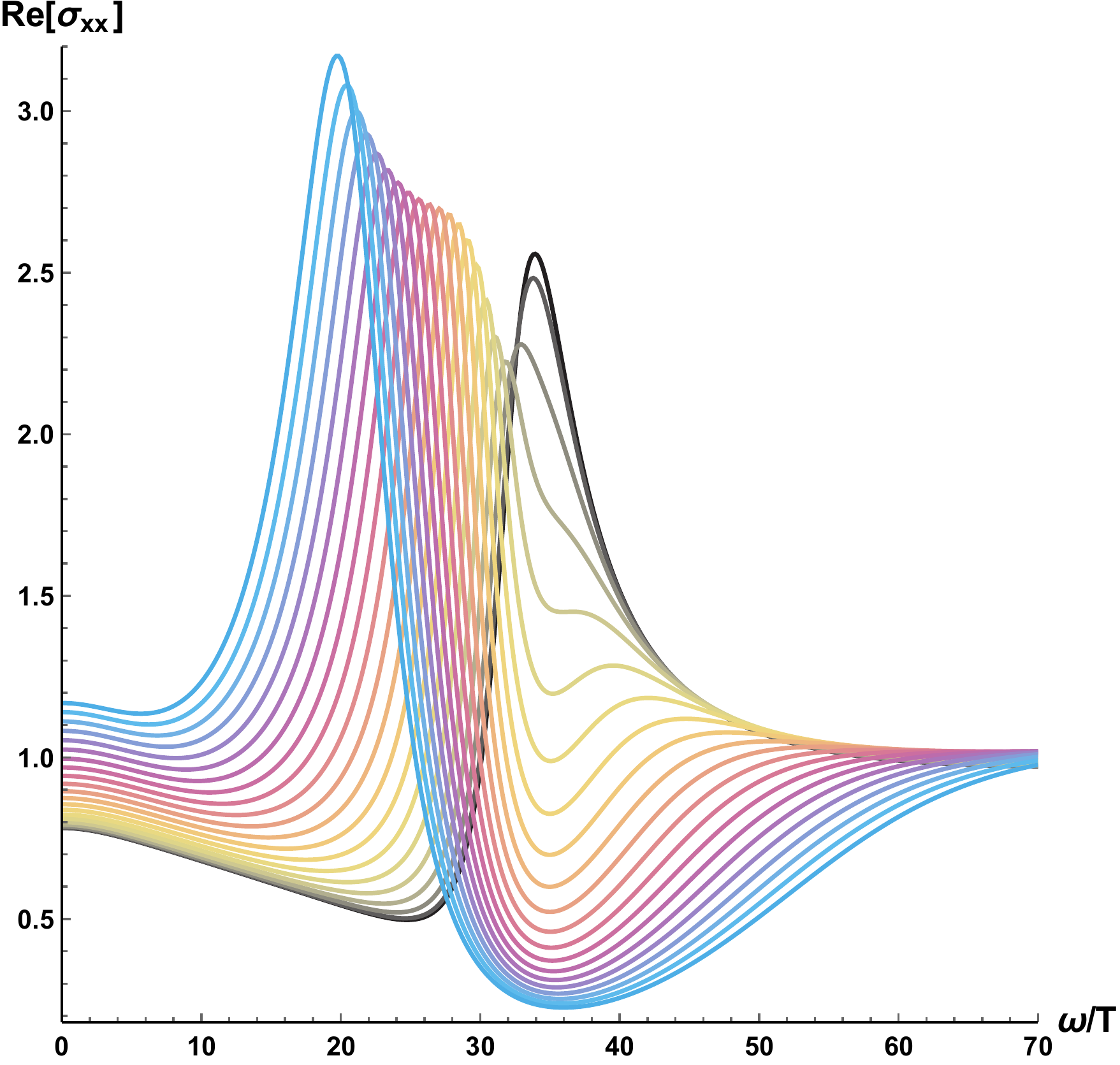}\quad \quad
    \includegraphics[width=0.45\linewidth]{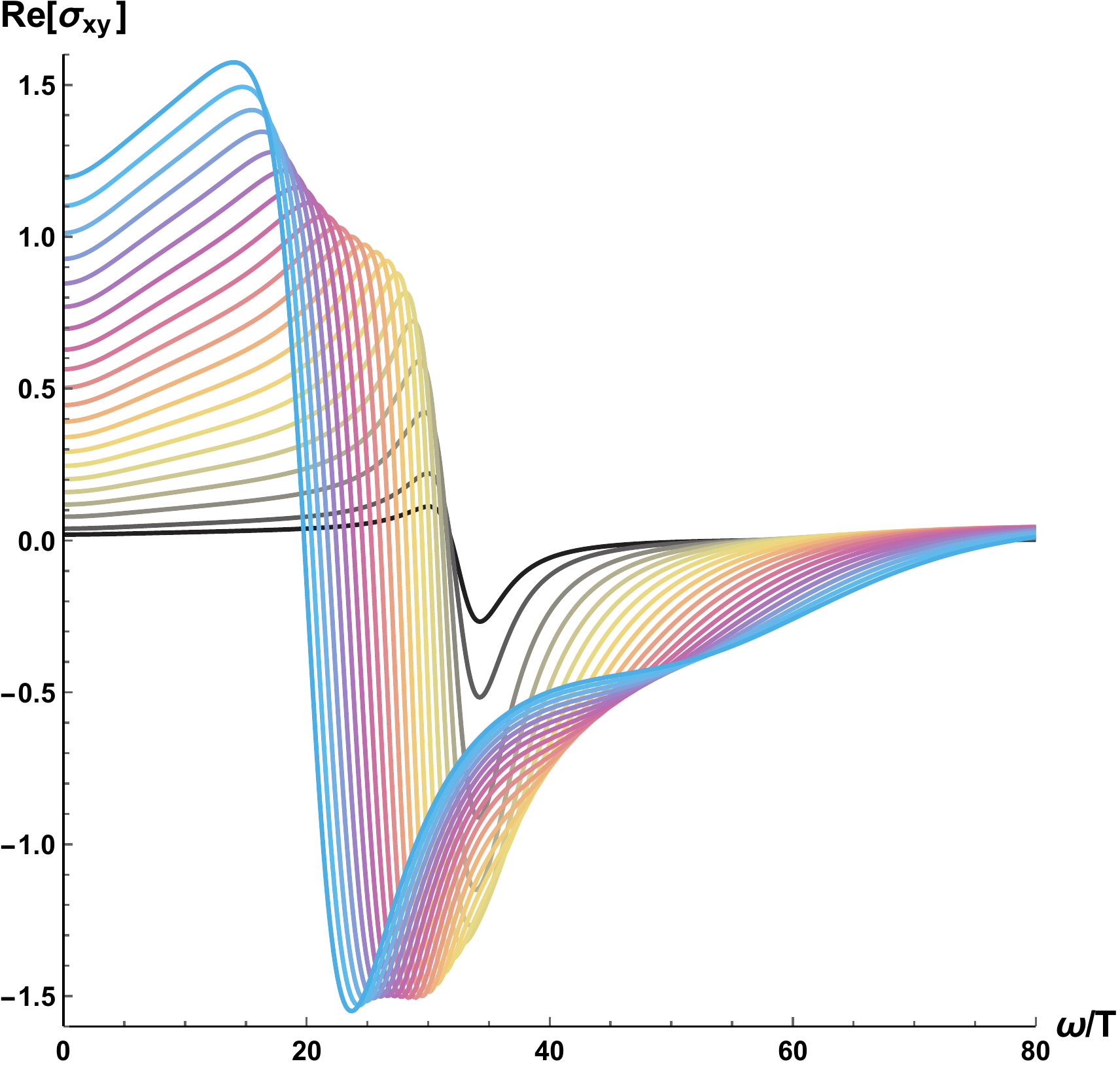}
    \caption{The evolution of the pseudo-phonon peak increasing the charge density. We move the parameter $\rho/T^2 \in [50,2000]$ (from black to light blue). Other parameters $\beta=1,\kappa/T=31.4,\alpha=0.1, B/T^2=10^3$ are kept fixed.}
    \label{fig6}
\end{figure}

Finally, we are going to investigate the behavior of the conductivity keeping the magnetic field $B$ and other parameters fixed, and dialing only the charge density of the system. The results for the AC conductivities are shown in figure~\ref{fig6} from small charge density (black line) to large values (light blue line). Two features are observed: (I) the DC conductivity grows as a function of the charge density, which is expected and in agreement with the DC formulas~\eqref{DCformula}; (II) the magnetophonon peak moves to lower frequencies.
%
%\begin{figure}
%    \centering
%    \includegraphics[width=0.45\linewidth]{peakbeta.pdf}
%    \caption{Dependence of the peak position $\omega_{pk}$ as a function of the parameter $\beta$, which determines the strength of the SSB -- the shear elastic modulus. The peak is extracted both from the real part of the longitudinal conductivity.}
%    \label{fig5}
%\end{figure}
%

\begin{figure}
    \centering
     \includegraphics[width=0.45\linewidth]{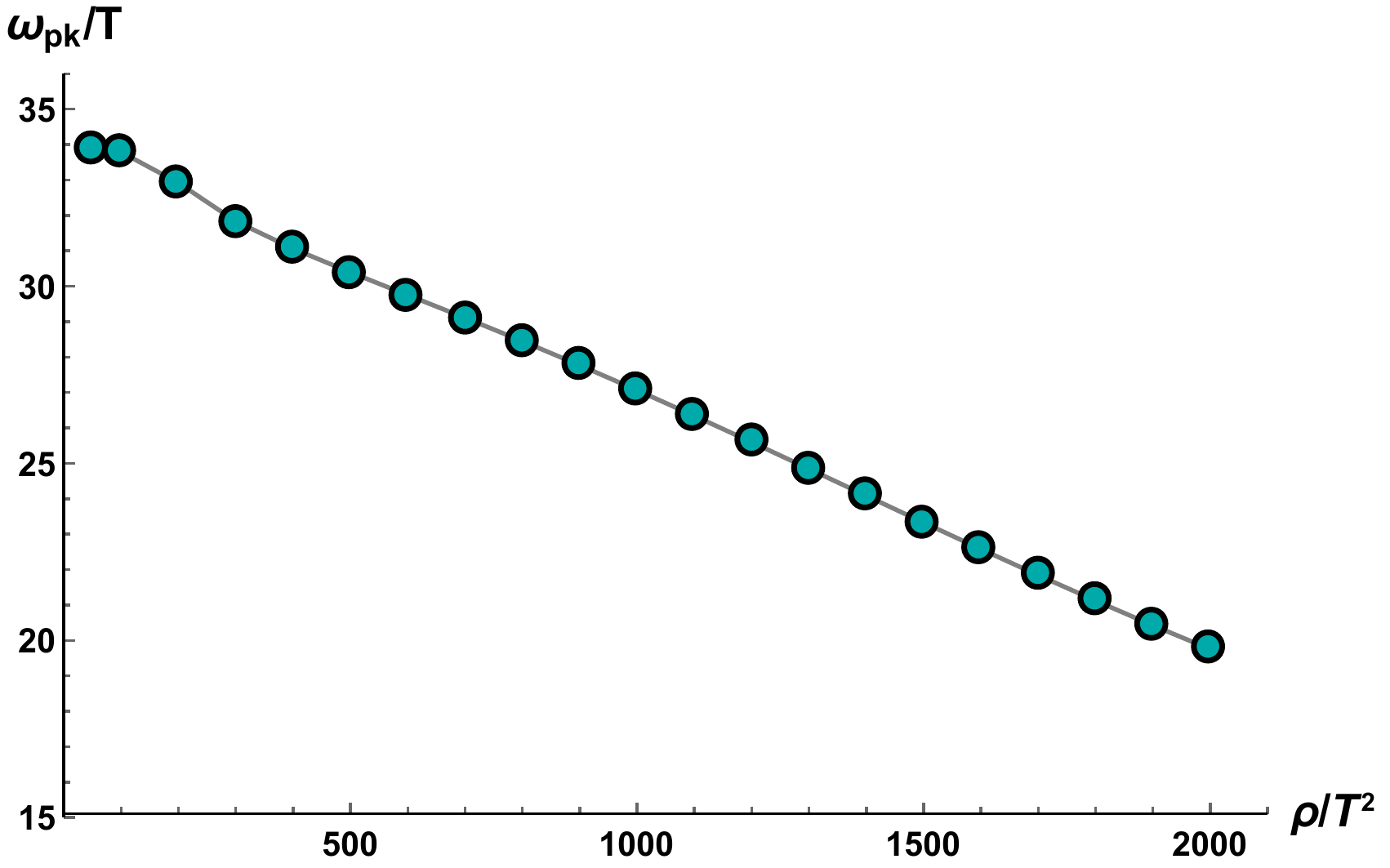}\qquad \includegraphics[width=0.45\linewidth]{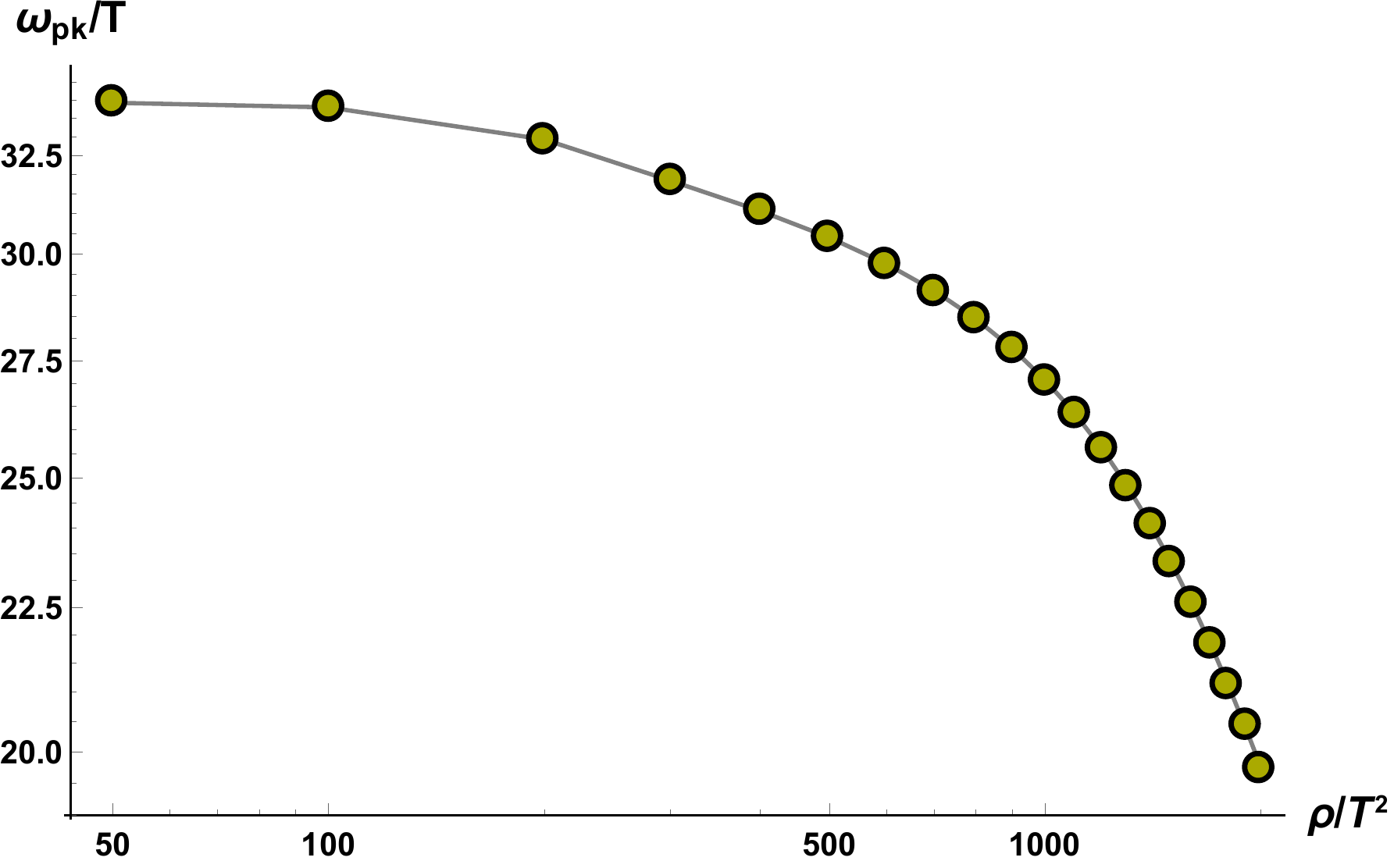}
    \caption{\textbf{Left: }Dependence of the peak position $\omega_{pk}$ as a function of the charge density $\rho/T^2$. The peak is extracted from the real part of the longitudinal conductivity. \textbf{Right: } A log-log plot for $\omega_{pk}$.}
    \label{fig7}
\end{figure}

The complete dynamics is shown in figure~\ref{fig7}. 
At large values of the charge density, the decrease is well approximated by a linear function. The result in figure~\ref{fig7} are consistent with the one showed in figure~2 of~\cite{Delacretaz:2019wzh}, where the frequency of the peak decreases by increasing the charge of the system $\rho \sim \nu$.
We have not been able to find a robust power-law scaling of $\omega_{pk}$ as a function of the charge density.

\section{Conclusions}\label{conc}
In this work, we have made an extensive analysis of a large class of holographic models with broken translations at finite charge density and magnetic field. Let us briefly summarize our main results:
\begin{itemize}
    \item In section \ref{sec:hydro}, we studied in detail the hydrodynamic excitations of our system at zero magnetic field and in the absence of explicit breaking of translations. We have identified two propagating sound modes and two coupled diffusive modes, and examined their dispersion relations as a function of the dimensionless parameters of the model. We have successfully verified the matching between our results and the hydrodynamic framework of~\cite{Armas:2020bmo}. We have checked explicitly the structure of all the correlators, the dispersion relations of the hydrodynamics modes, and the transport coefficients appearing therein. Finally, we have verified the validity of the universal bound required by the positivity of the entropy production.
    \item In section \ref{sec:typeB}, we analyzed the dynamics of the hydrodynamic modes in the presence of finite charge density, finite magnetic field and spontaneous breaking of translations. We identified a transition between two type-A Goldstone modes -- the transverse and longitudinal sound modes, to a single type-B quadratic Goldstone mode (the magnetophonon) and one gapped partner (the magnetoplasmon). To the best of our knowledge, ours is the first holographic example of a type-B Goldstone mode related to the breaking of spacetime symmetries (for internal symmetries, this was observed already in~\cite{Amado:2013xya}). Additionally, we performed an analysis of the dispersion relations at small momentum and successfully compared the results with hydrodynamics. Interestingly, as in~\cite{Amado:2013xya}, we observed that the imaginary part of the type-B Goldstone is not quartic as expected, but it follows a clear diffusive behavior $\mathrm{Im}\,\omega\,\sim\,k^2$. We are not aware of any explanation for this mechanism in the context of effective field theories for type-B Goldstone bosons. Finally, we showed the dynamics of the hydrodynamic modes as a function of the various parameters of the system, and compared to the hydrodynamic formulas. Even though the qualitative behavior is compatible with the expectations, the concrete numerical values do not coincide. The disagreement increases at large magnetic field, and it is simply the proofs that the hydrodynamics of~\cite{Armas:2020bmo} needs to be generalized at finite $B$ in order to match our numerical data. Nevertheless, we proved explicitly that the scalings found numerically for small parameters are totally consistent with the expectations from hydrodynamics given by~\cite{PhysRevB.18.6245,PhysRevB.46.3920}:
\begin{equation}
    \mathrm{Re}\,[\omega_{\text{magnetoplasmon}}]\,=\,\omega_c\,+\,\frac{(v_\parallel^2+v_\perp^2)}{2\,\omega_c}\,k^2\,+\,\dots\,,\quad \mathrm{Re}\,[\omega_{\text{magnetophonon}}]\,=\,\frac{v_\perp\,v_\parallel}{\omega_c}\,k^2\,+\,\dots\,.
\end{equation} 
We hope that our work will prompt the construction of a complete hydrodynamic framework at finite magnetic field to which our results may be compared.
    \item In section \ref{sec:peak}, we computed the transport properties of the model in the presence of a large magnetic field and the pseudo-spontaneous breaking of translation symmetry. We identified the presence of a pinned magneto-resonance peak in the optical electric conductivities and studied its dynamics. We observed that the pinning frequency of the peak grows with the magnetic, and at large magnetic field follows an approximate scaling
    \begin{equation}
        \omega_{pk}\,\sim\,B^{1/2}\,. \label{scale}
    \end{equation}
    \begin{figure}[h]
    \centering
    \includegraphics[width=0.43\linewidth]{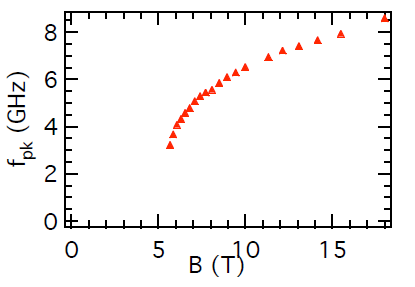}
    \quad 
     \includegraphics[width=0.43\linewidth]{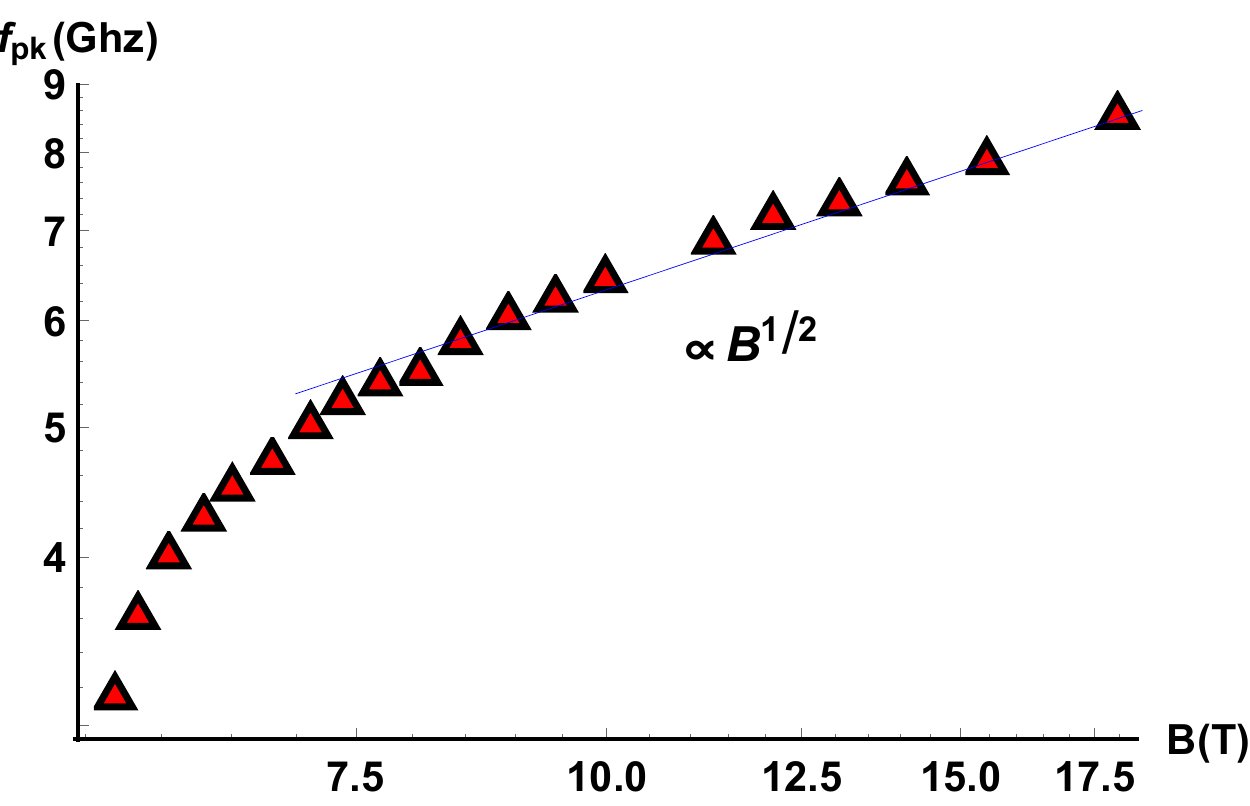}
\caption{Experimental data for a 15nm wide AlGaAs/GaAs/AlGaAs
QW sample taken from \cite{Chen2005QuantumSO}. The data are the same as in figure~\ref{fig2} and are taken with permission from~\cite{Chen2005QuantumSO}. \textbf{Left: }The motion of the magnetophonon resonance peak as a function of the external magnetic field. \textbf{Right: } A log-log plot which shows a qualitative $\sim B^{1/2}$ scaling at large magnetic field, in well agreement with our data shown in figure~\ref{fig3b}.}
    \label{figlast}
\end{figure}\\
    (I) This behavior indicates that our model does not obey the classical arguments of~\cite{PhysRevB.18.6245,PhysRevB.46.3920,Delacretaz:2019wzh}, for which the magneto-resonance peak should decrease with the magnetic field. On the contrary, it confirms that the pinning mechanism has a fundamental quantum nature which cannot be described by the naive classical theory giving $\omega_{pk}\sim 1/B$ (see~\cite{Chen2005QuantumSO} for more  details). (II) The observed scaling \eqref{scale} is consistent with the experimental observation in certain two-dimensional materials. In figure \ref{figlast} we have taken some experimental data from~\cite{Chen2005QuantumSO} and analyzed the scaling using a log-log plot.  The data are indeed qualitatively compatible with our scaling \eqref{scale}. Obviously, this does not imply that our model is describing any specific material, but rather that the scaling we find is consistent with realistic data, but at odds with the discussion in~\cite{Delacretaz:2019wzh}.
    Obviously, by generalizing the holographic action with additional couplings and fields (e.g. a dilaton-Maxwell coupling) it could be possible to fine tune the dynamics of the peak to become smaller by increasing $B$\,\footnote{In the same way,~\cite{Amoretti:2018tzw} fine tuned the system to have a pinning frequency increasing with $T$ in a very small regime of low temperatures.}.
\end{itemize}

In summary, we performed a complete analysis of the hydrodynamic and transport properties of a large class of holographic models with spontaneously and pseudo-spontaneously broken translations at finite charge, and to the best of our knowledge for the first time at finite magnetic field. This study has revealed the presence of interesting features both from a theoretical and phenomenological point of view. Finally, it represents a new step towards the understanding of the homogeneous holographic models with broken translations and their application to strange metals and metallic transport in the absence of quasiparticles.
In the future, it would be interesting to extend our studies at finite magnetic field to more complicated holographic systems which break translations without retaining the  homogeneity of the background such as~\cite{Cremonini:2019fzz,Cremonini:2017usb,Cremonini:2016rbd,Cai:2017qdz}. This would help to understand to which extent these simpler homogeneous models can be trusted, in which features they concretely differ from the inhomogeneous setups (e.g. commensurability \cite{Andrade:2015iyf,Andrade:2017leb,Andrade:2017ghg}) and which phases of matter they are actually describing \cite{Baggioli:2020nay,Esposito:2020wsn}.

\section*{Acknowledgements}
We thank M.~Ammon, S.~Cremonini, D.~Arean, Y.~P.~Chen, Y.~Hidaka, C.~Hoyos M.~Kaminski, M.~Landry, W.~Li and A.~Rebhan for discussions and useful comments.
We thank Y.~P.~Chen for permitting us to reproduce some of the figures in~\cite{Chen2005QuantumSO}.
M.B. acknowledges the support of the Spanish MINECO’s “Centro de Excelencia Severo Ochoa” Programme under grant SEV-2012-0249. S.G. gratefully acknowledges financial support by the DAAD (German Academic Exchange Service) for a \textit{Forschungsstipendium für promovierte Nachwuchswissenschaftler (Postdoc-Programm)} in 2020.
L.L. acknowledges the support of the National Natural Science Foundation of China Grants No.11947302 and No.11991052.

\appendix

\section{Hydrodynamics of spontaneously broken translations in external magnetic field}\label{app0}
In this appendix, we review the original hydrodynamic formulation proposed by Fukuyama and Lee in 1978~\cite{PhysRevB.18.6245}.
We consider a periodic charge distribution (charge density wave)
\begin{equation}
    \rho(\Vec{r})\,=\,n\,+\,\rho_0\,\,\left[\cos(Q x +\phi_x(\Vec{r}))\,+\,\cos(Q y +\phi_y(\Vec{r}))\right]\,,
\end{equation}
where $\rho_0$ is the amplitude of the charge density wave and $\phi_\alpha\, (\alpha=x, y)$ the dynamical phase. This structure is periodic with period $2\pi /Q$. 

We are interested in the dynamics at scales larger than the lattice spacing:
\begin{equation}
   | \nabla \phi_\alpha|\,\ll\,Q\,,
\end{equation}
which defines our hydrodynamic limit.
For simplicity, we neglect Coulomb interactions for the rest of this section. Under these assumptions, the Hamiltonian of the system reads
\begin{equation}
    H\,=\,H_K\,+\,U\,+\,V \label{ham}\,.
\end{equation}
The first term is simply the kinetic term for the phases:
\begin{equation}
    H_K\,=\,\frac{n\,m}{2\,Q^2}\,\int\,\left(\dot{\phi}_x^2\,+\,\dot{\phi}_y^2\right)\,d\Vec{r}\,,
\end{equation}
where $n\,m$ is the mass density and $\dot{\phi}_\alpha/Q$ the local velocity in the $\alpha_{th}$ direction.
The second term of \eqref{ham} is the elastic potential energy. Assuming that the longitudinal speed of sound is equivalent to the transverse one, such term takes the simple form:
\begin{equation}
    U\,=\,\frac{1}{2}\,C_0\,\sum_{\alpha,\beta}\int\,\left(\frac{d\phi_\alpha}{dx_\beta}\right)^2\,d\Vec{r}\,.
\end{equation}
In a more realistic case, there would be a non-trivial tensor $C_{\alpha \beta}$ taking into account the different shear moduli, and the speeds would be different.
Finally, the last term in \eqref{ham} is the external periodic potential responsible for the pinning, which can be written as
\begin{equation}
    V\,=\,e\,V_0\,\rho_0\,\sum_i\,\left[\cos(Q x_i +\phi_x(\Vec{R}_i))\,+\,\cos(Q y_i +\phi_y(\Vec{R}_i))\right]\,,
\end{equation}
where $V_0$ defines the strength of such potential and $\Vec{R}_i=(x_i, y_i)$ denotes the location of the $i_{th}$ ``atom''.

In terms of the momentum density operator $P_\alpha$, which is the canonical conjugate to $\phi_\alpha$, the algebra for our system reads
\begin{align}
   & \left[P_\alpha(\Vec{r})\,,\phi_\beta(\Vec{r}')\right]\,=\,-\,i\,Q\,\delta(\Vec{r}-\Vec{r}')\,\delta_{\alpha\beta}\,,\\
   &\left[P_\alpha(\Vec{r})\,,P_\beta(\Vec{r}')\right]\,=\,-\,i\,\frac{n}{l_B^2}\,\delta(\Vec{r}-\Vec{r}')\,\delta_{\alpha\beta}\,, \label{comm}
\end{align}
where the second commutation rule comes from the presence of an external magnetic field:
\begin{equation}
    l_B^2\,=\,\frac{c}{e\,B}\,.
\end{equation}
The non-commutation of the momentum density operators is the fundamental reason behind the quadratic dispersion relation of the magnetophonon modes. One can understand this by considering the free Hamiltonian for a system in an external gauge field:
\begin{equation}
    H_0\,=\,\sum_i\,\frac{1}{2\,m}\,\left(p_i\,+\,\frac{e}{c}\,A_i\right)^2\,.
\end{equation}
The momentum density can then be written as:
\begin{equation}
    P_{\alpha}\,=\,\left(p_\alpha\,+\,\frac{e}{c}\,A_\alpha\right)\,,
\end{equation}
and it is importantly not just $p_\alpha$. Given the standard commutation rule and remembering that $A_i\,=\,\frac{B}{2}\,\epsilon_{ij}\,x^j$, one can immediately derive the non-commutation of the $P_\alpha$, which indeed turns to be proportional to the magnetic field $B$.
The commutation rule \eqref{comm}, together with the Hamiltonian \eqref{ham}, determines the full dynamics of the system.

Going to Fourier space, the final equation of motions can be written as
\begin{equation}
    \omega_n^2\,D_{\alpha\beta}\,+\,\frac{C_0\,Q^2}{n\,m}\,q^2\,D_{\alpha\beta}\,-\,\omega_n\,\omega_c\,\epsilon_{\alpha\beta}\,D_{\alpha\beta}\,=\,\frac{Q^2}{n\,m}\,(2\pi^2)\,\delta(\Vec{q}-\Vec{q}')\delta_{\alpha\beta}\,,
\end{equation}
where for simplicity we have dropped also the pinning term proportional to the potential $V$. The complete computation can be found in~\cite{PhysRevB.18.6245}. In the absence of pinning, we can write down
\begin{equation}
    D_{\alpha\beta}(\Vec{q},\Vec{q}';i\,\omega_n)\,=\,(2\,\pi)^2\,\frac{Q^2}{n\,m}\,\frac{1}{(\omega_n^2\,+\,\omega_+)^2(\omega_n^2+\omega_-^2)}\,\begin{pmatrix}
  \omega_n^2\,+\,\omega_\perp^2 & \omega_n\,\omega_c\\ 
  -\,\omega_n\,\omega_c & \omega_n^2\,+\,\omega_\parallel^2
\end{pmatrix}\,,
\end{equation}
where $D_{\alpha \beta}$ is the Green's function matrix (in Matsubara formalism $\omega \rightarrow i\,\omega_n$).
Some comments are in order. (I) In the absence of magnetic field the matrix would be diagonal, giving rise to the usual longitudinal and transverse sound modes. (II) The off-diagonal terms are antisymmetric because of the breaking of parity induced by the magnetic field. (III) The diagonal entries are not the same because we have re-introduced a difference in the longitudinal and transverse speeds of sound.

At this point, the excitations of the system are determined by the poles of the Green's function matrix:
\begin{equation}
    \omega\,=\,\omega_\pm\,,
\end{equation}
where
\begin{equation}
    \omega_{\pm}^2\,=\,\frac{1}{2}\,\left(\omega_c^2\,+\,\omega_\perp^2\,+\,\omega_\parallel^2\right)\,\pm\,\frac{1}{2}\,\sqrt{\left(\omega_c^2\,+\,\omega_\perp^2\,+\,\omega_\parallel^2\right)^2\,-\,4\,\omega_\perp^2\,\omega_\parallel^2}\,.
\end{equation}
Note that this formula is slightly more  general than the one presented in the original paper of Fukujama and Lee~\cite{PhysRevB.18.6245} in the sense that it includes a different speed of propagation for transverse and longitudinal phonons. This formula can be found in~\cite{Chen2005QuantumSO}.
Expanding these two eigenfrequencies at small momentum and using $\omega_{\perp,\parallel}=v_{\perp,\parallel}\, k$, we obtain the expressions appearing in the main text:
\begin{equation}
    \omega_+\,=\,\omega_c\,+\,\frac{v_\parallel^2\,+\,v_\perp^2}{2\,\omega_c}\,k^2\,,\quad \quad \omega_-\,=\,\frac{v_\perp\,v_\parallel}{\omega_c}\,k^2\,,
\end{equation}
where the plus mode is the magnetoplasmon, and the minus one is the magnetophonon.
If one includes the Coulomb interaction, the gap of the magnetoplasmon will increase to
\begin{equation}
    \omega_+\,=\,\omega_c\,+\,\frac{\omega_p^2}{2\,\omega_c}\,,
\end{equation}
where $\omega_p$ is the plasma frequency.

In the presence of pinning, the equation for the frequencies at zero momentum can be easily derived and becomes:
\begin{equation}
    \omega_{\pm}^2\,=\,\frac{1}{2}\,\left[\omega_c^2\,+\,2\,\omega_0^2\,\,\pm\,\omega_c\,\sqrt{\omega_c^2\,+4\,\omega_0^2}\right]\,.
\end{equation}
In the large magnetic field regime, the two modes simplify to:
\begin{equation}
    \omega_+\,=\,\omega_c\,+\,\frac{\omega_0^2}{\omega_c}\,\sim\,\omega_c\,,\quad \quad \omega_-\,=\,\frac{\omega_0^2}{\omega_c}\,,
\end{equation}
which are the results mentioned in the main text.
Using the same techniques, one can derive the structure of the magneto-conductivity at low frequencies.

\section{Charged viscoelastic hydrodynamics}\label{appHydro}
In this appendix, we summarize the hydrodynamic framework of \cite{Armas:2020bmo}, which leads to the results presented in section \ref{sec:hydro}.

The theory starts with the definition of the scalar fields $\phi^I$ which can be understood as the Goldstone modes for spontaneously broken translations. The indices $I, J,\dots$ run over the spatial directions, while $\mu, \nu,\dots=0,\dots, d$ denote spacetime indices. We then write down the tensor
\begin{equation}
    h^{IJ}\,=\,g^{\mu\nu}\,\partial_\mu \phi^I\,\partial_\nu \phi^J\,,
\end{equation}
with its preferred background value $\bar{h}^{IJ}\sim \delta^{IJ}$. The mechanical deformations of the medium are described in terms of the strain tensor
\begin{equation}
    u_{IJ}\equiv \frac{h_{IJ}-\bar{h}_{IJ}}{2}\,.
\end{equation}
The free energy of the medium is given by
\begin{equation}
    \mathcal{F}\,=\,-\,\int d^dx \,\sqrt{-g}\;\mathrm{F}\,,
\end{equation}
with
\begin{equation}
    \mathrm{F}\,=\,p\,+\,\mathcal{P}\,\left({u^I}_I\,+\,u^{IJ}u_{IJ}\right)\,-\,\frac{1}{2}\,\mathcal{B}\,\left({u^I}_I\right)^2\,-\,G\,\left(u^{IJ}u_{IJ}\,-\,\frac{1}{d}\left({u^I}_I\right)^2\right)\,+\,\dots\,,
\end{equation}
where higher order terms in the strain tensor are neglected. Here, $p$ is the  thermodynamic pressure, $\mathcal{P}$ the crystal pressure, $\mathcal{B}$ the bulk modulus and $G$ the shear modulus.

In addition to the Goldstone modes $\phi^I$, the dynamics  of the system includes also the stress tensor $T^{\mu\nu}$ and the electric  current $J^\mu$, which satisfy the following equations
\begin{equation}
    \nabla_\mu T^{\mu\nu}\,=\,F^{\nu\rho}J_\rho\,,\quad \quad \nabla_\mu J^\mu\,=\,0
\end{equation}
where additional external forces are set to vanish.
The most general set of constitutive relations for a charged viscoelastic medium at one-derivative order in the Landau frame are given by:
\begin{align}
   & J^\mu\,=\,q\,v^\mu\,\,-\,\mathrm{P}^{I\nu}\,\sigma^q_{IJ}\mathrm{P}^{J\nu}\,\left(T\partial_\nu \frac{\mu}{T}-E_\nu\right)\,-\,\mathrm{P}^{I\mu}\gamma_{IJ}v^\nu \partial_\nu \phi^J\,,\\
     & T^{\mu\nu}\,=\,\left(\epsilon+p\right)v^\mu v^\nu\,+\,p\,g^{\mu\nu}\,-\,r_{IJ}\partial_\mu \phi^I \partial_\nu \phi^J\,-\,\mathrm{P}^{I(\mu}\mathrm{P}^{J\nu)}\,\eta_{IJKL}\,\mathrm{P}^{K(\rho}\mathrm{P}^{L\sigma)}\,\nabla_\rho\,v_\sigma\,,
\end{align}
where we have defined
\begin{equation}
   \mathrm{P}^{\mu\nu}\equiv g^{\mu\nu}+v^\mu v^\nu\,,\quad \mathrm{P}^{I\mu}\equiv  \mathrm{P}^{\mu\nu} \partial_\nu \phi^I\,,\quad E_{\mu}\equiv F_{\mu\nu} v^\nu\,.
\end{equation}
Here $q$ is the charge density, $v^\mu$ the four-velocity, $\epsilon$ the energy density, $p$ the thermodynamic pressure, $r_{IJ}$ the elastic tensor, $\mu$ the chemical potential, and $\sigma^q_{IJ},\eta_{IJKL},\gamma_{IJ}$ dissipative tensors.
In addition, we also have the dynamical equation for the  Goldstone modes
\begin{equation}
    \sigma^\phi_{IJ}\,v^\mu\,\partial_\mu \phi^I\,+\,\gamma'_{JK}\mathrm{P}^{K\mu}\left(T\partial_\nu \frac{\mu}{T}-E_\nu\right)\,+\,\nabla_\mu \left(r_{JK}\partial^\mu \phi^K\right)\,=\,0\,,
\end{equation}
where new dissipative  tensors $\sigma^\phi_{IJ},\gamma'_{IJ}$ appear. Expanding the last equation around a background value $\phi^I=x^I-\delta \phi^I$, one obtains the more standard Josephson relation:
\begin{equation}
    v^t\,\partial_t \delta \phi^I\,=\,v^I\,-\,v^i \partial_i \delta \phi^I\,+\,\dots\,.
\end{equation}

Given the most generic expressions, we now focus on the linear regime in which
\begin{align}
    &\sigma^q_{IJ}=\sigma_q\,h_{IJ}\,,\quad \sigma^\phi_{IJ}=\sigma\,h_{IJ}\,,\quad  \gamma_{IJ}=\gamma\,h_{IJ}\,,\quad \gamma'_{IJ}=\gamma'\,h_{IJ}\,, \\
    &\eta_{IJKL}\,=\,\left(\zeta\,-\,\frac{2}{d}\,\eta\right)h_{IJ}h_{KL}\,+\,2\,\eta\,h_{IK}h_{JL}\,,
\end{align}
where $\sigma_q$ is the incoherent conductivity, $\zeta$ the bulk viscosity and $\eta$ the shear viscosity, and the rest are new dissipative parameters.
In the linear regime, the constitutive relations and the Josephson equation become:
\begin{align}
    & J^\mu\,=\,\left(q_f+q_l\, {u^\lambda}_\lambda\right)v^\mu-\sigma_q\, \mathrm{P}^{\mu\nu}\,\left(T\partial_\nu \frac{\mu}{T}-E_\nu\right)- \gamma\, \mathrm{P}^\mu_I\,v^v\, \partial_\nu \phi^I\,,\\
    &T^{\mu\nu}=\left(\epsilon+\epsilon_l\,{u^\lambda}_\lambda\right)v^\mu v^\nu+\left(p\,+\mathcal{P} {u^\lambda}_\lambda\right)\,\mathrm{P}^{\mu\nu}+\mathcal{P}\,h^{\mu\nu}-\eta \sigma^{\mu\nu}-\zeta \mathrm{P}^{\mu\nu}\partial_\rho v^\rho-2 G u^{\mu\nu}-\left(\mathcal{B}-\frac{2}{d}G\right)\,{u^\lambda}_\lambda\,h^{\mu\nu}\,,\\
    & \sigma\,v^\mu \partial_\mu \phi^I-h^{IJ}\nabla_\mu \left(\mathcal{P}\partial^\mu \phi_J-\left(\mathcal{B}-\frac{2}{d}G\right){u^\lambda}_\lambda\partial^\mu \phi_J-2 G u^{\mu\nu}\partial_\nu \phi_J\right)+\gamma' \mathrm{P}^{I\mu}\left(T\partial_\nu \frac{\mu}{T}-E_\nu\right)=0\,,
\end{align}
where $q_f=\rho$ is the charge density of the system and $h_{\mu\nu}=h_{IJ}\partial_\mu \phi^I \partial_\nu \phi^J$, $u_{\mu\nu}=u_{IJ}\partial_\mu \phi^I \partial_\nu \phi^J$.
Solving the linear equations in momentum space, we
can find the complete set of linear modes admitted by
the theory presented in the main text. Notice that conformality and $\mathcal{PT}$ invariance require the additional constraints:
\begin{align}
    \epsilon\,=\,d\,p\,-\,r_{IJ}\,h^{IJ}\,,\quad h^{IJ}\,\eta_{IJKL}=\,\eta_{IJKL}h^{KL}=0\,,\quad \gamma=-\gamma'\,. 
\end{align}
We conclude here our summary and refer to~\cite{Armas:2020bmo} for more details.

\section{Technical details}\label{app1}
In this appendix, we provide more technical details about the computations done in this work.

\subsection{Equations of motion}\label{app:eoms}
First, we start by displaying the equations of motions for our system. We define the following perturbations:
\begin{equation}
    g_{\mu\nu}\,=\,\bar g_{\mu\nu}\,+\,u^{-2}\,\delta g_{\mu\nu}\,,\quad A_\mu\,=\,\bar A_\mu\,+\,\delta A_\mu\,,\quad \phi^I\,=\,\bar{\phi}^I\,+\,\delta \phi^I\,,
\end{equation}
where the bar quantities are the background values for the fields. We will adopt the radial gauge:
\begin{equation}
    \delta A_u\,=\,0\,,\quad\delta g_{au}\,=\,0\,,\quad \delta g_{uu}\,=\,0\,,
\end{equation}
with $u$ being the radial direction. 
To simplify the equations, we also introduce
\begin{equation}
    \delta g_{xx}\,=\,\frac{1}{2}\left(\delta g_{22}\,+\,\delta g_{33}\right)\,,\quad \quad \delta g_{yy}\,=\,\frac{1}{2}\left(\delta g_{22}\,-\,\delta g_{33}\right)\,.
\end{equation}
Moreover, we perform a Fourier decomposition of every fluctuation as
\begin{equation}
    \delta \zeta (u,t,x,y)\,=\,e^{-\,i\,\omega\,t\,+\,i\,k\,y}\,\delta \zeta (u)
\end{equation}

Using these conventions, the equations of motions for the  perturbations are:
\begin{align}
    &\text{\underline{Maxwell equations} :}\nonumber\\
    &-\,i \,\mu \, \delta g_{22}'\,+\,k\, \delta A_y'\,+\,i\, \delta A_t''\,=\,0\,,\\[0.2cm]
    &i \,B\, k\, \delta g_{22}\,-\,B\, \delta g_{ty}'\,+\,\delta A_x' \,\left(f'\,+\,2 \,i\, \omega \right)\,+\,f\, \delta A_x''\, -\,\mu\,  \delta g_{tx}'\,-\,k^2\, \delta A_x\,=\,0\,,\\[0.2cm]
    &B\, \delta g_{tx}'\,+\,\delta A_y'\, \left(f'\,+\,2\, i\, \omega \right)\,+\,f\, \delta A_y''\,-\,\mu \, \delta g_{ty}'\,+\,i\, k\, \delta A_t'\,=\,0\,,\\[0.2cm]
    &-\,\mu\,  \omega \, \delta g_{22}\,+\,B\, k\, \delta g_{tx}\,+\,k\, f\, \delta A_y'\,-\,k \,\mu\,  \delta g_{ty}\,+\,i\, k^2 \,\delta A_t\,+\,i\, k\, \omega  \delta A_y\,+\,\omega \, \delta A_t'\,=\,0\,,\\[0.2cm]
    &\text{\underline{Scalar equations} :}\nonumber\\
    &u \,f'\, \delta \phi_x'\,+\,u\, f\, \delta \phi_x''\,+\,2\, f \,\delta \phi_x'\,+\,u\, \delta g_{tx}'\,+\,2\, \delta g_{tx}\,-\,i\, k\, u\, \delta g_{xy}\,-\,k^2\, u\, \delta \phi_x \,+\,\nonumber \\ &\,+\,2\, i\, u\,
   \omega \, \delta \phi_x'\,+\,2\, i\, \omega\,  \delta \phi_x\,=\,0\,,\\[0.2cm]
    &i \,k \,u \,\delta g_{33}\,-\,2\, i\, k\, u\, \delta g_{22}\,+\,u\, f'\, \delta \phi_y'\,+\,u\, f \,\delta \phi_y''\,+\,2\, f\, \delta \phi_y'\,+\,u\, \delta g_{ty}'\,+\nonumber\\&\,+\,2\, \delta g_{ty}\,-\,3\, k^2\,
   u\, \delta \phi_y\,+\,2\, i\, u\, \omega\,  \delta \phi_y'\,+\,2\, i\, \omega\,  \delta \phi_y\,=\,0\,,\\[0.2cm]
    &\text{\underline{Einstein equations} :}\nonumber\\
   &f\, \Big(u \,\Big(2 \,\Big(B^2\, u^3\,+\,6\, m^2\, u^5\,-\,i\, \omega \Big) \delta g_{22}\,+\,u\, f'\, \delta g_{22}'\,-\,u\, \delta g_{tt}''\,+\,4\, \delta g_{tt}'\,-\,2\, i\, k\, \delta g_{ty}\,\nonumber\\
   &-\,12\, i\, k\, m^2\, u^5\,
   \delta \phi_y\,+\,2\, \mu \, u^3\, \delta A_t'\Big)\,+\,2\, i\, B\, k\, u^4\, \delta A_x\,-\,12\, \delta g_{tt}\Big)\,+\,u \,\Big(u\, \Big(2 \,\omega \,+\,i\, f'\Big)\nonumber\\& \,\Big(\omega\, \delta g_{22}\,+\,k\, \delta g_{ty}\Big)\,+\,\delta g_{tt}\, \Big(u^3\, \Big(B^2\,+\,\mu ^2\,+\,4 m^2 \,u^2\Big)\,-\,u\, f''\,+\,4\, f'\,+\,k^2\, u\,-\,2\, i\, \omega \Big)\Big)\,\nonumber \\ &\,-\,2\, u\, f^2\, \delta g_{22}'\,+\,6\,  \delta g_{tt}\,=\,0\,,\\[0.2cm]
    &B^2\, u^4\, \delta g_{tx}\,-\,2\, B\, \mu\,  u^4 \,\delta g_{ty}\,+\,2\, i\, B\, k\, u^4\, \delta A_t\,+\,2\, i\, B\, u^4\, \omega  \,\delta A_y\,+\,2\, u\, \delta g_{tx}\, f'\,-\,u^2\, f\,
   \delta g_{tx}''\,+\nonumber\\&+\,2 \,u \,f\, \delta g_{tx}'\,-\,6\, f\, \delta g_{tx}\,+\,2 \,\mu\,  u^4\, f\, \delta A_x'\,-\,i\, u^2\, \omega  \,\delta g_{tx}'\,+\,k^2\, u^2\, \delta g_{tx}\,+\,4\, m^2\, u^6\, \delta g_{tx}\,-\,\mu ^2\, u^4\,
   \delta g_{tx}\,\nonumber\\&+\,6\, \delta g_{tx}\,+\,k\, u^2\, \omega \, \delta g_{xy}\,+\,6\, i\, m^2\, u^6\, \omega\,  \delta \phi_x\,+\,2\, i\, \mu\,  u^4\, \omega\,  \delta A_x\,=\,0\,,\\[0.2cm]
    &i \,\Big(i\, k\, u^2\, \omega \, \delta g_{33}\,+\,i \,k \,u^2\, \omega \, \delta g_{22}\,-\,i\, B^2\, u^4\, \delta g_{ty}\,-\,2\, i\, B\, \mu \, u^4\, \delta g_{tx}\,-\,2\, B\, u^4 \,\omega \, \delta A_x\,-\,2\,
   i\, u\, \delta g_{ty}\, f'\,\nonumber\\&+\,i\, u^2\, f\, \delta g_{ty}''\,-\,2\, i\, u\, f\, \delta g_{ty}'\,+\,6\, i\, f\, \delta g_{ty}\,-\,2\, i\, \mu \, u^4\, f\, \delta A_y'\,-\,k\, u^2\, \delta g_{tt}'\,+\,2\, k\, u\, \delta g_{tt}\,-\,u^2\, \omega \,
   \delta g_{ty}'\,\nonumber\\ &-\,4\, i\, m^2\, u^6 \,\delta g_{ty}\,+\,i\, \mu ^2\, u^4\, \delta g_{ty}\,-\,6\, i\, \delta g_{ty}\,+\,2\, k\, \mu\,  u^4\, \delta A_t\,+\,6 \,m^2\, u^6 \,\omega\,  \delta \phi_y\,+\,2\, \mu \, u^4 \,\omega 
\,   \delta A_y\Big)\,=\,0\,,\\[0.2cm]
&2\, B^2\, u^4\, \delta g_{22}\,+\,u^2\, f'\, \delta g_{22}'\,-\,2 \,u \,f\, \delta g_{22}'\,+\,12\, m^2\, u^6\, \delta g_{22}\,+\,2\, i\, u^2\, \omega\,  \delta g_{22}'\,-\,2\, i\, u\, \omega\, 
   \delta g_{22}\,+\nonumber\\&+\,2\, i\, B\, k\, u^4\, \delta A_x\,-\,u^2\, \delta g_{tt}''\,+\,4\, u \,\delta g_{tt}'\,-\,6\, \delta g_{tt}\,+\,i\, k\, u^2\, \delta g_{ty}'\,\nonumber\\ &-\,2\, i\, k\, u\, \delta g_{ty}\,-\,12 \,i\, k\, m^2\, u^6\, \delta \phi_y\,+\,2\, \mu\,  u^4\, \delta A_t'\,=\,0\,,\\[0.2cm]
    &B^2 \,u^4\, \delta g_{tx}\,-\,2 \,B \,\mu \, u^4\, \delta g_{ty}\,+\,2\, i\, B\, k\, u^4\, \delta A_t\,+\,2\, i\, B \,u^4\, \omega \, \delta A_y\,+\,2\, u\, \delta g_{tx}\, f'\,-\,u^2\, f\,
   \delta g_{tx}''\,+\nonumber\\&+\,2\, u\, f\, \delta g_{tx}'\,-\,6\, f\, \delta g_{tx}\,+\,2\, \mu \, u^4\, f\, \delta A_x'\,-\,i\, u^2\, \omega\,  \delta g_{tx}'\,+\,k^2\, u^2\, \delta g_{tx}\,+\,4\, m^2\, u^6\, \delta g_{tx}\,-\,\mu ^2\, u^4\,
   \delta g_{tx}\,\nonumber\\&+\,6\, \delta g_{tx}\,+\,k\, u^2\, \omega  \,\delta g_{xy}\,+\,6\, i\, m^2 \,u^6 \,\omega \, \delta \phi_x\,+\,2\, i\, \mu \, u^4 \,\omega\,  \delta A_x\,=\,0\,,\\[0.2cm]
    &-\,B^2 \,u^4\, \delta g_{33}\-\,u^2\, f'\, \delta g_{33}'\,+\,2\, u\, f'\, \delta g_{33}\,-\,u^2\, f\, \delta g_{33}''\,+\,2\, u\, f\, \delta g_{33}'\,-\,6\, f\,
   \delta g_{33}\,+\,k^2\, u^2\, \delta g_{33}\,+\nonumber\\&+\,4\, m^2\, u^6 \,\delta g_{33}\,-\,\mu ^2\, u^4\, \delta g_{33}\,-\,2\, i\, u^2\, \omega \, \delta g_{33}'\,+\,2\, i\, u\, \omega \,
   \delta g_{33}\,+\,6\, \delta g_{33}\,+\,B^2\, u^4\, \delta g_{22}\,-\,u^2\, f'\, \delta g_{22}'\,\nonumber\\&+\,2\, u \,f'\, \delta g_{22}\,-\,u^2\, f\, \delta g_{22}''\,+\,4\, u \,f\,
   \delta g_{22}'\,-\,6\, f\, \delta g_{22}\,+\,k^2\, u^2\, \delta g_{22}\,+\,4\, m^2\, u^6 \,\delta g_{22}\,-\,\mu ^2 \,u^4 \,\delta g_{22}\,\nonumber\\&-\,2\, i\, u^2\, \omega \, \delta g_{22}'\,+\,4\, i\, u \,\omega \,
   \delta g_{22}\,+\,6\, \delta g_{22}\,+\,2\, i\, B\, k\, u^4\, \delta A_x\,-\,2\, u\, \delta g_{tt}'\,+\,6\, \delta g_{tt}\,+\,2\, i\, k\, u\, \delta g_{ty}\,+\nonumber\\&+\,2\, \mu  \,u^4\, \delta A_t'\,=\,0\,,\\[0.2cm]
    &-\,B^2 \,u^4 \,\delta g_{xy}\,-\,u^2\, f'\, \delta g_{xy}'\,+\,2\, u\, \delta g_{xy}\,f'\,-\,u^2\, f\, \delta g_{xy}''\,+\,2\, u\, f\, \delta g_{xy}'\,-\,6\, f\, \delta g_{xy}\,-\,i\, k\, u^2\, \delta g_{tx}'\,+\nonumber\\&\,2\, i\, k\, u\, \delta g_{tx}\,-\,2\, i\,
   u^2\, \omega \, \delta g_{xy}'\,+\,4\, m^2\, u^6\, \delta g_{xy}\,-\,\mu ^2\, u^4\, \delta g_{xy}\,+\,2\, i\, u\, \omega\,  \delta g_{xy}\,+\,6\, \delta g_{xy}\,\nonumber\\&-\,6\, i\, k\, m^2\, u^6\, \delta \phi_x\,=\,0\,,\\[0.2cm]
    &-\,2 \,B \,u^3\, \delta A_y'\,-\,u \,\delta g_{tx}''\,+\,2\, \delta g_{tx}'\,+\,i\, k \,u\, \delta g_{xy}'\,-\,6\, m^2 \,u^5\, \delta \phi_x'\,+\,2\, \mu\,  u^3 \,\delta A_x'\,=\,0\,,\\[0.2cm]
    &\,i \,\Big(i\, k\, u^2\, \omega\,  \delta g_{33}\,+\,i\, k\, u^2\, \omega \, \delta g_{22}\,-\,i\, B^2\, u^4\, \delta g_{ty}\,-\,2\, i\, B\, \mu\,  u^4\, \delta g_{tx}\,-\,2\, B\, u^4\, \omega \, \delta A_x\,-\,2\,
   i\, u\, \delta g_{ty}\, f'\,+\nonumber\\&\,i\, u^2\, f\, \delta g_{ty}''\,-\,2\, i\, u\, f\, \delta g_{ty}'\,+\,6\, i\, f\, \delta g_{ty}\,-\,2\, i\, \mu\,  u^4\, f\, \delta A_y'\,-\,k\, u^2\, \delta g_{tt}'\,+\,2 \,k \,u \,\delta g_{tt}\,-\,u^2\, \omega\, 
   \delta g_{ty}'\,\nonumber\\&-\,4\, i\, m^2\, u^6\, \delta g_{ty}\,+\,i\, \mu ^2 \,u^4\, \delta g_{ty}\,-\,6\, i\, \delta g_{ty}\,+\,2\, k\, \mu \, u^4 \,\delta A_t\,+\,6 \,m^2\, u^6 \,\omega \, \delta \phi_y\,+\,2\, \mu\,  u^4 \,\omega\, \delta A_y\Big)\,=\,0\,,\\[0.2cm]
    &\,-\,i \,k \,u \,\delta g_{33}'\,-\,i\, k\, u \,\delta g_{22}'\,+\,2\, B\, u^3\, \delta A_x'\,-\,u\, \delta g_{ty}''\,+\,2\, \delta g_{ty}'\,-\,6\, m^2\, u^5\, \delta \phi_y'\,+\,2\, \mu\,  u^3 \delta A_y'\,=\,0\,,\\[0.2cm]
    &\delta g_{22}''\,=\,0\,,
\end{align}
where we have shown for simplicity only the case $V(X)=m^2 X^3$ and $\phi^I=x^I$. The full set of equations for a generic potential $V(X)$ is not shown, since it is rather lengthy and not particular illuminating.
We remind the reader that these equations are written in Eddington–Finkelstein (EF) coordinates. We also note that one can consistently set $\delta g_{22}=0$.

\subsection{Asymptotics and Green's functions}
In Poincar\'e coordinates, the asymptotic expansion of the various fields at the boundary $u\rightarrow 0$ are given by:
\begin{align}
    & \delta \phi_x\,=\, \delta \phi_x^{(1)}+\dots\,+\,\delta \phi_x^{(2)}\,u^{5\,-\,2\,N}+\dots\,,&&
     \delta \phi_y\,=\, \delta \phi_y^{(1)}+\dots\,+\,\delta \phi_y^{(2)}\,u^{5\,-\,2\,N}+\dots\,,\\
    &\delta A_x\,=\, \delta A_x^{(L)}\,+\,\delta A_x^{(S)}\,u+\dots\,,\label{eq:axax}&
      &\delta A_y\,=\, \delta A_y^{(L)}\,+\,\delta A_y^{(S)}\,u+\dots\,,\\
        &\delta A_t\,=\, \delta A_t^{(L)}\,+\,\delta A_t^{(S)}\,u+\dots&
          &\delta g_{tt}\,=\,\delta g_{tt}^{(L)}+\dots\,+\,\delta g_{tt}^{(S)}\,u^3+\dots\,,\\
                &\delta g_{33}\,=\, \delta g_{33}^{(L)}+\dots\,+\,\delta g_{33}^{(S)}\,u^3+\dots\,,&
                   &\delta g_{tx}\,=\, \delta g_{tx}^{(L)}+\dots\,+\,\delta g_{tx}^{(S)}\,u^3+\dots\,,\\
                      &\delta g_{ty}\,=\, \delta g_{ty}^{(L)}+\dots\,+\,\delta g_{ty}^{(S)}\,u^3+\dots\,,&
                         &\delta g_{xy}\,=\, \delta g_{xy}^{(L)}+\dots\,+\,\delta g_{xy}^{(S)}\,u^3+\dots\,,
\end{align}
where we have assumed a potential $V(X)=m^2\,X^N$\,\footnote{Notice that all potentials of the type:
\begin{equation}
    V(X)=X^N\,+\,a\,X^M\,,
\end{equation}
with $M>N$ give the same asymptotic expansion. In other words, the smallest power in the potential controls the dynamics in the UV.}. In EF coordinates, the constraint equations require the absence of the leading term for the $tt$ component of the metric, $\delta g_{tt}^{(L)}=0$, as explained in~\cite{coming}. This is an artifact of the choice of radial gauge in EF coordinates and it can be relaxed in a more general gauge.

For the case with $N<5/2$, for which the translational symmetry is broken explicitly, the indexes $^{(S)}$ and $^{(L)}$ stand respectively for ``subleading'' and ``leading'' terms, corresponding to response and source of the corresponding operator. For the scalar fields we have used $^{(1)}$ and $^{(2)}$ since the role of the two coefficients change depending on $N$. Using the holographic dictionary~\cite{Skenderis:2002wp}, the various retarded Green’s functions defined in Poincar\'e coordinates are given by:
\begin{align}
    &\mathcal{G}^R_{\Phi_I\Phi_I}\,=\,\frac{1}{N\,\left(2N-5\right)\,m^2}\,\frac{\delta \phi_I^{(2)}}{\delta \phi_I^{(1)}}\,\Big|_{u=0}\,, \quad \mathcal{G}^R_{J_xJ_x}\,=\,\frac{\delta A_x^{(S)}}{\delta A_x^{(L)}}\,\Big|_{u=0}\,, \quad \mathcal{G}^R_{J_xJ_y}\,=\,\,\frac{\delta A_x^{(S)}}{\delta A_y^{(L)}}\,\Big|_{u=0}\,,\\
    &\mathcal{G}^R_{J_j\Phi_I}\,=\,\frac{1}{N\,\left(2N-5\right)\,m^2}\,\frac{\delta A_j^{(S)}}{\delta \phi_I^{(1)}}\,\Big|_{u=0}\,,\quad \mathcal{G}^R_{\Phi_I J_j}\,=\,\frac{\delta \phi_I^{(S)}}{\delta A_j^{(L)}}\,\Big|_{u=0}\,,\quad \mathcal{G}^R_{T_{xy}T_{xy}}\,=\,-\,\frac{3}{2}\,\frac{\delta g_{xy}^{(S)}}{\delta g_{xy}^{(L)}}\,\Big|_{u=0}\,.
\end{align}

It should be pointed out that when $N>5/2$, $\delta \phi_x^{(2)}$ and $\delta \phi_y^{(2)}$ in the UV expansion become the leading source terms, and the translational invariance does not break explicitly but rather spontaneously. In this case, the corresponding retarded Green’s functions are given by
\begin{align}
 &\mathcal{G}^R_{\Phi_I\Phi_I}\,=\,\frac{1}{N\,\left(2N-5\right)\,m^2}\,\frac{\delta \phi_I^{(1)}}{\delta \phi_I^{(2)}}\,\Big|_{u=0}\,,  \\
 &\mathcal{G}^R_{J_j\Phi_I}\,=\,\frac{1}{N\,\left(2N-5\right)\,m^2}\,\frac{\delta A_j^{(2)}}{\delta \phi_I^{(2)}}\,\Big|_{u=0}\,,\quad\quad
 \mathcal{G}^R_{\Phi_I J_j}\,=\,\frac{\delta \phi_I^{(1)}}{\delta A_j^{(L)}}\,\Big|_{u=0}\,.
\end{align}

Notice that reading off the Green's functions in EF coordinates requires a careful treatment of the holographic renormalization procedure in order to identify the source terms and expectation values of the dual field theory operators correctly. For example, in the language of eq. \eqref{eq:axax}, the current-current correlator translates to

\begin{equation}
    \mathcal{G}^{R\,\,(EF)}_{J_xJ_x}(k=0)\,=\,-\,i\,\omega\,+\,\frac{\delta A_x^{(S)\,\,\,(EF)}}{\delta A_x^{(L)\,\,\,(EF)}}\,\Big|_{u=0}\,.
\end{equation}
Finally, let us remark that all the retarded Green's functions are computed by imposing ingoing boundary conditions at the horizon which is set to $u_h=1$ without loss of generality.

\section{Numerical methods}\label{app2}
In this appendix, we briefly outline the numerical methods and check the quality of our numerics. A more detailed introduction may be found in~\cite{dbt_mods,Ammon:2019apj,Baggioli:2019abx}. The numerical methods used for computing the QNMs and Green's functions are based on so-called pseudo-spectral methods (see~\cite{Boyd1989ChebyshevAF,Grandclement:2007sb,canuto2007spectral,trefethen2000spectral,dutykh2016brief} for an introduction). Note that we choose Chebychev polynomials as basis functions and discretize all functions on a Chebychev-Lobatto grid. 
\begin{figure}[h]
    \centering
    \includegraphics[width=6.5cm]{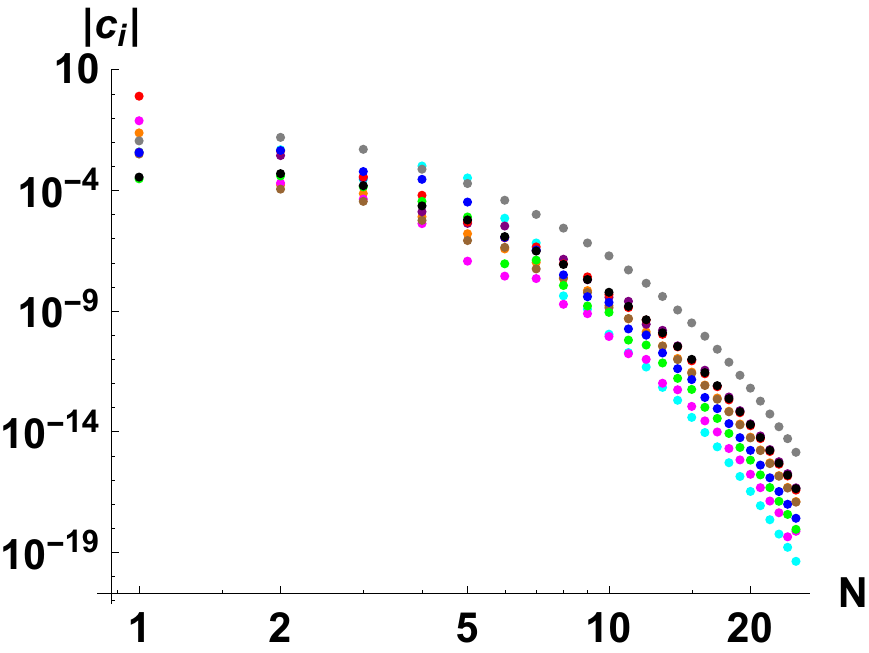}\hspace{0.6cm}\includegraphics[width=6.5cm]{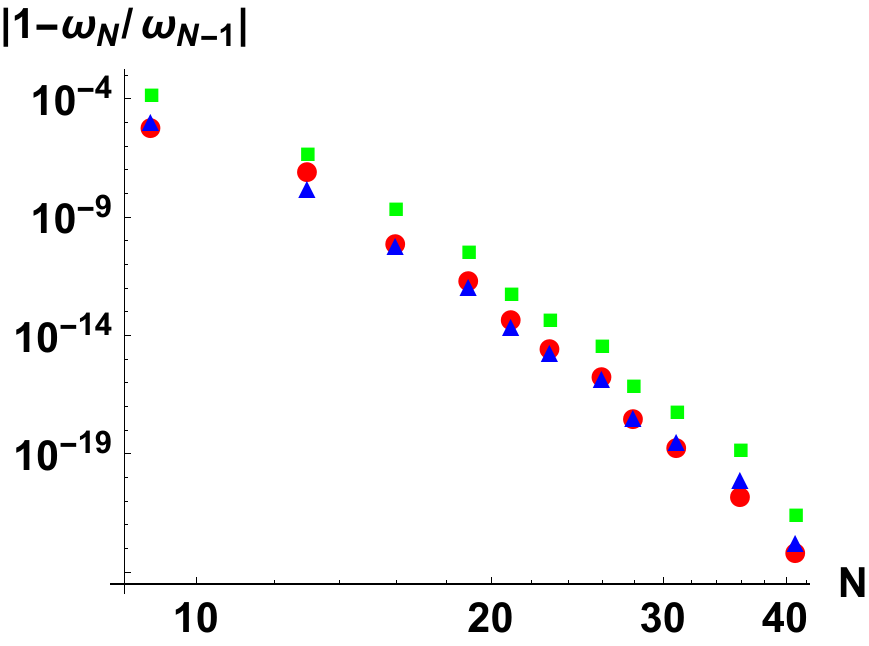}
    \caption{The convergence of our numerics for the parameters $k/T=1.676,\,m/T=4.189,\,B/T^2=105.276,\,\mu/T=4.189$. The corresponding lowest three QNMs are $\omega_{1,N=41}/T=\pm0.0804 - 0.0661\,i,\,\omega_{2,N=41}/T=-0.2045\,i$ and $\omega_{3,N=41}/T=-0.3926\,i$. \textbf{Left:} Decay of the absolute value of the Chebychev coefficients corresponding to the eigenfunctions of the lowest QNM $\omega_1$ (\{$\delta\phi_x,\delta\phi_y,\delta A_t,\,\delta A_x,\,\delta A_y,\,\delta h_{tt},\,\delta h_{tx},\,\delta h_{ty},\,\delta h_{33},\,\delta h_{xy}$\}=\{red, orange, magenta, brown, purple, cyan, blue, green, black, gray\}). \textbf{Right:} Moving of the lowest three QNMs with increasing gridsize (\{red, blue, green\}=\{$\omega_1,\,\omega_2,\omega_3$\}).}
    \label{fig:my_labelconv}
\end{figure}

In order to prove that our numerical procedure is convergent, we show the decay of the Chebychev-coefficients and the moving of the QNMs at large magnetic fields (see figure\,\ref{fig:my_labelconv}). We see that even for only 25 gridpoints, the coefficients fall off sufficiently. In the right side, we show that for $N=25$, the QNMs move less than $10^{-16}$, for increasing the gridsize. Increasing the gridsize makes the error exponentially smaller as we expect for spectral convergence.  We furthermore checked that all equations of motion including the constraint equations outlined in appendix \ref{app:eoms} are satisfied for the obtained numerical solutions.
%\newpage
\bibliographystyle{JHEP}
\bibliography{magneto}

\providecommand{\href}[2]{#2}\begingroup\raggedright\begin{thebibliography}{100}

\bibitem{Beekman:2019pmi}
A.~J. Beekman, L.~Rademaker and J.~van Wezel, \emph{{An Introduction to
  Spontaneous Symmetry Breaking}},
  \href{https://arxiv.org/abs/1909.01820}{{\tt 1909.01820}}.

\bibitem{Burgess:1998ku}
C.~P. Burgess, \emph{{Goldstone and pseudoGoldstone bosons in nuclear, particle
  and condensed matter physics}},
  \href{http://dx.doi.org/10.1016/S0370-1573(99)00111-8}{\emph{Phys. Rept.}
  {\bf 330} (2000) 193--261}, [\href{https://arxiv.org/abs/hep-th/9808176}{{\tt
  hep-th/9808176}}].

\bibitem{PhysRev.127.965}
J.~Goldstone, A.~Salam and S.~Weinberg, \emph{Broken symmetries},
  \href{http://dx.doi.org/10.1103/PhysRev.127.965}{\emph{Phys. Rev.} {\bf 127}
  (Aug, 1962) 965--970}.

\bibitem{PhysRev.117.648}
Y.~Nambu, \emph{Quasi-particles and gauge invariance in the theory of
  superconductivity},
  \href{http://dx.doi.org/10.1103/PhysRev.117.648}{\emph{Phys. Rev.} {\bf 117}
  (Feb, 1960) 648--663}.

\bibitem{Cho141}
A.~Cho, \emph{Higgs boson makes its debut after decades-long search},
  \href{http://dx.doi.org/10.1126/science.337.6091.141}{\emph{Science} {\bf
  337} (2012) 141--143},
  [\href{https://arxiv.org/abs/https://science.sciencemag.org/content/337/6091/141.full.pdf}{{\tt
  https://science.sciencemag.org/content/337/6091/141.full.pdf}}].

\bibitem{Higgs:1964ia}
P.~W. Higgs, \emph{{Broken symmetries, massless particles and gauge fields}},
  \href{http://dx.doi.org/10.1016/0031-9163(64)91136-9}{\emph{Phys. Lett.} {\bf
  12} (1964) 132--133}.

\bibitem{PhysRevLett.13.321}
F.~Englert and R.~Brout, \emph{Broken symmetry and the mass of gauge vector
  mesons}, \href{http://dx.doi.org/10.1103/PhysRevLett.13.321}{\emph{Phys. Rev.
  Lett.} {\bf 13} (Aug, 1964) 321--323}.

\bibitem{Low:2001bw}
I.~Low and A.~V. Manohar, \emph{{Spontaneously broken space-time symmetries and
  Goldstone's theorem}},
  \href{http://dx.doi.org/10.1103/PhysRevLett.88.101602}{\emph{Phys. Rev.
  Lett.} {\bf 88} (2002) 101602},
  [\href{https://arxiv.org/abs/hep-th/0110285}{{\tt hep-th/0110285}}].

\bibitem{Nicolis:2013sga}
A.~Nicolis, R.~Penco, F.~Piazza and R.~A. Rosen, \emph{{More on gapped
  Goldstones at finite density: More gapped Goldstones}},
  \href{http://dx.doi.org/10.1007/JHEP11(2013)055}{\emph{JHEP} {\bf 11} (2013)
  055}, [\href{https://arxiv.org/abs/1306.1240}{{\tt 1306.1240}}].

\bibitem{Endlich:2013vfa}
S.~Endlich, A.~Nicolis and R.~Penco, \emph{{Ultraviolet completion without
  symmetry restoration}},
  \href{http://dx.doi.org/10.1103/PhysRevD.89.065006}{\emph{Phys. Rev.} {\bf
  D89} (2014) 065006}, [\href{https://arxiv.org/abs/1311.6491}{{\tt
  1311.6491}}].

\bibitem{Alberte:2020eil}
L.~Alberte and A.~Nicolis, \emph{{Spontaneously broken boosts and the Goldstone
  continuum}},  \href{https://arxiv.org/abs/2001.06024}{{\tt 2001.06024}}.

\bibitem{Watanabe:2011ec}
H.~Watanabe and T.~Brauner, \emph{{On the number of Nambu-Goldstone bosons and
  its relation to charge densities}},
  \href{http://dx.doi.org/10.1103/PhysRevD.84.125013}{\emph{Phys. Rev.} {\bf
  D84} (2011) 125013}, [\href{https://arxiv.org/abs/1109.6327}{{\tt
  1109.6327}}].

\bibitem{PhysRevLett.108.251602}
H.~Watanabe and H.~Murayama, \emph{Unified description of nambu-goldstone
  bosons without lorentz invariance},
  \href{http://dx.doi.org/10.1103/PhysRevLett.108.251602}{\emph{Phys. Rev.
  Lett.} {\bf 108} (Jun, 2012) 251602}.

\bibitem{PhysRevLett.110.091601}
Y.~Hidaka, \emph{Counting rule for nambu-goldstone modes in nonrelativistic
  systems}, \href{http://dx.doi.org/10.1103/PhysRevLett.110.091601}{\emph{Phys.
  Rev. Lett.} {\bf 110} (Feb, 2013) 091601}.

\bibitem{Lenz:2020bxk}
J.~Lenz, L.~Pannullo, M.~Wagner, B.~Wellegehausen and A.~Wipf,
  \emph{{Inhomogeneous phases in the Gross-Neveu model in 1+1 dimensions at
  finite number of flavors}},  \href{https://arxiv.org/abs/2004.00295}{{\tt
  2004.00295}}.

\bibitem{Pannullo:2019prx}
L.~Pannullo, J.~Lenz, M.~Wagner, B.~Wellegehausen and A.~Wipf in \emph{{37th
  International Symposium on Lattice Field Theory}}, 9, Lattice investigation
  of the phase diagram of the 1+1 dimensional Gross-Neveu model at finite
  number of fermion flavors, 2019.
\newblock \href{https://arxiv.org/abs/1909.11513}{{\tt 1909.11513}}.

\bibitem{Watanabe:2014fva}
H.~Watanabe and H.~Murayama, \emph{{Effective Lagrangian for Nonrelativistic
  Systems}}, \href{http://dx.doi.org/10.1103/PhysRevX.4.031057}{\emph{Phys.
  Rev.} {\bf X4} (2014) 031057}, [\href{https://arxiv.org/abs/1402.7066}{{\tt
  1402.7066}}].

\bibitem{Watanabe:2019xul}
H.~Watanabe, \emph{{Counting Rules of Nambu-Goldstone Modes}},
  \href{http://dx.doi.org/10.1146/annurev-conmatphys-031119-050644}{\emph{Ann.
  Rev. Condensed Matter Phys.} {\bf 11} (2020) 169},
  [\href{https://arxiv.org/abs/1904.00569}{{\tt 1904.00569}}].

\bibitem{Minami:2018oxl}
Y.~Minami and Y.~Hidaka, \emph{{Spontaneous symmetry breaking and
  Nambu-Goldstone modes in dissipative systems}},
  \href{http://dx.doi.org/10.1103/PhysRevE.97.012130}{\emph{Phys. Rev.} {\bf
  E97} (2018) 012130}, [\href{https://arxiv.org/abs/1509.05042}{{\tt
  1509.05042}}].

\bibitem{Hidaka:2019irz}
Y.~Hidaka and Y.~Minami, \emph{{Spontaneous symmetry breaking and
  Nambu-Goldstone modes in open classical and quantum systems}},
  \href{https://arxiv.org/abs/1907.08241}{{\tt 1907.08241}}.

\bibitem{Landry:2019iel}
M.~J. Landry, \emph{{The coset construction for non-equilibrium systems}},
  \href{https://arxiv.org/abs/1912.12301}{{\tt 1912.12301}}.

\bibitem{PhysRevLett.75.4326}
J.~Toner and Y.~Tu, \emph{Long-range order in a two-dimensional dynamical
  $\mathrm{XY}$ model: How birds fly together},
  \href{http://dx.doi.org/10.1103/PhysRevLett.75.4326}{\emph{Phys. Rev. Lett.}
  {\bf 75} (Dec, 1995) 4326--4329}.

\bibitem{Donos:2019txg}
A.~Donos, D.~Martin, C.~Pantelidou and V.~Ziogas, \emph{{Hydrodynamics of
  broken global symmetries in the bulk}},
  \href{http://dx.doi.org/10.1007/JHEP10(2019)218}{\emph{JHEP} {\bf 10} (2019)
  218}, [\href{https://arxiv.org/abs/1905.00398}{{\tt 1905.00398}}].

\bibitem{Amoretti:2018tzw}
A.~Amoretti, D.~Areán, B.~Goutéraux and D.~Musso, \emph{{Universal relaxation
  in a holographic metallic density wave phase}},
  \href{http://dx.doi.org/10.1103/PhysRevLett.123.211602}{\emph{Phys. Rev.
  Lett.} {\bf 123} (2019) 211602},
  [\href{https://arxiv.org/abs/1812.08118}{{\tt 1812.08118}}].

\bibitem{Ammon:2020xyv}
M.~Ammon, M.~Baggioli, S.~Gray, S.~Grieninger and A.~Jain, \emph{{On the
  Hydrodynamic Description of Holographic Viscoelastic Models}},
  \href{https://arxiv.org/abs/2001.05737}{{\tt 2001.05737}}.

\bibitem{Baggioli:2020nay}
M.~Baggioli, \emph{{Are The Homogeneous Holographic Viscoelastic Models
  Quasicrystals ?}},  \href{https://arxiv.org/abs/2001.06228}{{\tt
  2001.06228}}.

\bibitem{Gromov:2020yoc}
A.~Gromov, A.~Lucas and R.~M. Nandkishore, \emph{{Fracton hydrodynamics}},
  \href{https://arxiv.org/abs/2003.09429}{{\tt 2003.09429}}.

\bibitem{doi:10.1143/JPSJ.75.111002}
Y.~Endoh and P.~Böni, \emph{Magnetic excitations in metallic ferro- and
  antiferromagnets},
  \href{http://dx.doi.org/10.1143/JPSJ.75.111002}{\emph{Journal of the Physical
  Society of Japan} {\bf 75} (2006) 111002},
  [\href{https://arxiv.org/abs/https://doi.org/10.1143/JPSJ.75.111002}{{\tt
  https://doi.org/10.1143/JPSJ.75.111002}}].

\bibitem{TONER2005170}
J.~Toner, Y.~Tu and S.~Ramaswamy, \emph{Hydrodynamics and phases of flocks},
  \href{http://dx.doi.org/https://doi.org/10.1016/j.aop.2005.04.011}{\emph{Annals
  of Physics} {\bf 318} (2005) 170 -- 244}.

\bibitem{Baggioli:2019rrs}
M.~Baggioli, \emph{{Applied Holography}}.
\newblock PhD thesis, Madrid, IFT, 2019.
\newblock \href{https://arxiv.org/abs/1908.02667}{{\tt 1908.02667}}.
\newblock 10.1007/978-3-030-35184-7.

\bibitem{Hartnoll:2008vx}
S.~A. Hartnoll, C.~P. Herzog and G.~T. Horowitz, \emph{{Building a Holographic
  Superconductor}},
  \href{http://dx.doi.org/10.1103/PhysRevLett.101.031601}{\emph{Phys. Rev.
  Lett.} {\bf 101} (2008) 031601}, [\href{https://arxiv.org/abs/0803.3295}{{\tt
  0803.3295}}].

\bibitem{Alberte:2017oqx}
L.~Alberte, M.~Ammon, A.~Jiménez-Alba, M.~Baggioli and O.~Pujolàs,
  \emph{{Holographic Phonons}},
  \href{http://dx.doi.org/10.1103/PhysRevLett.120.171602}{\emph{Phys. Rev.
  Lett.} {\bf 120} (2018) 171602},
  [\href{https://arxiv.org/abs/1711.03100}{{\tt 1711.03100}}].

\bibitem{Amado:2013xya}
I.~Amado, D.~Arean, A.~Jimenez-Alba, K.~Landsteiner, L.~Melgar and I.~S.
  Landea, \emph{{Holographic Type II Goldstone bosons}},
  \href{http://dx.doi.org/10.1007/JHEP07(2013)108}{\emph{JHEP} {\bf 07} (2013)
  108}, [\href{https://arxiv.org/abs/1302.5641}{{\tt 1302.5641}}].

\bibitem{PhysRevB.18.6245}
H.~Fukuyama and P.~A. Lee, \emph{Pinning and conductivity of two-dimensional
  charge-density waves in magnetic fields},
  \href{http://dx.doi.org/10.1103/PhysRevB.18.6245}{\emph{Phys. Rev. B} {\bf
  18} (Dec, 1978) 6245--6252}.

\bibitem{PhysRevB.46.3920}
B.~G.~A. Normand, P.~B. Littlewood and A.~J. Millis, \emph{Pinning and
  conductivity of a two-dimensional charge-density wave in a strong magnetic
  field}, \href{http://dx.doi.org/10.1103/PhysRevB.46.3920}{\emph{Phys. Rev. B}
  {\bf 46} (Aug, 1992) 3920--3934}.

\bibitem{PhysRev.175.2195}
M.~Gell-Mann, R.~J. Oakes and B.~Renner, \emph{Behavior of current divergences
  under
  ${\mathrm{su}}_{3}\ifmmode\times\else\texttimes\fi{}{\mathrm{su}}_{3}$},
  \href{http://dx.doi.org/10.1103/PhysRev.175.2195}{\emph{Phys. Rev.} {\bf 175}
  (Nov, 1968) 2195--2199}.

\bibitem{RevModPhys.60.1129}
G.~Gr\"uner, \emph{The dynamics of charge-density waves},
  \href{http://dx.doi.org/10.1103/RevModPhys.60.1129}{\emph{Rev. Mod. Phys.}
  {\bf 60} (Oct, 1988) 1129--1181}.

\bibitem{Delacretaz:2019wzh}
L.~V. Delacrétaz, B.~Goutéraux, S.~A. Hartnoll and A.~Karlsson, \emph{{Theory
  of collective magnetophonon resonance and melting of a field-induced Wigner
  solid}}, \href{http://dx.doi.org/10.1103/PhysRevB.100.085140}{\emph{Phys.
  Rev.} {\bf B100} (2019) 085140},
  [\href{https://arxiv.org/abs/1904.04872}{{\tt 1904.04872}}].

\bibitem{Chen2005QuantumSO}
Y.~P. Chen, \emph{Quantum Solids of Two Dimensional Electrons in Magnetic
  Fields}.
\newblock PhD thesis, Princeton U, Dept. of Electrical Engineering,
  \href{https://search.proquest.com/docview/305420029}{https://search.proquest.com/docview/305420029},
  2005.

\bibitem{PhysRevLett.89.176802}
P.~D. Ye, L.~W. Engel, D.~C. Tsui, R.~M. Lewis, L.~N. Pfeiffer and K.~West,
  \emph{Correlation lengths of the wigner-crystal order in a two-dimensional
  electron system at high magnetic fields},
  \href{http://dx.doi.org/10.1103/PhysRevLett.89.176802}{\emph{Phys. Rev.
  Lett.} {\bf 89} (Oct, 2002) 176802}.

\bibitem{PhysRevLett.93.206805}
Y.~P. Chen, R.~M. Lewis, L.~W. Engel, D.~C. Tsui, P.~D. Ye, Z.~H. Wang et~al.,
  \emph{Evidence for two different solid phases of two-dimensional electrons in
  high magnetic fields},
  \href{http://dx.doi.org/10.1103/PhysRevLett.93.206805}{\emph{Phys. Rev.
  Lett.} {\bf 93} (Nov, 2004) 206805}.

\bibitem{Chen_2006}
Y.~P. Chen, G.~Sambandamurthy, Z.~H. Wang, R.~M. Lewis, L.~W. Engel, D.~C. Tsui
  et~al., \emph{Melting of a 2d quantum electron solid in high magnetic field},
  \href{http://dx.doi.org/10.1038/nphys322}{\emph{Nature Physics} {\bf 2} (Jun,
  2006) 452–455}.

\bibitem{2007IJMPB..21.1379C}
Y.~P. {Chen}, G.~{Sambandamurthy}, L.~W. {Engel}, D.~C. {Tsui}, L.~N.
  {Pfeiffer} and K.~W. {West}, \emph{{Microwave Resonance Study of Melting in
  High Magnetic Field Wigner Solid}},
  \href{http://dx.doi.org/10.1142/S0217979207042860}{\emph{International
  Journal of Modern Physics B} {\bf 21} (Jan., 2007) 1379--1385}.

\bibitem{PhysRevB.89.075310}
B.-H. Moon, L.~W. Engel, D.~C. Tsui, L.~N. Pfeiffer and K.~W. West,
  \emph{Pinning modes of high-magnetic-field wigner solids with controlled
  alloy disorder},
  \href{http://dx.doi.org/10.1103/PhysRevB.89.075310}{\emph{Phys. Rev. B} {\bf
  89} (Feb, 2014) 075310}.

\bibitem{PhysRevB.92.035121}
B.-H. Moon, L.~W. Engel, D.~C. Tsui, L.~N. Pfeiffer and K.~W. West,
  \emph{Microwave pinning modes near landau filling $\ensuremath{\nu}=1$ in
  two-dimensional electron systems with alloy disorder},
  \href{http://dx.doi.org/10.1103/PhysRevB.92.035121}{\emph{Phys. Rev. B} {\bf
  92} (Jul, 2015) 035121}.

\bibitem{Baggioli:2014roa}
M.~Baggioli and O.~Pujolas, \emph{{Electron-Phonon Interactions,
  Metal-Insulator Transitions, and Holographic Massive Gravity}},
  \href{http://dx.doi.org/10.1103/PhysRevLett.114.251602}{\emph{Phys. Rev.
  Lett.} {\bf 114} (2015) 251602}, [\href{https://arxiv.org/abs/1411.1003}{{\tt
  1411.1003}}].

\bibitem{Alberte:2015isw}
L.~Alberte, M.~Baggioli, A.~Khmelnitsky and O.~Pujolas, \emph{{Solid Holography
  and Massive Gravity}},
  \href{http://dx.doi.org/10.1007/JHEP02(2016)114}{\emph{JHEP} {\bf 02} (2016)
  114}, [\href{https://arxiv.org/abs/1510.09089}{{\tt 1510.09089}}].

\bibitem{Andrade:2013gsa}
T.~Andrade and B.~Withers, \emph{{A simple holographic model of momentum
  relaxation}}, \href{http://dx.doi.org/10.1007/JHEP05(2014)101}{\emph{JHEP}
  {\bf 05} (2014) 101}, [\href{https://arxiv.org/abs/1311.5157}{{\tt
  1311.5157}}].

\bibitem{Alberte:2016xja}
L.~Alberte, M.~Baggioli and O.~Pujolas, \emph{{Viscosity bound violation in
  holographic solids and the viscoelastic response}},
  \href{http://dx.doi.org/10.1007/JHEP07(2016)074}{\emph{JHEP} {\bf 07} (2016)
  074}, [\href{https://arxiv.org/abs/1601.03384}{{\tt 1601.03384}}].

\bibitem{Baggioli:2016oqk}
M.~Baggioli and O.~Pujolas, \emph{{On holographic disorder-driven
  metal-insulator transitions}},
  \href{http://dx.doi.org/10.1007/JHEP01(2017)040}{\emph{JHEP} {\bf 01} (2017)
  040}, [\href{https://arxiv.org/abs/1601.07897}{{\tt 1601.07897}}].

\bibitem{Amoretti:2016cad}
A.~Amoretti, M.~Baggioli, N.~Magnoli and D.~Musso, \emph{{Chasing the cuprates
  with dilatonic dyons}},
  \href{http://dx.doi.org/10.1007/JHEP06(2016)113}{\emph{JHEP} {\bf 06} (2016)
  113}, [\href{https://arxiv.org/abs/1603.03029}{{\tt 1603.03029}}].

\bibitem{Baggioli:2016rdj}
M.~Baggioli, \emph{{Gravity, holography and applications to condensed matter}}.
\newblock PhD thesis, Barcelona U., 2016.
\newblock \href{https://arxiv.org/abs/1610.02681}{{\tt 1610.02681}}.

\bibitem{Baggioli:2016pia}
M.~Baggioli, B.~Goutéraux, E.~Kiritsis and W.-J. Li, \emph{{Higher derivative
  corrections to incoherent metallic transport in holography}},
  \href{http://dx.doi.org/10.1007/JHEP03(2017)170}{\emph{JHEP} {\bf 03} (2017)
  170}, [\href{https://arxiv.org/abs/1612.05500}{{\tt 1612.05500}}].

\bibitem{Baggioli:2017ojd}
M.~Baggioli and W.-J. Li, \emph{{Diffusivities bounds and chaos in holographic
  Horndeski theories}},
  \href{http://dx.doi.org/10.1007/JHEP07(2017)055}{\emph{JHEP} {\bf 07} (2017)
  055}, [\href{https://arxiv.org/abs/1705.01766}{{\tt 1705.01766}}].

\bibitem{Cremonini:2017qwq}
S.~Cremonini, A.~Hoover and L.~Li, \emph{{Backreacted DBI Magnetotransport with
  Momentum Dissipation}},
  \href{http://dx.doi.org/10.1007/JHEP10(2017)133}{\emph{JHEP} {\bf 10} (2017)
  133}, [\href{https://arxiv.org/abs/1707.01505}{{\tt 1707.01505}}].

\bibitem{Alberte:2017cch}
L.~Alberte, M.~Ammon, M.~Baggioli, A.~Jiménez and O.~Pujolàs, \emph{{Black
  hole elasticity and gapped transverse phonons in holography}},
  \href{http://dx.doi.org/10.1007/JHEP01(2018)129}{\emph{JHEP} {\bf 01} (2018)
  129}, [\href{https://arxiv.org/abs/1708.08477}{{\tt 1708.08477}}].

\bibitem{Blauvelt:2017koq}
E.~Blauvelt, S.~Cremonini, A.~Hoover, L.~Li and S.~Waskie, \emph{{Holographic
  model for the anomalous scalings of the cuprates}},
  \href{http://dx.doi.org/10.1103/PhysRevD.97.061901}{\emph{Phys. Rev. D} {\bf
  97} (2018) 061901}, [\href{https://arxiv.org/abs/1710.01326}{{\tt
  1710.01326}}].

\bibitem{Cremonini:2018kla}
S.~Cremonini, A.~Hoover, L.~Li and S.~Waskie, \emph{{Anomalous scalings of
  cuprate strange metals from nonlinear electrodynamics}},
  \href{http://dx.doi.org/10.1103/PhysRevD.99.061901}{\emph{Phys. Rev. D} {\bf
  99} (2019) 061901}, [\href{https://arxiv.org/abs/1812.01040}{{\tt
  1812.01040}}].

\bibitem{Baggioli:2018vfc}
M.~Baggioli and K.~Trachenko, \emph{{Low frequency propagating shear waves in
  holographic liquids}},
  \href{http://dx.doi.org/10.1007/JHEP03(2019)093}{\emph{JHEP} {\bf 03} (2019)
  093}, [\href{https://arxiv.org/abs/1807.10530}{{\tt 1807.10530}}].

\bibitem{Baggioli:2018nnp}
M.~Baggioli and K.~Trachenko, \emph{{Maxwell interpolation and close
  similarities between liquids and holographic models}},
  \href{http://dx.doi.org/10.1103/PhysRevD.99.106002}{\emph{Phys. Rev.} {\bf
  D99} (2019) 106002}, [\href{https://arxiv.org/abs/1808.05391}{{\tt
  1808.05391}}].

\bibitem{Baggioli:2019abx}
M.~Baggioli and S.~Grieninger, \emph{{Zoology of solid \& fluid holography —
  Goldstone modes and phase relaxation}},
  \href{http://dx.doi.org/10.1007/JHEP10(2019)235}{\emph{JHEP} {\bf 10} (2019)
  235}, [\href{https://arxiv.org/abs/1905.09488}{{\tt 1905.09488}}].

\bibitem{Esposito:2017qpj}
A.~Esposito, S.~Garcia-Saenz, A.~Nicolis and R.~Penco, \emph{{Conformal solids
  and holography}},
  \href{http://dx.doi.org/10.1007/JHEP12(2017)113}{\emph{JHEP} {\bf 12} (2017)
  113}, [\href{https://arxiv.org/abs/1708.09391}{{\tt 1708.09391}}].

\bibitem{Amoretti:2019kuf}
A.~Amoretti, D.~Areán, B.~Goutéraux and D.~Musso, \emph{{Gapless and gapped
  holographic phonons}},
  \href{http://dx.doi.org/10.1007/JHEP01(2020)058}{\emph{JHEP} {\bf 01} (2020)
  058}, [\href{https://arxiv.org/abs/1910.11330}{{\tt 1910.11330}}].

\bibitem{Amoretti:2017frz}
A.~Amoretti, D.~Areán, B.~Goutéraux and D.~Musso, \emph{{Effective
  holographic theory of charge density waves}},
  \href{http://dx.doi.org/10.1103/PhysRevD.97.086017}{\emph{Phys. Rev. D} {\bf
  97} (2018) 086017}, [\href{https://arxiv.org/abs/1711.06610}{{\tt
  1711.06610}}].

\bibitem{Amoretti:2019cef}
A.~Amoretti, D.~Areán, B.~Goutéraux and D.~Musso, \emph{{Diffusion and
  universal relaxation of holographic phonons}},
  \href{http://dx.doi.org/10.1007/JHEP10(2019)068}{\emph{JHEP} {\bf 10} (2019)
  068}, [\href{https://arxiv.org/abs/1904.11445}{{\tt 1904.11445}}].

\bibitem{Donos:2018kkm}
A.~Donos, J.~P. Gauntlett, T.~Griffin and V.~Ziogas, \emph{{Incoherent
  transport for phases that spontaneously break translations}},
  \href{http://dx.doi.org/10.1007/JHEP04(2018)053}{\emph{JHEP} {\bf 04} (2018)
  053}, [\href{https://arxiv.org/abs/1801.09084}{{\tt 1801.09084}}].

\bibitem{Donos:2019tmo}
A.~Donos and C.~Pantelidou, \emph{{Holographic transport and density waves}},
  \href{http://dx.doi.org/10.1007/JHEP05(2019)079}{\emph{JHEP} {\bf 05} (2019)
  079}, [\href{https://arxiv.org/abs/1903.05114}{{\tt 1903.05114}}].

\bibitem{Donos:2020viz}
A.~Donos, J.~P. Gauntlett and C.~Pantelidou, \emph{{Holographic Abrikosov
  Lattices}},  \href{https://arxiv.org/abs/2001.11510}{{\tt 2001.11510}}.

\bibitem{Gouteraux:2014hca}
B.~Goutéraux, \emph{{Charge transport in holography with momentum
  dissipation}}, \href{http://dx.doi.org/10.1007/JHEP04(2014)181}{\emph{JHEP}
  {\bf 04} (2014) 181}, [\href{https://arxiv.org/abs/1401.5436}{{\tt
  1401.5436}}].

\bibitem{Ammon:2019apj}
M.~Ammon, M.~Baggioli, S.~Gray and S.~Grieninger, \emph{{Longitudinal Sound and
  Diffusion in Holographic Massive Gravity}},
  \href{http://dx.doi.org/10.1007/JHEP10(2019)064}{\emph{JHEP} {\bf 10} (2019)
  064}, [\href{https://arxiv.org/abs/1905.09164}{{\tt 1905.09164}}].

\bibitem{Armas:2020bmo}
J.~Armas and A.~Jain, \emph{{Hydrodynamics for charge density waves and their
  holographic duals}},  \href{https://arxiv.org/abs/2001.07357}{{\tt
  2001.07357}}.

\bibitem{Armas:2019sbe}
J.~Armas and A.~Jain, \emph{{Viscoelastic hydrodynamics and holography}},
  \href{http://dx.doi.org/10.1007/JHEP01(2020)126}{\emph{JHEP} {\bf 01} (2020)
  126}, [\href{https://arxiv.org/abs/1908.01175}{{\tt 1908.01175}}].

\bibitem{1975JETPL..22...11L}
Y.~E. {Lozovik} and V.~I. {Yudson}, \emph{{Crystallization of a two-dimensional
  electron gas in a magnetic field}}, {\emph{Soviet Journal of Experimental and
  Theoretical Physics Letters} {\bf 22} (July, 1975) 11}.

\bibitem{PhysRevB.19.5211}
H.~Fukuyama, P.~M. Platzman and P.~W. Anderson, \emph{Two-dimensional electron
  gas in a strong magnetic field},
  \href{http://dx.doi.org/10.1103/PhysRevB.19.5211}{\emph{Phys. Rev. B} {\bf
  19} (May, 1979) 5211--5217}.

\bibitem{Kapustin:2012cr}
A.~Kapustin, \emph{{Remarks on nonrelativistic Goldstone bosons}},
  \href{https://arxiv.org/abs/1207.0457}{{\tt 1207.0457}}.

\bibitem{Moroz:2018noc}
S.~Moroz, C.~Hoyos, C.~Benzoni and D.~T. Son, \emph{{Effective field theory of
  a vortex lattice in a bosonic superfluid}},
  \href{http://dx.doi.org/10.21468/SciPostPhys.5.4.039}{\emph{SciPost Phys.}
  {\bf 5} (2018) 039}, [\href{https://arxiv.org/abs/1803.10934}{{\tt
  1803.10934}}].

\bibitem{PhysRevLett.79.1353}
C.-C. Li, L.~W. Engel, D.~Shahar, D.~C. Tsui and M.~Shayegan, \emph{Microwave
  conductivity resonance of two-dimensional hole system},
  \href{http://dx.doi.org/10.1103/PhysRevLett.79.1353}{\emph{Phys. Rev. Lett.}
  {\bf 79} (Aug, 1997) 1353--1356}.

\bibitem{PhysRevB.59.2120}
H.~A. Fertig, \emph{Electromagnetic response of a pinned wigner crystal},
  \href{http://dx.doi.org/10.1103/PhysRevB.59.2120}{\emph{Phys. Rev. B} {\bf
  59} (Jan, 1999) 2120--2141}.

\bibitem{PhysRevB.62.7553}
M.~M. Fogler and D.~A. Huse, \emph{Dynamical response of a pinned
  two-dimensional wigner crystal},
  \href{http://dx.doi.org/10.1103/PhysRevB.62.7553}{\emph{Phys. Rev. B} {\bf
  62} (Sep, 2000) 7553--7570}.

\bibitem{PhysRevB.65.035312}
R.~Chitra, T.~Giamarchi and P.~Le~Doussal, \emph{Pinned wigner crystals},
  \href{http://dx.doi.org/10.1103/PhysRevB.65.035312}{\emph{Phys. Rev. B} {\bf
  65} (Dec, 2001) 035312}.

\bibitem{Kim_2012}
Y.~Kim, Y.~Ma, A.~Imambekov, N.~G. Kalugin, A.~Lombardo, A.~C. Ferrari et~al.,
  \emph{Magnetophonon resonance in graphite: High-field raman measurements and
  electron-phonon coupling contributions},
  \href{http://dx.doi.org/10.1103/physrevb.85.121403}{\emph{Physical Review B}
  {\bf 85} (Mar, 2012) }.

\bibitem{Goerbig_2007}
M.~O. Goerbig, J.-N. Fuchs, K.~Kechedzhi and V.~I. Fal’ko,
  \emph{Filling-factor-dependent magnetophonon resonance in graphene},
  \href{http://dx.doi.org/10.1103/physrevlett.99.087402}{\emph{Physical Review
  Letters} {\bf 99} (Aug, 2007) }.

\bibitem{PhysRevB.88.165407}
C.~Qiu, X.~Shen, B.~Cao, C.~Cong, R.~Saito, J.~Yu et~al., \emph{Strong
  magnetophonon resonance induced triple g-mode splitting in graphene on
  graphite probed by micromagneto raman spectroscopy},
  \href{http://dx.doi.org/10.1103/PhysRevB.88.165407}{\emph{Phys. Rev. B} {\bf
  88} (Oct, 2013) 165407}.

\bibitem{PhysRevLett.110.227402}
Y.~Kim, J.~M. Poumirol, A.~Lombardo, N.~G. Kalugin, T.~Georgiou, Y.~J. Kim
  et~al., \emph{Measurement of filling-factor-dependent magnetophonon
  resonances in graphene using raman spectroscopy},
  \href{http://dx.doi.org/10.1103/PhysRevLett.110.227402}{\emph{Phys. Rev.
  Lett.} {\bf 110} (May, 2013) 227402}.

\bibitem{Ploch_2007}
D.~Ploch, E.~Sheregii, M.~Marchewka and G.~Tomaka, \emph{Magnetophonon
  resonance in multimode lattices and two-dimensional structures ({DQW})},
  \href{http://dx.doi.org/10.1088/1742-6596/92/1/012066}{\emph{Journal of
  Physics: Conference Series} {\bf 92} (dec, 2007) 012066}.

\bibitem{HAMAGUCHI199085}
C.~Hamaguchi and N.~Mori, \emph{Magnetophonon resonance in semiconductors},
  \href{http://dx.doi.org/https://doi.org/10.1016/0921-4526(90)90065-3}{\emph{Physica
  B: Condensed Matter} {\bf 164} (1990) 85 -- 96}.

\bibitem{Greenaway_2019}
M.~T. Greenaway, R.~Krishna~Kumar, P.~Kumaravadivel, A.~K. Geim and L.~Eaves,
  \emph{Magnetophonon spectroscopy of dirac fermion scattering by transverse
  and longitudinal acoustic phonons in graphene},
  \href{http://dx.doi.org/10.1103/physrevb.100.155120}{\emph{Physical Review B}
  {\bf 100} (Oct, 2019) }.

\bibitem{Kumaravadivel_2019}
P.~Kumaravadivel, M.~T. Greenaway, D.~Perello, A.~Berdyugin, J.~Birkbeck,
  J.~Wengraf et~al., \emph{Strong magnetophonon oscillations in extra-large
  graphene}, \href{http://dx.doi.org/10.1038/s41467-019-11379-3}{\emph{Nature
  Communications} {\bf 10} (Jul, 2019) }.

\bibitem{Ammon:2015wua}
M.~Ammon and J.~Erdmenger, \emph{{Gauge/gravity duality}: {Foundations and
  applications}}.
\newblock Cambridge University Press, Cambridge, 4, 2015.

\bibitem{Andrade:2019zey}
T.~Andrade, M.~Baggioli and O.~Pujolàs, \emph{{Linear viscoelastic dynamics in
  holography}},
  \href{http://dx.doi.org/10.1103/PhysRevD.100.106014}{\emph{Phys. Rev.} {\bf
  D100} (2019) 106014}, [\href{https://arxiv.org/abs/1903.02859}{{\tt
  1903.02859}}].

\bibitem{Baggioli:2019elg}
M.~Baggioli, V.~C. Castillo and O.~Pujolas, \emph{{Scale invariant solids}},
  \href{https://arxiv.org/abs/1910.05281}{{\tt 1910.05281}}.

\bibitem{Baggioli:2019mck}
M.~Baggioli, S.~Grieninger and H.~Soltanpanahi, \emph{{Nonlinear Oscillatory
  Shear Tests in Viscoelastic Holography}},
  \href{http://dx.doi.org/10.1103/PhysRevLett.124.081601}{\emph{Phys. Rev.
  Lett.} {\bf 124} (2020) 081601},
  [\href{https://arxiv.org/abs/1910.06331}{{\tt 1910.06331}}].

\bibitem{Baggioli:2018bfa}
M.~Baggioli and A.~Buchel, \emph{{Holographic Viscoelastic Hydrodynamics}},
  \href{http://dx.doi.org/10.1007/JHEP03(2019)146}{\emph{JHEP} {\bf 03} (2019)
  146}, [\href{https://arxiv.org/abs/1805.06756}{{\tt 1805.06756}}].

\bibitem{Ammon:2019wci}
M.~Ammon, M.~Baggioli and A.~Jiménez-Alba, \emph{{A Unified Description of
  Translational Symmetry Breaking in Holography}},
  \href{http://dx.doi.org/10.1007/JHEP09(2019)124}{\emph{JHEP} {\bf 09} (2019)
  124}, [\href{https://arxiv.org/abs/1904.05785}{{\tt 1904.05785}}].

\bibitem{Kim:2015wba}
K.-Y. Kim, K.~K. Kim, Y.~Seo and S.-J. Sin, \emph{{Thermoelectric
  Conductivities at Finite Magnetic Field and the Nernst Effect}},
  \href{http://dx.doi.org/10.1007/JHEP07(2015)027}{\emph{JHEP} {\bf 07} (2015)
  027}, [\href{https://arxiv.org/abs/1502.05386}{{\tt 1502.05386}}].

\bibitem{PhysRevB.22.2514}
A.~Zippelius, B.~I. Halperin and D.~R. Nelson, \emph{Dynamics of
  two-dimensional melting},
  \href{http://dx.doi.org/10.1103/PhysRevB.22.2514}{\emph{Phys. Rev. B} {\bf
  22} (Sep, 1980) 2514--2541}.

\bibitem{chaikin2000principles}
P.~Chaikin and T.~Lubensky, \emph{Principles of Condensed Matter Physics}.
\newblock Cambridge University Press, 2000.

\bibitem{PhysRevA.6.2401}
P.~C. Martin, O.~Parodi and P.~S. Pershan, \emph{Unified hydrodynamic theory
  for crystals, liquid crystals, and normal fluids},
  \href{http://dx.doi.org/10.1103/PhysRevA.6.2401}{\emph{Phys. Rev. A} {\bf 6}
  (Dec, 1972) 2401--2420}.

\bibitem{Delacretaz:2017zxd}
L.~V. Delacrétaz, B.~Goutéraux, S.~A. Hartnoll and A.~Karlsson, \emph{{Theory
  of hydrodynamic transport in fluctuating electronic charge density wave
  states}}, \href{http://dx.doi.org/10.1103/PhysRevB.96.195128}{\emph{Phys.
  Rev.} {\bf B96} (2017) 195128}, [\href{https://arxiv.org/abs/1702.05104}{{\tt
  1702.05104}}].

\bibitem{Ge:2008ak}
X.-H. Ge, Y.~Matsuo, F.-W. Shu, S.-J. Sin and T.~Tsukioka, \emph{{Density
  Dependence of Transport Coefficients from Holographic Hydrodynamics}},
  \href{http://dx.doi.org/10.1143/PTP.120.833}{\emph{Prog. Theor. Phys.} {\bf
  120} (2008) 833--863}, [\href{https://arxiv.org/abs/0806.4460}{{\tt
  0806.4460}}].

\bibitem{Amoretti:2017axe}
A.~Amoretti, D.~Areán, B.~Goutéraux and D.~Musso, \emph{{DC resistivity of
  quantum critical, charge density wave states from gauge-gravity duality}},
  \href{http://dx.doi.org/10.1103/PhysRevLett.120.171603}{\emph{Phys. Rev.
  Lett.} {\bf 120} (2018) 171603},
  [\href{https://arxiv.org/abs/1712.07994}{{\tt 1712.07994}}].

\bibitem{PhysRevB.100.085140}
L.~V. Delacr\'etaz, B.~Gout\'eraux, S.~A. Hartnoll and A.~Karlsson,
  \emph{Theory of collective magnetophonon resonance and melting of a
  field-induced wigner solid},
  \href{http://dx.doi.org/10.1103/PhysRevB.100.085140}{\emph{Phys. Rev. B} {\bf
  100} (Aug, 2019) 085140}.

\bibitem{Hartnoll:2007ip}
S.~A. Hartnoll and C.~P. Herzog, \emph{{Ohm's Law at strong coupling: S duality
  and the cyclotron resonance}},
  \href{http://dx.doi.org/10.1103/PhysRevD.76.106012}{\emph{Phys. Rev.} {\bf
  D76} (2007) 106012}, [\href{https://arxiv.org/abs/0706.3228}{{\tt
  0706.3228}}].

\bibitem{Ammon:2016fru}
M.~Ammon, S.~Grieninger, A.~Jimenez-Alba, R.~P. Macedo and L.~Melgar,
  \emph{{Holographic quenches and anomalous transport}},
  \href{http://dx.doi.org/10.1007/JHEP09(2016)131}{\emph{JHEP} {\bf 09} (2016)
  131}, [\href{https://arxiv.org/abs/1607.06817}{{\tt 1607.06817}}].

\bibitem{Grieninger:2017jxz}
S.~Grieninger, \emph{{Holographic quenches and anomalous transport}},  Master's
  thesis, Jena U., TPI, 2016,
  \href{https://arxiv.org/abs/1711.08422}{[1711.08422]}.

\bibitem{Ammon:2017ded}
M.~Ammon, M.~Kaminski, R.~Koirala, J.~Leiber and J.~Wu, \emph{{Quasinormal
  modes of charged magnetic black branes \& chiral magnetic transport}},
  \href{http://dx.doi.org/10.1007/JHEP04(2017)067}{\emph{JHEP} {\bf 04} (2017)
  067}, [\href{https://arxiv.org/abs/1701.05565}{{\tt 1701.05565}}].

\bibitem{Hayata:2014yga}
T.~Hayata and Y.~Hidaka, \emph{{Dispersion relations of Nambu-Goldstone modes
  at finite temperature and density}},
  \href{http://dx.doi.org/10.1103/PhysRevD.91.056006}{\emph{Phys. Rev.} {\bf
  D91} (2015) 056006}, [\href{https://arxiv.org/abs/1406.6271}{{\tt
  1406.6271}}].

\bibitem{valentinis2020optical}
D.~Valentinis, \emph{Optical signatures of shear collective modes in strongly
  interacting fermi liquids}, {\emph{arXiv preprint arXiv:2003.06619} (2020) }.

\bibitem{HoyosBadajoz:2010kd}
C.~Hoyos-Badajoz, A.~O'Bannon and J.~M.~S. Wu, \emph{{Zero Sound in Strange
  Metallic Holography}},
  \href{http://dx.doi.org/10.1007/JHEP09(2010)086}{\emph{JHEP} {\bf 09} (2010)
  086}, [\href{https://arxiv.org/abs/1007.0590}{{\tt 1007.0590}}].

\bibitem{Krikun:2018agd}
A.~Romero-Bermúdez, A.~Krikun, K.~Schalm and J.~Zaanen, \emph{{Anomalous
  attenuation of plasmons in strange metals and holography}},
  \href{http://dx.doi.org/10.1103/PhysRevB.99.235149}{\emph{Phys. Rev.} {\bf
  B99} (2019) 235149}, [\href{https://arxiv.org/abs/1812.03968}{{\tt
  1812.03968}}].

\bibitem{Baggioli:2019sio}
M.~Baggioli, U.~Gran and M.~Tornsö, \emph{{Transverse Collective Modes in
  Interacting Holographic Plasmas}},
  \href{http://dx.doi.org/10.1007/JHEP04(2020)106}{\emph{JHEP} {\bf 04} (2020)
  106}, [\href{https://arxiv.org/abs/1912.07321}{{\tt 1912.07321}}].

\bibitem{Baumgartner:2017kme}
A.~Baumgartner, A.~Karch and A.~Lucas, \emph{{Magnetoresistance in relativistic
  hydrodynamics without anomalies}},
  \href{http://dx.doi.org/10.1007/JHEP06(2017)054}{\emph{JHEP} {\bf 06} (2017)
  054}, [\href{https://arxiv.org/abs/1704.01592}{{\tt 1704.01592}}].

\bibitem{Donos:2014cya}
A.~Donos and J.~P. Gauntlett, \emph{{Thermoelectric DC conductivities from
  black hole horizons}},
  \href{http://dx.doi.org/10.1007/JHEP11(2014)081}{\emph{JHEP} {\bf 11} (2014)
  081}, [\href{https://arxiv.org/abs/1406.4742}{{\tt 1406.4742}}].

\bibitem{Blake:2015ina}
M.~Blake, A.~Donos and N.~Lohitsiri, \emph{{Magnetothermoelectric Response from
  Holography}}, \href{http://dx.doi.org/10.1007/JHEP08(2015)124}{\emph{JHEP}
  {\bf 08} (2015) 124}, [\href{https://arxiv.org/abs/1502.03789}{{\tt
  1502.03789}}].

\bibitem{Hartnoll:2007ih}
S.~A. Hartnoll, P.~K. Kovtun, M.~Muller and S.~Sachdev, \emph{{Theory of the
  Nernst effect near quantum phase transitions in condensed matter, and in
  dyonic black holes}},
  \href{http://dx.doi.org/10.1103/PhysRevB.76.144502}{\emph{Phys. Rev.} {\bf
  B76} (2007) 144502}, [\href{https://arxiv.org/abs/0706.3215}{{\tt
  0706.3215}}].

\bibitem{Cremonini:2019fzz}
S.~Cremonini, L.~Li and J.~Ren, \emph{{Spectral Weight Suppression and Fermi
  Arc-like Features with Strong Holographic Lattices}},
  \href{http://dx.doi.org/10.1007/JHEP09(2019)014}{\emph{JHEP} {\bf 09} (2019)
  014}, [\href{https://arxiv.org/abs/1906.02753}{{\tt 1906.02753}}].

\bibitem{Cremonini:2017usb}
S.~Cremonini, L.~Li and J.~Ren, \emph{{Intertwined Orders in Holography: Pair
  and Charge Density Waves}},
  \href{http://dx.doi.org/10.1007/JHEP08(2017)081}{\emph{JHEP} {\bf 08} (2017)
  081}, [\href{https://arxiv.org/abs/1705.05390}{{\tt 1705.05390}}].

\bibitem{Cremonini:2016rbd}
S.~Cremonini, L.~Li and J.~Ren, \emph{{Holographic Pair and Charge Density
  Waves}}, \href{http://dx.doi.org/10.1103/PhysRevD.95.041901}{\emph{Phys.
  Rev.} {\bf D95} (2017) 041901}, [\href{https://arxiv.org/abs/1612.04385}{{\tt
  1612.04385}}].

\bibitem{Cai:2017qdz}
R.-G. Cai, L.~Li, Y.-Q. Wang and J.~Zaanen, \emph{{Intertwined Order and
  Holography: The Case of Parity Breaking Pair Density Waves}},
  \href{http://dx.doi.org/10.1103/PhysRevLett.119.181601}{\emph{Phys. Rev.
  Lett.} {\bf 119} (2017) 181601},
  [\href{https://arxiv.org/abs/1706.01470}{{\tt 1706.01470}}].

\bibitem{Andrade:2015iyf}
T.~Andrade and A.~Krikun, \emph{{Commensurability effects in holographic
  homogeneous lattices}},
  \href{http://dx.doi.org/10.1007/JHEP05(2016)039}{\emph{JHEP} {\bf 05} (2016)
  039}, [\href{https://arxiv.org/abs/1512.02465}{{\tt 1512.02465}}].

\bibitem{Andrade:2017leb}
T.~Andrade and A.~Krikun, \emph{{Commensurate lock-in in holographic
  non-homogeneous lattices}},
  \href{http://dx.doi.org/10.1007/JHEP03(2017)168}{\emph{JHEP} {\bf 03} (2017)
  168}, [\href{https://arxiv.org/abs/1701.04625}{{\tt 1701.04625}}].

\bibitem{Andrade:2017ghg}
T.~Andrade, A.~Krikun, K.~Schalm and J.~Zaanen, \emph{{Doping the holographic
  Mott insulator}},
  \href{http://dx.doi.org/10.1038/s41567-018-0217-6}{\emph{Nature Phys.} {\bf
  14} (2018) 1049--1055}, [\href{https://arxiv.org/abs/1710.05791}{{\tt
  1710.05791}}].

\bibitem{Esposito:2020wsn}
A.~Esposito, R.~Krichevsky and A.~Nicolis, \emph{{Solidity without
  inhomogeneity: Perfectly homogeneous, weakly coupled, UV-complete solids}},
  \href{https://arxiv.org/abs/2004.11386}{{\tt 2004.11386}}.

\bibitem{coming}
M.~Ammon, S.~Grieninger, J.~Hernandez, M.~Kaminski, R.~Koirala, J.~Leiber
  et~al., \emph{{Chiral transport in strong magnetic fields from hydrodynamics
  \& holography}},  \href{https://arxiv.org/abs/to appear}{{\tt to appear}}.

\bibitem{Skenderis:2002wp}
K.~Skenderis, \emph{{Lecture notes on holographic renormalization}},
  \href{http://dx.doi.org/10.1088/0264-9381/19/22/306}{\emph{Class. Quant.
  Grav.} {\bf 19} (2002) 5849--5876},
  [\href{https://arxiv.org/abs/hep-th/0209067}{{\tt hep-th/0209067}}].

\bibitem{dbt_mods}
S.~Grieninger, \emph{Non-equilibrium dynamics in holography}.
\newblock PhD thesis, Jena, 2020, to appear.

\bibitem{Boyd1989ChebyshevAF}
J.~P. Boyd, \emph{Chebyshev and Fourier Spectral Methods}.
\newblock Dover Publications Inc., 2003.

\bibitem{Grandclement:2007sb}
P.~Grandclement and J.~Novak, \emph{{Spectral methods for numerical
  relativity}}, \href{http://dx.doi.org/10.12942/lrr-2009-1}{\emph{Living Rev.
  Rel.} {\bf 12} (2009) 1}, [\href{https://arxiv.org/abs/0706.2286}{{\tt
  0706.2286}}].

\bibitem{canuto2007spectral}
C.~Canuto, M.~Hussaini, A.~Quarteroni and T.~Zang, \emph{Spectral Methods:
  Fundamentals in Single Domains}.
\newblock Scientific Computation. Springer Berlin Heidelberg, 2007.

\bibitem{trefethen2000spectral}
L.~Trefethen, \emph{Spectral Methods in MATLAB}.
\newblock Software, Environments, and Tools. Society for Industrial and Applied
  Mathematics, 2000.

\bibitem{dutykh2016brief}
D.~Dutykh, \emph{A brief introduction to pseudo-spectral methods: application
  to diffusion problems},  2016.

\end{thebibliography}\endgroup
\end{document}